\documentclass[11pt,a4paper]{article}
\usepackage{amssymb}

\usepackage{graphicx}
\usepackage{amsmath}
\usepackage{makeidx}
\usepackage{graphicx}
\usepackage{graphicx}
\usepackage{indentfirst}

\renewcommand{\theequation}{\thesection.\arabic{equation}}

\newcounter{resultnum}[section]

\newcounter{conclusionnum}[section]

\newcounter{conditionnum}[section]

\newcounter{conjecturenum}[section]

\newcounter{examplenum}[section]

\newcounter{exercisenum}[section]

\newcounter{lemmanum}[section]

\newcounter{notationnum}[section]

\newcounter{theoremnum}[section]

\newcounter{definitionnum}[section]

\newcounter{corollarynum}[section]

\newcounter{remarknum}[section]

\newcounter{propositionnum}[section]

\newcounter{acknowledgementnum}[section]

\newcounter{algorithmnum}[section]

\newcounter{axiomnum}[section]

\newcounter{casenum}[section]

\newcounter{claimnum}[section]

\newcounter{summarynum}[section]

\newcounter{problemnum}[section]

\begin{document}

\title{Parametric Nonholonomic Frame Transforms\\
and Exact Solutions in Gravity }
\date{September 19, 2007}
\author{Sergiu I. Vacaru\thanks{%
sergiu$_{-}$vacaru@yahoo.com, svacaru@fields.utoronto.ca } \\
{\quad} \\
\textsl{The Fields Institute for Research in Mathematical Science} \\
\textsl{222 College Street, 2d Floor, } \textsl{Toronto \ M5T 3J1, Canada} }
\maketitle

\begin{abstract}
A generalized geometric method is developed for constructing exact
solutions of gravitational field equations in Einstein theory and
generalizations. First, we apply the formalism of nonholonomic frame
deformations (formally considered for nonholonomic manifolds and Finsler
spaces) when the gravitational field equations transform into systems of
nonlinear partial differential equations which can be integrated in general
form. The new classes of solutions are defined by generic off--diagonal
metrics depending on integration functions on one, two and three (or three
and four) variables if we consider four (or five) dimensional spacetimes.
Second, we use a general scheme when one (two) parameter families of exact
solutions are defined by any source--free solutions of Einstein's equations
with one (two) Killing vector field(s). A successive iteration procedure
results in new classes of solutions characterized by an infinite number of
parameters for a non--Abelian group involving arbitrary functions on one
variable. Five classes of exact off--diagonal solutions are constructed
 in vacuum Einstein and in string gravity describing solitonic
pp--wave interactions. We explore possible physical consequences of such
solutions derived from primary Schwarzschild or pp--wave metrics.
\vskip3pt


\end{abstract}


\section{ Introduction}

Even through, a large number of exact solutions were found
in various models of
gravity theory \cite{kramer,bic,vsgg}, there are available only a few
general methods for generating new solutions from a given metric describing
a physical situation to certain new physical properties and geometric
configurations. In quantum field theory, (although approximated) some
 methods where formulated by using the formalism of Green's
functions, or quantum integrals, the solutions are constructed to represent
a linear or nonlinear prescribed physical situation. Perhaps it is unlikely
that similar computation techniques can be elaborated in general form in
gravity theories. Nevertheless, such approaches where developed when
new classes of exact solutions are constructed following some general
geometric/ group principles and ideas \cite%
{geroch1,geroch2,vhep,vt,vs,vesnc,vp}. Although many of the solutions
resulting from such methods have no obvious physical interpretation, one can
be formulated some criteria selecting explicit classes of solutions with
prescribed symmetries and physical properties.

The technique proposed in works \cite{geroch1,geroch2} generates exact
source--free solutions of Einstein equations and treats spacetimes having
one, or two, Killing vectors. The scheme introduced in \cite{geroch1} begins
with any source--free solution of Einstein's equation with a Killing vector
and defines a one parameter family (possessing a nontrivial group structure)
of exact solutions. Even through starting from a quite simple solution like the
Schwarzschild one, the resulting metrics were considered too sophisticated
to admit any simple interpretation. In the second work \cite{geroch2}, the
author proved that the case of two Killing vectors\footnote{%
for instance, a Weyl solution, which is a space with two commuting Killing
vectors} is more appropriate for physical interpretation. In  such a case,
one must specify an arbitrary curve (up to parametrization) on a
three--dimensional vector spaces associated to an exact four dimensional
solution. The so--called parametric  transform (forming a
non--Abelian group) was defined generating, from a single solution, a
family of new
solutions involving two arbitrary functions of one variable. A successive
iteration of such transforms results in a class of exact solutions
characterized by an infinite number of parameters.

Almost 20 years after formulation of the parametric method, a new
approach to constructing exact solutions in gravity (the so--called,
anholonomic frame method) was proposed and developed in works \cite%
{vhep,vt,vs,vesnc,vsgg,vp,vap2,vjmp1,vap1,vnp1,vlodz}. One of its
distinguished properties is that the existence of Killing symmetry is not
crucial for definition of moving anholonomic frames\footnote{%
we shall use both equivalent terms anholonomic and/or nonholonomic; here we
note that a local basis is nonholonomic if its vectors do not commute like
for the coordinate bases but satisfy some anholonomy relations, see section 2%
}. The first publication \cite{vhep} contained certain examples of generic
off--diagonal exact solutions, in three and four dimensional gravity (in
brief, we shall write respectively 3D and 4D). The idea was to take any
well--known exact solution (of black hole, instanton or monopole ... type)
which
can be diagonalized with respect to a corresponding coordinate frame and
then to deform it by introducing generic off--diagonal metric terms\footnote{%
parametrizing, for instance, certain 2D or 3D solitonic waves; such metrics
can not be diagonalized by coordinate transforms} in a manner that generates
 new classes of exact solutions. We note that one could be
constructed source--free solutions and more general ones with matter fields,
or with string gravity corrections, when extra dimensions and nontrivial
torsion fields are considered. Various classes of such solutions were
analyzed \cite{vt,vs,vesnc,vsgg,vp} (they describe nonholonomic deformations
of Taub -- NUT spaces, locally anisotropic wormholes, black ellipsoid and
toroidal configurations, self--consistent interactions of (non)commutative
Dirac and/or solitonic gravitational waves...).

The anholonomic frame method works as follows. We take a 'primary' metric in
a 5D (or 4D/ 3D) spacetime. The constructions are more simple if this metric
is at least a conformal transform of a well known exact solution with
diagonal metric. As a matter of principle, we can consider that the primary
metric is a general one on a Riemann--Cartan manifold, not being obligatory
a solution of gravitational field equations. By anholonomic frame (vielbein;
or vierbein/ tetradic, in 4D) deformations, the primary metric and linear
connection structures are transformed into the corresponding 'target' ones
for which the Einstein equations are exactly integrable. We note that the
target metrics are generic off--diagonal, depend on classes of integration
constants and arbitrary functions on one, two and three/four local
coordinates (respectively for 4D/5D spacetimes).

It should be emphasized that the nonholonomic deformations induce nontrivial
torsion structures, which can be effectively exploited in string/ brane
gravity where the antisymmetric torsion plays a corn--stone role. The method
can be applied in a straightforward form to some general classes of generic
off--diagonal metrics and linear connections with nontrivial torsion. Haven
being constructed certain classes of exact solutions with integral varieties
parametrized by a some classes of integration functions, it is possible to
constrain the set of such functions when the so--called canonical
d--connection (with nontrivial torsion) transforms into the Levi--Civita
connection (with vanishing torsion). This way, some more general classes of
'nonholonomic' solutions can be restricted to define exact solutions in
Einstein gravity. Nevertheless, even the metrics defining Einstein spaces
depend on various types of integration functions and possess general
nonlinear symmetries.

Sure, many of the off--diagonal solutions generated following the
anholonomic frame method have no obvious physical interpretation. It is
quite a cumbersome task to define the nonlinear symmetries of such spacetimes
and decide what kind of physical interpretation may be adequate.
Nevertheless, if any initial physical situations were given, it is possible
to analyze if such nonholonomic deformations can preserve certain similarity
and admit nonlinear superpositions and any new prescribed properties. We
distinguish here five special cases (preliminary analyzed in our previous
works): 1) The generic off--diagonal metric terms effectively polarize the
constants of the primary metric (for instance, the point mass and/or
electric, or cosmological constants). 2) The existing horizons (if any) are
slightly deformed, for instance, from the spherical to an ellipsoidal
symmetry. 3) The symmetry of former solutions can be broken in a spacetime
region. 4) One can be changing of topological structure but certain former
physical properties and analogy are preserved. 5) The primary solution is
imbedded into a nontrivial background (for instance, consisting from a
superposition of solitonic and pp--waves).

A rigorous analysis is necessary in order to state what kind of
prescribed spacetimes can be generated by a class of nonholonomic transforms
from a primary metric. Nevertheless, at least for certain classes of 'small'
smooth deformations, we can conclude that the singular properties (of the
curvature scalar and tensor) and topology are preserved even additional
nonlinear interactions are present and the symmetries are deformed. In such
cases, it is possible to preserve the former physical interpretation but
with modified constants, deformed horizons and nontrivial backgrounds.

In order to decide if a new class of generic off--diagonal solutions have
nontrivial physical limits to the Einstein gravity, we must take into
account various type of black hole 'uniqueness' theorems and cosmic
censorship criteria \cite{isr,cart,rob,heu,haw}. The strategy to deal with
such solutions is to chose certain type of integration functions and
boundary conditions when 'far away' from the 'slightly' deformed horizons
and finite spacetime regions with nonlinear polarizations of constants and
nontrivial backgrounds the Minkowski asymptotic and spherical topology hold
true. Here we note that the off--diagonal metric terms, for vacuum
solutions, may model certain effective matter field interactions (like in
the Kaluza--Klein gravity but, in our case, without linearization for
inducing electromagnetic fields and compactification on extra dimension
coordinate). In this case, there are introduced the so--called geometric
spacetime distorsions \cite{vbe1,vbe2} (like matter field distorsions for
black holes \cite{mys,gerh,fair}) and the restrictions of the uniqueness
theorems and censorship criteria may be avoided. In modern gravity, the
solutions with possible violation of mentioned type theorems and criteria
and even of local Lorentz symmetry also present a special interest.

The parametric method can be applied if the (peseudo) Riemannian spacetime
possesses at least one Killing vector. In the case when there are two Killing
vectors, one can be defined an iteration procedure of generating classes of
exact solutions involving arbitrary functions on spacetime coordinates
labelled by an infinite family of parameters (such parameters are not
spacetime coordinates). The set of such parameters can be treated as a
specific space of internal symmetries of the solutions of vacuum Einstein
equations but there is not clear the complete physical significance of such
symmetries. For application of the anholonomic frame method, it is not
crucial that the primary metric is a solution with Killing symmetries. The
most important point is to define a nonholonomic frame deformation to a
special type, off--diagonal, ansatz solving the Einstein equations for a
connection deforming 'minimally' the Livi--Civita connection in order to
include the contribution of anholonomy coefficients. Certain constraint on
integral varieties of such solutions allow to generate usual Einstein spaces
and their generalizations with matter sources and, for instance, string
contributions. Such classes of solutions, in general, are not characterized
by a a group of parameters. Nevertheless, a number of commutative and
noncommutative, Lie algebroid\footnote{%
Lie algebroids can be considered as certain generalizations of spaces with
generalized Lie algebra symmetries when, roughly speaking, the structure
constants depend on basic manifold coordinates and certain singular maps,
anchors, are introduced into consideration, see Ref. \cite{valg} for a
detailed discussion on definition of such geometric structures as exact
solutions in gravity} and Clifford algebroid or other nonlinear symmetries
can be prescribed for such metrics.

Because the target off--diagonal metrics generated by applying the
anholonomic frame method positively do not depend on one spacetime
coordinate (but certain coefficients of metrics depend, for instance, on
four/three coordinates of 5D/4D spacetimes), for sure, the generated
nonolonomic spacetimes\footnote{
A manifold is nonholonomic if it is provided with a nonintegrable
distribution, for instance, with a preferred frame structure with associated
nonlinear connection (such spacetimes are also called locally anisotropic),
see Refs. \cite{vesnc,bejf,esv} for basic references and applications in
modern gravity.} possesses a Killing vector symmetry. In this case, a
parametric transform can be applied after an anholonomic frame generation of
a vacuum spacetime and the resulting vacuum solution will be characterized
both by nonholonomic and parametric group structures. If one of the primary
or target solutions is at least a conformal transform, or a small
nonholonomic deformation of a well known exact solution of physical
importance, we can formulate the criteria when certain prescribed
geometrical and physical properties are preserved or may be induced. For
instance, we can generate black hole, wormhole,... solutions with locally
anisotropic parameters, deformed horizons and propagating in nontrivial
solitonic/pp--wave backgrounds when the physical parameters and geometrical
objects are parametrized by an infinite number of group parameters and
possess generalized (for instance, Lie algebroid) symmetries.

The goals of this work is to show how new classes of exact solutions can be
constructed by superpositions of the parametric and anholonomic frame
transform and to carry out a program of extracting physically valuable
solutions. We shall emphasize the possibility to select physically important
solutions in Einstein and string gravity.

The paper has the following structure: in Sec. 2,
 we outline two geometric methods of constructing exact
solutions in modern gravity. We begin with new geometric conventions
necessary for a common description both of the parametric and anholonomic
frame methods. The constructions distinguish the approaches related to
Killing vectors and to the formalism of anholonomic frames with associated
nonlinear connection structure. Then we formulate the techniques of
constructing solutions for five and four dimensional (generic off--diagonal)
metric ansatz and analyze the conditions when Einstein foliations can be
defined by such solutions.

Section 3 is devoted to the main goal of this paper:\ elaboration of a
unified formalism both for the parametric and anholonomic frame methods of
constructing solutions in modern gravity. We start with the geometry of
nonholonomic deformations of metrics resulting in exact solutions. Then we
study superpositions of parametric transforms and anholonomic maps. Finally,
there are proposed two alternative constructions when a class of solutions
generated by the parametric method is deformed nonholonomically to other
ones and, inversely, when the parametric transform is applied to
nonholonomic Einstein spacetimes.

In Sec. 4, we construct five classes of exact solutions of vacuum Einstein
equations and with sources from string gravity, generated by superpositions
of nonholonomic frame and parametric transforms. We briefly explain the
computations and emphasize the conditions when physically valuable solutions
can be extracted. In explicit form, such metrics are defined by
superpositions of solitonic pp--waves interacting in nonholonomically
deformed black hole solutions.

We conclude the paper with some comments and remarks in Sec. 5. The Appendix
contains some necessary formulas on curvature, Ricci and Einstein tensors
for the so--called canonical distinguished connection. An ansatz with
antisymmetric torsion in string gravity is considered and a general solution
for nonholonomically constrained components of Einstein equations is
considered.

\section{Outline of the Methods}

In this section, we outline and compare both the parametric and the
anholonomic frame methods of constructing exact solutions in gravity, see
details in Refs. \cite{geroch1,geroch2} and \cite{vesnc,valg,vsgg}.

\subsection{Geometric conventions}

Let us begin with some general notations to be used in this work. We
consider a spacetime as a manifolds of necessary smooth class $V$ of
dimension $n+m,$ with $n\geq 2$ and $m\geq 1$ (the meaning of conventional
splitting of dimensions will be explained in section \ref{sah}), provided
with a metric
\begin{equation}
g=g_{\alpha \beta }e^{\alpha }\otimes e^{\alpha }
\end{equation}%
of any (pseudo) Euclidean signature and a linear connection $D=\{\Gamma _{\
\beta \gamma }^{\alpha }e^{\beta }\}$ satisfying the metric compatibility
condition $Dg=0.$\footnote{%
in this work, the Einstein's summation rule on repeating ''upper--lower''
indices will be applied if the contrary will not be stated} The components
of geometrical objects, for instance, $g_{\alpha \beta }$ and $\Gamma _{\
\beta \gamma }^{\alpha },$ are defined with respect to a local base (frame) $%
e_{\alpha }$ and its dual base (co--base, or co--frame) $e^{\alpha }$ for
which $e_{\alpha }\rfloor \ e^{\beta }=\delta _{\alpha }^{\beta },$ where ''$%
\rfloor "$ denotes the interior product induced by $g$ and $\delta _{\alpha
}^{\beta }$ is the Kronecker symbol. For a local system of coordinates $%
u^{\alpha }=(x^{i},y^{a})$ on $V$ (in brief, $u=(x,y)),$ we can write
respectively
\begin{equation}
e_{\alpha }=(e_{i}=\partial _{i}=\frac{\partial }{\partial x^{i}}%
,e_{a}=\partial _{a}=\frac{\partial }{\partial y^{a}})
\end{equation}%
and
\begin{equation}
c^{\beta }=(c^{j}=dx^{j},c^{b}=dy^{b}),
\end{equation}%
for $e_{\alpha }\rfloor c^{\tau }=\delta _{\alpha }^{\tau };$ the indices
run correspondingly values of type:\ $i,j,...=1,2,...,n$ and $%
a,b,...=n+1,n+2,....,n+m$ for any conventional splitting $\alpha
=(i,a),\beta =(j,b),...$

Any local (vector) basis $e_{\alpha }$ can be decomposed with respect to any
other basis $e_{\underline{\alpha }}$ and $c^{\underline{\beta }}$ by
considering frame transforms,%
\begin{equation}
e_{\alpha }=A_{\alpha }^{\ \underline{\alpha }}(u)e_{\underline{\alpha }}%
\mbox{ and }c^{\beta }=A_{\ \underline{\beta }}^{\beta \ }(u)c^{\underline{%
\beta }}  \label{ft}
\end{equation}%
where the matrix $A_{\ \underline{\beta }}^{\beta \ }$ is the inverse to $%
A_{\alpha }^{\ \underline{\alpha }}.$ It should be noted that an arbitrary
basis $e_{\alpha }$ is nonholonomic (equivalently, anholonomic) because, in
general, it satisfies certain anholonomy conditions
\begin{equation}
e_{\alpha }e_{\beta }-e_{\beta }e_{\alpha }=W_{\alpha \beta }^{\gamma }\
e_{\gamma }  \label{anhr}
\end{equation}%
with nontrivial anholonomy coefficients $W_{\alpha \beta }^{\gamma
}=W_{\alpha \beta }^{\gamma }(u).$ For $W_{\alpha \beta }^{\gamma }=0,$ we
get holonomic frames: for instance, if we fix a local coordinate basis, $%
e_{\alpha }=\partial _{\alpha }.$

Denoting by $D_{X}=X\rfloor D$ the covariant derivative along a vector field
$X=X^{\alpha }e_{\alpha },$ we can define the torsion $\mathcal{T}=\{T_{\
\beta \gamma }^{\alpha }\},$
\begin{equation}
\mathcal{T}(X,Y)\doteqdot D_{X}Y-D_{Y}X-[X,Y],  \label{tors}
\end{equation}%
and the curvature $\mathcal{R}=\{R_{\ \beta \gamma \tau }^{\alpha }\},$%
\begin{equation}
\mathcal{R}(X,Y)Z\doteqdot D_{X}D_{Y}Z-D_{Y}D_{X}Z-D_{[X,Y]}Z,  \label{curv}
\end{equation}%
tensors of connection $D,$ where we use ''by definition'' symbol ''$%
\doteqdot $'' and $[X,Y]\doteqdot XY-YX.$ The components $T_{\ \beta \gamma
}^{\alpha }$ and $R_{\ \beta \gamma \tau }^{\alpha }$ are computed by
introducing $X\rightarrow e_{\alpha },Y\rightarrow e_{\beta },Z\rightarrow
e_{\gamma }$ into respective formulas (\ref{tors}) and (\ref{curv}), see %
\cite{vesnc} and \cite{valg} for details and computations related to the
system of denotations considered in this paper.

The Ricci tensor
\begin{equation}
Ric(D)=\{R_{\beta \gamma }\doteqdot R_{\ \beta \gamma \alpha }^{\alpha }\}
\end{equation}
is constructed by contracting the first (upper) and the last (lower) indices
of the curvature tensor. The scalar curvature $R$ is by definition the
contraction with the inverse metric $g^{\alpha \beta }$ (being the inverse
to the matrix $g_{\alpha \beta }$),
\begin{equation}
R\doteqdot g^{\alpha \beta }R_{\alpha \beta }
\end{equation}%
and the Einstein tensor $\mathcal{E}$ is introduced as
\begin{equation}
\mathcal{E}=\{E_{\alpha \beta }\doteqdot R_{\alpha \beta }-\frac{1}{2}%
g_{\alpha \beta }R\}.
\end{equation}%
The vacuum (source--free) Einstein equations are postulated
\begin{equation}
\mathcal{E}=\{E_{\alpha \beta }=R_{\alpha \beta }\}=0.  \label{enstet}
\end{equation}

The four dimensional (4D) general relativity theory is distinguished by the
property that the connection $D=\nabla $ is uniquely defined by the
coefficients $g_{\alpha \beta }$ following the conditions of metric
compatibility, $\nabla g=0,$ and of zero torsion, $_{\shortmid }\mathcal{T}%
=0.$ This defines the so--called Levi--Civita connection $\ _{\shortmid
}D=\nabla ;$ we respectively label its curvature tensor, Ricci tensor,
scalar curvature and Einstein tensor in the form $\ _{\shortmid }\mathcal{R}%
=\{\ _{\shortmid }R_{\ \beta \gamma \tau }^{\alpha }\},$ $\ _{\shortmid
}Ric(\nabla )=\{\ _{\shortmid }R_{\alpha \beta }\doteqdot \ _{\shortmid
}R_{\ \beta \gamma \alpha }^{\alpha }\},$ $\ _{\shortmid }R\doteqdot
g^{\alpha \beta }\ _{\shortmid }R_{\alpha \beta }$ and $\ _{\shortmid }%
\mathcal{E}=\{\ _{\shortmid }E_{\alpha \beta }\}.$ Modern gravity theories
consider extra dimensions and connections with nontrivial torsion. We note
that in this work we shall consider both nontrivial and trivial torsion
configurations. The aim is to show not only how our methods can be applied
to various types of theories (string/ gauge/ Einstein--Cartan/ Finsler
gravity models) but also to follow a simplified computational formalism
related to spaces with effective torsions $\mathbf{T} $ induced by
nonholonomic frame deformations.\footnote{%
The difference between ''boldfaced'' and ''calligraphic'' labels,
respectively for operators on spaces provided with nonlinear connection
structure and certain differential forms, will be explained in Sec. \ref%
{sah}.} Certain nontrivial limits to the vacuum Einstein gravity can be
selected if we impose the conditions
\begin{equation}
\mathbf{E}=\ _{\shortmid }\mathcal{E}  \label{cond1}
\end{equation}%
even, in general, we have $\mathbf{D}\neq \nabla .$ Such cases are
considered in Refs \cite{vesnc,valg,vsgg}. In this paper, we shall follow
more restrictive conditions when $\mathbf{D}$ and $\nabla $ have the same
components with respect to certain preferred bases, when the equality (\ref%
{cond1}) can be satisfied for some very general classes of metric ansatz.

We shall use a left--up label ''$\circ $'' for a metric
\begin{equation}
\ ^{\circ }g=\ ^{\circ }g_{\alpha \beta }\ c^{\alpha }\otimes c^{\beta }
\end{equation}%
a metric being a solution of the Einstein equations $\mathcal{E}=0$ (\ref%
{enstet}) for a linear connection $D$ with possible torsion $\mathcal{T}\neq
0.$ In order to emphasize that a metric is a solution of the vacuum Einstein
equations, in any dimension $n+m\geq 3,$ for the Levi--Civita connection $%
\nabla ,$ or for any metric compatible connection $\mathbf{D}\neq \nabla $
satisfying the conditions (\ref{cond1}), we shall write
\begin{equation}
\ _{\shortmid }^{\circ }g=\ _{\shortmid }^{\circ }g_{\alpha \beta }\
c^{\alpha }\otimes c^{\beta },
\end{equation}%
where the left--low label $"_{\shortmid }"$ will distinguish the geometric
objects for the Ricci flat space defined by a Levi--Civita connection $\nabla
.$

\subsection{The formalism related to Killing vectors}

The first parametric method \cite{geroch1} proposes a scheme of constructing
a one--para\-met\-er family of vacuum exact solutions (labelled by tilde '' $%
\widetilde{}\ "$ and depending on a real parameter $\theta $)
\begin{equation}
\ _{\shortmid }^{\circ }\widetilde{g}(\theta )=\ _{\shortmid }^{\circ }%
\widetilde{g}_{\alpha \beta }\ c^{\alpha }\otimes c^{\beta }  \label{germ1}
\end{equation}%
beginning with any source--free solution $\ \ _{\shortmid }^{\circ }g=\{\
_{\shortmid }^{\circ }g_{\alpha \beta }\}$ with Killing vector $\xi =\{\xi
_{\alpha }\}$ symmetry satisfying the conditions $\ _{\shortmid }\mathcal{E}%
=0$ (Einstein equations) and $\nabla _{\xi }(\ _{\shortmid }^{\circ }g)=0$
(Killing equations). We denote this 'primary' spacetime as $(V,\ _{\shortmid
}^{\circ }g,\xi _{\alpha }).$ One has to follow the rule \footnote{%
For our purposes, in order to elaborate and unified approach to the
parametric and the anholono\-mic frame methods, we introduce a new system of
denotations.}: The class of metrics $\ _{\shortmid }^{\circ }\widetilde{g}$
is generated by the transforms
\begin{equation}
\ _{\shortmid }^{\circ }\widetilde{g}_{\alpha \beta }=\widetilde{B}_{\alpha
}^{\ \alpha ^{\prime }}(u,\theta )\ \widetilde{B}_{\beta }^{\ \beta ^{\prime
}}(u,\theta )\ _{\shortmid }^{\circ }g_{\alpha ^{\prime }\beta ^{\prime }}
\label{ger1t}
\end{equation}%
where the matrix $\widetilde{B}_{\alpha }^{\ \alpha ^{\prime }}$ is
parametrized in the form when
\begin{equation}
\ _{\shortmid }^{\circ }\widetilde{g}_{\alpha \beta }=\lambda \widetilde{%
\lambda }^{-1}(\ _{\shortmid }^{\circ }g_{\alpha \beta }-\lambda ^{-1}\xi
_{\alpha }\xi _{\beta })+\widetilde{\lambda }\mu _{\alpha }\mu _{\beta }
\label{germ1c}
\end{equation}%
for
\begin{eqnarray}
\widetilde{\lambda } &=&\lambda \lbrack (\cos \theta -\omega \sin \theta
)^{2}+\lambda ^{2}\sin ^{2}\theta ]^{-1} \\
\mu _{\tau } &=&\widetilde{\lambda }^{-1}\xi _{\tau }+\alpha _{\tau }\sin
2\theta -\beta _{\tau }\sin ^{2}\theta . \nonumber
\end{eqnarray}%
A rigorous proof \ \cite{geroch1} states that the metrics (\ref{germ1}) also
define exact vacuum solutions with $\ _{\shortmid }\widetilde{\mathcal{E}%
}=0$ if and only if the values $\xi _{\alpha },\alpha _{\tau },$ $\mu _{\tau
}$ from (\ref{germ1c}), subjected to the conditions
\begin{equation*}
\lambda =\xi _{\alpha }\xi _{\beta }\ _{\shortmid }^{\circ }g^{\alpha \beta
},\ \omega =\xi ^{\gamma }\alpha _{\gamma },\xi ^{\gamma }\mu _{\gamma
}=\lambda ^{2}+\omega ^{2}-1,
\end{equation*}%
solve the equations%
\begin{eqnarray}
\nabla _{\alpha }\omega &=&\epsilon _{\alpha \beta \gamma \tau }\xi ^{\beta
}\ \nabla ^{\gamma }\xi ^{\tau } \nonumber    \\
\nabla _{\lbrack \alpha }\alpha _{\beta ]} &=&\frac{1}{2}\epsilon _{\alpha
\beta \gamma \tau }\ \nabla ^{\gamma }\xi ^{\tau }  \nonumber
\\
\nabla _{\lbrack \alpha }\mu _{\beta ]} &=&2\lambda \ \nabla _{\alpha }\xi
_{\beta }+\omega \epsilon _{\alpha \beta \gamma \tau }\ \nabla ^{\gamma }\xi
^{\tau }   \label{eq01}
\end{eqnarray}%
where the Levi--Civita connection $\nabla $ is defined by $\ _{\shortmid
}^{\circ }g$ and $\epsilon _{\alpha \beta \gamma \tau }$ is the absolutely
antisymmetric tensor. The existence of solutions for (\ref{eq01}) (Geroch's
equations) is guaranteed by the Einstein's and Killing equations.

The first type of parametric transforms (\ref{ger1t}) can parametrized by a
matrix $\widetilde{B}_{\alpha }^{\ \alpha ^{\prime }}$ with the coefficients
depending functionally on solutions for (\ref{eq01}). Fixing a signature $g_{%
\underline{\alpha }\underline{\beta }}=diag[\pm 1,\pm 1,....\pm 1]$ and a
local coordinate system on $(V,\ _{\shortmid }^{\circ }g,\xi _{\alpha }),$
one can define a local frame of reference%
\begin{equation}
e_{\alpha ^{\prime }}=A_{\alpha ^{\prime }}^{\ \underline{\alpha }%
}(u)\partial _{\underline{\alpha }},
\end{equation}%
like in (\ref{ft}), for which
\begin{equation}
_{\shortmid }^{\circ }g_{\alpha ^{\prime }\beta ^{\prime }}=A_{\alpha
^{\prime }}^{\ \underline{\alpha }}A_{\beta ^{\prime }}^{\ \underline{\beta }%
}g_{\underline{\alpha }\underline{\beta }}.  \label{qe}
\end{equation}%
We note that $A_{\alpha ^{\prime }}^{\ \underline{\alpha }}$ have to be
constructed as a solution of a system of quadratic algebraic equations (\ref%
{qe}) for given values $g_{\underline{\alpha }\underline{\beta }}$ and $%
_{\shortmid }^{\circ }g_{\alpha ^{\prime }\beta ^{\prime }}.$ In a similar
form, we can determine
\begin{equation}
\widetilde{e}_{\alpha }=\widetilde{A}_{\alpha }^{\ \underline{\alpha }%
}(\theta ,u)\partial _{\underline{\alpha }}
\end{equation}%
when%
\begin{equation}
_{\shortmid }^{\circ }\widetilde{g}_{\alpha \beta }=\widetilde{A}_{\alpha
}^{\ \underline{\alpha }}\widetilde{A}_{\beta }^{\ \underline{\beta }}g_{%
\underline{\alpha }\underline{\beta }}.  \label{qef}
\end{equation}%
The method guarantees that the family of spacetimes $(V,\ _{\shortmid
}^{\circ }\widetilde{g})$ is also vacuum Einstein but for the corresponding
families of Levi--Civita connections $\widetilde{\nabla }.$ In explicit form,
the matrix $\widetilde{B}_{\alpha }^{\ \alpha ^{\prime }}(u,\theta )$ of
parametric transforms can be computed by introducing the relations (\ref{qe}%
), (\ref{qef}) into (\ref{ger1t}),
\begin{equation}
\widetilde{B}_{\alpha }^{\ \alpha ^{\prime }}=\widetilde{A}_{\alpha }^{\
\underline{\alpha }}\ A_{\underline{\alpha }}^{\ \alpha ^{\prime }}
\label{mgt}
\end{equation}%
where $A_{\underline{\alpha }}^{\ \alpha ^{\prime }}$ is inverse to $%
A_{\alpha ^{\prime }}^{\ \underline{\alpha }}.$

The second parametric method \cite{geroch2} was similarly developed which
yields a family of new exact solutions involving two arbitrary functions on
one variables, beginning with any two commuting Killing fields for which a
certain pair of constants vanish (for instance, the exterior field of a
rotating star). By successive iterating such parametric transforms, one
generates a class of exact solutions characterized by an infinite number of
parameters and involving arbitrary functions. For simplicity, in this work
we shall apply only the first parametric method in order to generate other
nonholonomically deformed vacuum Einstein spaces. The case with
off--diagonal metrics and two Killing vectors is more special; it will be
analyzed in our further works.

\subsection{The anholonomic frame method}

\label{sah} We outline the results necessary for elaborating an approach
containing both the parametric transforms and nonholonomic frame
deformations. In details, the anholonomic frame method is reviewed in Refs. %
\cite{vesnc,vsgg}, see also Appendix to \cite{valg} containing proofs of
basic theorems and formulas.

Let us consider a $(n+m)$--dimensional manifold $\mathbf{V}$\ enabled with a
prescribed frame structure (\ref{ft}) when frame transform coefficients
depend linearly on values $N_{i}^{b}(u),$
\begin{eqnarray}
\mathbf{A}_{\alpha }^{\ \underline{\alpha }}(u) &=&\left[
\begin{array}{cc}
e_{i}^{\ \underline{i}}(u) & -N_{i}^{b}(u)e_{b}^{\ \underline{a}}(u) \\
0 & e_{a}^{\ \underline{a}}(u)%
\end{array}%
\right] ,  \label{vt1} \\
\mathbf{A}_{\ \underline{\beta }}^{\beta }(u) &=&\left[
\begin{array}{cc}
e_{\ \underline{i}}^{i\ }(u) & N_{k}^{b}(u)e_{\ \underline{i}}^{k\ }(u) \\
0 & e_{\ \underline{a}}^{a\ }(u)%
\end{array}%
\right] ,  \label{vt2}
\end{eqnarray}%
where $i,j,..=1,2,...,n$ and $a,b,...=n+1,n+2,...n+m$ and $u=\{u^{\alpha
}=(x^{i},y^{a})\}$ are local coordinates. The geometric constructions will
be adapted to a conventional $n+m$ splitting stated by a set of coefficients
$\mathbf{N}=\{N_{i}^{a}(u)\}$ defining a nonlinear connection
(N--connection) structure as a nonintergrable distribution%
\begin{equation}
T\mathbf{V=}h\mathbf{V\oplus }v\mathbf{V}  \label{distr}
\end{equation}%
with a conventional horizontal (h) subspace, $h\mathbf{V,}$ (with geometric
objects labelled by ''horizontal'' indices $i,j,..)$ and vertical (v)
subspace $v\mathbf{V}$ (with geometric objects labelled by indices $a,b,..)$ .%
\footnote{%
For simplicity, in this work, we shall not enter in the details of the
formalism of N--connections and (pseudo) Riemannian and Riemann--Cartan
spaces, and of the so--called N--anholonomic manifolds, considered in Refs. %
\cite{vesnc,esv,valg} and in the Introduction section of Ref.
 \cite{vsgg}. In an
alternative way, for different classes of connections not related to
solutions of the Einstein equations, the theory of nonholonomic manifolds
and (pseudo) Riemannian foliations is considered in Ref. \cite{bejf}.} We
shall use ''boldfaced'' symbols in order to emphasize that certain spaces
(geometric objects) are provided (adapted) with (to) a N--connection
structure $\mathbf{N.}$

The transforms (\ref{vt1}) and (\ref{vt2}) define a N--adapted frame
(vielbein) structure
\begin{equation}
\mathbf{e}_{\nu }=\left( \mathbf{e}_{i}=\partial _{i}-N_{i}^{a}(u)\partial
_{a},e_{a}=\partial _{a}\right) ,  \label{dder}
\end{equation}%
and the dual frame (coframe) structure%
\begin{equation}
\mathbf{c}^{\mu }=\left( e^{i}=dx^{i},\mathbf{e}%
^{a}=dy^{a}+N_{i}^{a}(u)dx^{i}\right) .  \label{ddif}
\end{equation}%
The vielbeins (\ref{ddif}) satisfy the corresponding nonholonomy
(equivalently, anholonomy) relations of type (\ref{anhr}),
\begin{equation}
\lbrack \mathbf{e}_{\alpha },\mathbf{e}_{\beta }]=\mathbf{e}_{\alpha }%
\mathbf{e}_{\beta }-\mathbf{e}_{\beta }\mathbf{e}_{\alpha }=W_{\alpha \beta
}^{\gamma }\mathbf{e}_{\gamma },  \label{nanhrel}
\end{equation}%
with (antisymmetric, $W_{\alpha \beta }^{\gamma }=-W_{\beta \alpha }^{\gamma
}$) anholonomy coefficients
\begin{equation}
W_{ia}^{b}=\partial _{a}N_{i}^{b}\mbox{ and }W_{ji}^{a}=\Omega _{ij}^{a}=%
\mathbf{e}_{j}(N_{i}^{a})-\mathbf{e}_{j}(N_{i}^{a}).  \label{anhncc}
\end{equation}%
We note that a distribution (\ref{distr}) is integrable, i.e. $\mathbf{V}$
is a foliation, if and only if the coefficients defined by $\mathbf{N}%
=\{N_{i}^{a}(u)\}$ satisfy the condition $\Omega _{ij}^{a}=0.$ In general, a
spacetime with prescribed nonholonomic splitting into h- and v--subspaces
can be considered as a nonholonomic manifold \cite{vesnc,bejf,esv}.

Let us consider a metric structure on $\mathbf{V},$%
\begin{equation}
\ \breve{g}=\underline{g}_{\alpha \beta }\left( u\right) du^{\alpha }\otimes
du^{\beta }  \label{metr}
\end{equation}%
defined by coefficients%
\begin{equation}
\underline{g}_{\alpha \beta }=\left[
\begin{array}{cc}
g_{ij}+N_{i}^{a}N_{j}^{b}h_{ab} & N_{j}^{e}h_{ae} \\
N_{i}^{e}h_{be} & h_{ab}%
\end{array}%
\right] .  \label{ansatz}
\end{equation}%
This metric is generic off--diagonal, i.e. it can not be diagonalized by
any coordinate transforms if $N_{i}^{a}(u)$ are any general functions.%
\footnote{%
We note that our $N$--coefficients depending nonlinearly on all coordinates $%
u^{\alpha }$ are not those from Kaluza--Klein theories which consist a
particular case when $N_{i}^{a}=A_{ib}^{a}(x^{k})y^{b}$ with further
compactifications on coordinates $y^{b}$.} We can adapt the metric (\ref%
{metr}) to a N--connection structure $\mathbf{N}=\{N_{i}^{a}(u)\}$ induced
by the off--diagonal coefficients in (\ref{ansatz}) if we impose that the
conditions
\begin{equation}
\breve{g}(e_{i},\ e_{a})=0,\mbox{ equivalently, }\underline{g}%
_{ia}-N_{i}^{b}h_{ab}=0,
\end{equation}%
where $\underline{g}_{ia}$ $\doteqdot g(\partial /\partial x^{i},\partial
/\partial y^{a}),$ are satisfied for the corresponding local basis (\ref%
{dder}). In this case $N_{i}^{b}=h^{ab}\underline{g}_{ia},$ where $h^{ab}$
is inverse to $h_{ab},$ and we can write the metric $\breve{g}$ (\ref{ansatz}%
)\ in equivalent form, as a distinguished metric (d--metric) adapted to a
N--connection structure\footnote{%
We shall call some geometric objects, like tensors, connections,..., to be
distinguished by a N--connection structure, in brief, d--tensors,
d--connection, if they are stated by components computed with respect to
N--adapted frames (\ref{dder}) and (\ref{ddif}). In this case, the geometric
constructions are elaborated in N--adapted form, i.e. they are adapted to
the nonholonomic distribution (\ref{distr}).},
\begin{equation}
\mathbf{g}=\mathbf{g}_{\alpha \beta }\left( u\right) \mathbf{c}^{\alpha
}\otimes \mathbf{c}^{\beta }=g_{ij}\left( u\right) c^{i}\otimes
c^{j}+h_{ab}\left( u\right) \ \mathbf{c}^{a}\otimes \ \mathbf{c}^{b},
\label{dmetr}
\end{equation}%
where $g_{ij}\doteqdot \mathbf{g}\left( \mathbf{e}_{i},\mathbf{e}_{j}\right)
$ and $h_{ab}\doteqdot \mathbf{g}\left( e_{a},e_{b}\right) .$ The
coefficients $\mathbf{g}_{\alpha \beta }$ and $\underline{g}_{\alpha \beta
}=g_{\underline{\alpha }\underline{\beta }}$ are related by formulas%
\begin{equation}
\mathbf{g}_{\alpha \beta }=\mathbf{A}_{\alpha }^{\ \underline{\alpha }}%
\mathbf{A}_{\beta }^{\ \underline{\beta }}g_{\underline{\alpha }\underline{%
\beta }},  \label{fmt}
\end{equation}%
or
\begin{equation}
g_{ij}=e_{i}^{\ \underline{i}}e_{j}^{\ \underline{j}}g_{\underline{i}%
\underline{j}}\mbox{ and }h_{ab}=e_{a}^{\ \underline{a}}e_{b}^{\ \underline{b%
}}g_{\underline{a}\underline{b}},
\end{equation}%
where the vielbein transform is given by matrices (\ref{vt1}) with $e_{i}^{\
\underline{i}}=\delta _{i}^{\ \underline{i}}$ and $e_{a}^{\ \underline{a}%
}=\delta _{a}^{\ \underline{a}}.$

Any vector field $\mathbf{X=(}hX\mathbf{,}\ vX\mathbf{)}$ on $T\mathbf{V}$
can be written in N--adapted form as a d--vector%
\begin{equation}
\mathbf{X=}X^{\alpha }\mathbf{e}_{\alpha }=\mathbf{(}hX=X^{i}\mathbf{e}_{i}%
\mathbf{,\ }vX=X^{a}e_{a}).
\end{equation}%
In a similar form we can 'N--adapt' any tensor object and call it to be a
d--tensor.

By definition, a d--connection is adapted to the distribution (\ref{distr})
and splits into h-- and v--covariant derivatives, $\mathbf{D}=hD+\ vD,$
where $hD=\{\mathbf{D}_{k}=\left( L_{jk}^{i},L_{bk\;}^{a}\right) \}$ and $\
vD=\{\mathbf{D}_{c}=\left( C_{jk}^{i},C_{bc}^{a}\right) \}$ are
correspondingly introduced as h- and v--parametrizations of the coefficients%
\begin{equation}
L_{jk}^{i}=\left( \mathbf{D}_{k}\mathbf{e}_{j}\right) \rfloor c^{i},\quad
L_{bk}^{a}=\left( \mathbf{D}_{k}e_{b}\right) \rfloor \mathbf{c}%
^{a},~C_{jc}^{i}=\left( \mathbf{D}_{c}\mathbf{e}_{j}\right) \rfloor
c^{i},\quad C_{bc}^{a}=\left( \mathbf{D}_{c}e_{b}\right) \rfloor \mathbf{c}%
^{a}.
\end{equation}%
The components $\mathbf{\Gamma }_{\ \alpha \beta }^{\gamma }=\left(
L_{jk}^{i},L_{bk}^{a},C_{jc}^{i},C_{bc}^{a}\right) ,$ with the coefficients
defined with respect to (\ref{ddif}) and (\ref{dder}), completely define a
d--connection $\mathbf{D}$ on a N--anholonomic manifold $\mathbf{V}.$

The simplest way to perform a local covariant calculus by applying
d--connecti\-ons is to use N--adapted differential forms and to introduce
the d--connection 1--form $\mathbf{\Gamma }_{\ \beta }^{\alpha }=\mathbf{%
\Gamma }_{\ \beta \gamma }^{\alpha }\mathbf{c}^{\gamma },$ when the
N--adapted components of d-connection $\mathbf{D}_{\alpha }=(\mathbf{e}%
_{\alpha }\rfloor \mathbf{D})$ are computed following formulas
\begin{equation}
\mathbf{\Gamma }_{\ \alpha \beta }^{\gamma }\left( u\right) =\left( \mathbf{D%
}_{\alpha }\mathbf{e}_{\beta }\right) \rfloor \mathbf{e}^{\gamma },
\label{cond2}
\end{equation}%
where ''$\rfloor "$ denotes the interior product. This allows us to define
in N--adapted form the torsion $\mathbf{T=\{\mathcal{T}^{\alpha }\}}$ (\ref%
{tors}),
\begin{equation}
\mathcal{T}^{\alpha }\doteqdot \mathbf{Dc}^{\alpha }=d\mathbf{c}^{\alpha }+%
\mathbf{\Gamma }_{\ \beta }^{\alpha }\wedge \mathbf{e}_{\alpha },
\label{torsa}
\end{equation}%
and curvature $\mathbf{R}=\{\mathcal{R}_{\ \beta }^{\alpha }\}$ (\ref{curv}%
),
\begin{equation}
\mathcal{R}_{\ \beta }^{\alpha }\doteqdot \mathbf{D\Gamma }_{\beta }^{\alpha
}=d\mathbf{\Gamma }_{\beta }^{\alpha }-\mathbf{\Gamma }_{\ \beta }^{\gamma
}\wedge \mathbf{\Gamma }_{\ \gamma }^{\alpha }.  \label{curva}
\end{equation}

The coefficients of torsion $\mathbf{T}$ (\ref{torsa}) of a d--connection $%
\mathbf{D}$ (in\ brief, d--torsion) are computed with respect to N--adapted
frames (\ref{ddif}) and (\ref{dder}),
\begin{eqnarray}
T_{\ jk}^{i} &=&L_{\ jk}^{i}-L_{\ kj}^{i},\ T_{\ ja}^{i}=-T_{\ aj}^{i}=C_{\
ja}^{i},\ T_{\ ji}^{a}=\Omega _{\ ji}^{a},\   \nonumber \\
T_{\ bi}^{a} &=&T_{\ ib}^{a}=\frac{\partial N_{i}^{a}}{\partial y^{b}}-L_{\
bi}^{a},\ T_{\ bc}^{a}=C_{\ bc}^{a}-C_{\ cb}^{a},  \label{dtors}
\end{eqnarray}%
where, for instance, $T_{\ jk}^{i}$ and $T_{\ bc}^{a}$ are respectively the
coefficients of the $h(hh)$--torsion $hT(hX,hY)$ and $v(vv)$--torsion $%
\mathbf{\ }vT(\mathbf{\ }vX,\mathbf{\ }vY).$ In a similar form, we can
compute the coefficients of a curvature $\mathbf{R,}$ d--curvatures (see
Appendix for the formulas for coefficients, proved in Refs. \cite{vesnc,vsgg}%
).

There is a preferred, canonical d--connection structure,$\ \widehat{\mathbf{D%
}}\mathbf{,}$ $\ $on a N--anholonomic manifold $\mathbf{V}$ constructed only
from the metric and N--con\-nec\-ti\-on coefficients $%
[g_{ij},h_{ab},N_{i}^{a}]$ and satisfying the conditions $\widehat{\mathbf{D}%
}\mathbf{g}=0$ and $\widehat{T}_{\ jk}^{i}=0$ and $\widehat{T}_{\ bc}^{a}=0.$
It should be noted that, in general, the components $\widehat{T}_{\
ja}^{i},\ \widehat{T}_{\ ji}^{a}$ and $\widehat{T}_{\ bi}^{a}$ are not zero.
This is an anholonomic frame (equivalently, off--diagonal metric) effect.
Hereafter, we consider only geometric constructions with the canonical
d--connection which allow, for simplicity, to omit ''hats'' on d--objects.%
\footnote{%
The preference to the canonical d---connection is motivated also by the fact
that it is possible to solve the vacuum Einstein equations for very general
ansatz for metric and N--connection structure just for this linear
connection. Usually, we can restrict the integral varieties in order to
generate solutions satisfying the conditions (\ref{cond1}), i.e. to
construct generic off--diagonal solutions in general relativity .} We can
verify by straightforward calculations that the linear connection $\mathbf{%
\Gamma }_{\ \alpha \beta }^{\gamma }=\left(
L_{jk}^{i},L_{bk}^{a},C_{jc}^{i},C_{bc}^{a}\right) $ with the coefficients
defined
\begin{equation}
\mathbf{D}_{\mathbf{e}_{k}}(\mathbf{e}_{j})=L_{jk}^{i}\mathbf{e}_{i},\
\mathbf{D}_{\mathbf{e}_{k}}(e_{b})=L_{bk}^{a}e_{a},\ \mathbf{D}_{e_{b}}(%
\mathbf{e}_{j})=C_{jb}^{i}\mathbf{e}_{i},\ \mathbf{D}%
_{e_{c}}(e_{b})=C_{bc}^{a}e_{a},
\end{equation}%
where
\begin{eqnarray}
L_{jk}^{i} &=&\frac{1}{2}g^{ir}\left( \mathbf{e}_{k}g_{jr}+\mathbf{e}%
_{j}g_{kr}-\mathbf{e}_{r}g_{jk}\right) ,  \nonumber \\
L_{bk}^{a} &=&e_{b}(N_{k}^{a})+\frac{1}{2}h^{ac}\left( \mathbf{e}%
_{k}h_{bc}-h_{dc}\ e_{b}N_{k}^{d}-h_{db}\ e_{c}N_{k}^{d}\right) ,
\label{candcon} \\
C_{jc}^{i} &=&\frac{1}{2}g^{ik}e_{c}g_{jk},\ C_{bc}^{a}=\frac{1}{2}%
h^{ad}\left( e_{c}h_{bd}+e_{c}h_{cd}-e_{d}h_{bc}\right),  \nonumber
\end{eqnarray}%
uniquely solve the conditions stated for the canonical d--connection.

The Levi--Civita linear connection $\bigtriangledown =\{\ _{\shortmid }\Gamma
_{\beta \gamma }^{\alpha }\},$ uniquely defined by the conditions $^{\nabla
}T=0$ and $\bigtriangledown \breve{g}=0,$ is not adapted to the distribution
(\ref{distr}). Let us parametrize its coefficients in the form
\begin{equation}
_{\shortmid }\Gamma _{\beta \gamma }^{\alpha }=\left( \ _{\shortmid
}L_{jk}^{i},\ _{\shortmid }L_{jk}^{a},\ _{\shortmid }L_{bk}^{i},\
_{\shortmid }L_{bk}^{a},\ _{\shortmid }C_{jb}^{i},\ _{\shortmid
}C_{jb}^{a},\ _{\shortmid }C_{bc}^{i},\ _{\shortmid }C_{bc}^{a}\right),
\end{equation}
where with respect to N--adapted bases (\ref{ddif}) and (\ref{dder})
\begin{eqnarray*}
\bigtriangledown _{\mathbf{e}_{k}}(\mathbf{e}_{j}) &=&\ _{\shortmid
}L_{jk}^{i}\mathbf{e}_{i}+\ _{\shortmid }L_{jk}^{a}e_{a},\ \bigtriangledown
_{\mathbf{e}_{k}}(e_{b})=\ _{\shortmid }L_{bk}^{i}\mathbf{e}_{i}+\
_{\shortmid }L_{bk}^{a}e_{a}, \\
\bigtriangledown _{e_{b}}(\mathbf{e}_{j}) &=&\ _{\shortmid }C_{jb}^{i}%
\mathbf{e}_{i}+\ _{\shortmid }C_{jb}^{a}e_{a},\ \bigtriangledown
_{e_{c}}(e_{b})=\ _{\shortmid }C_{bc}^{i}\mathbf{e}_{i}+\ _{\shortmid
}C_{bc}^{a}e_{a}.
\end{eqnarray*}%
A straightforward calculus\footnote{%
Such results were originally considered by R. Miron and M. Anastasiei for
vector bundles provided with N--connection and metric structures, see Ref. %
\cite{ma} . Similar proofs hold true for any nonholonomic manifold with
prescribed N--connection.} shows that the coefficients of the Levi--Civita
connection can be expressed in the form%
\begin{eqnarray}
\ _{\shortmid }L_{jk}^{i} &=&L_{jk}^{i},\ _{\shortmid
}L_{jk}^{a}=-C_{jb}^{i}g_{ik}h^{ab}-\frac{1}{2}\Omega _{jk}^{a},
\label{lccon} \\
\ _{\shortmid }L_{bk}^{i} &=&\frac{1}{2}\Omega _{jk}^{c}h_{cb}g^{ji}-\frac{1%
}{2}(\delta _{j}^{i}\delta _{k}^{h}-g_{jk}g^{ih})C_{hb}^{j},  \nonumber \\
\ _{\shortmid }L_{bk}^{a} &=&L_{bk}^{a}+\frac{1}{2}(\delta _{c}^{a}\delta
_{d}^{b}+h_{cd}h^{ab})\left[ L_{bk}^{c}-e_{b}(N_{k}^{c})\right] ,  \nonumber \\
\ _{\shortmid }C_{kb}^{i} &=&C_{kb}^{i}+\frac{1}{2}\Omega
_{jk}^{a}h_{cb}g^{ji}+\frac{1}{2}(\delta _{j}^{i}\delta
_{k}^{h}-g_{jk}g^{ih})C_{hb}^{j},  \nonumber \\
\ _{\shortmid }C_{jb}^{a} &=&-\frac{1}{2}(\delta _{c}^{a}\delta
_{b}^{d}-h_{cb}h^{ad})\left[ L_{dj}^{c}-e_{d}(N_{j}^{c})\right] ,\
_{\shortmid }C_{bc}^{a}=C_{bc}^{a},  \nonumber \\
\ _{\shortmid }C_{ab}^{i} &=&-\frac{g^{ij}}{2}\left\{ \left[
L_{aj}^{c}-e_{a}(N_{j}^{c})\right] h_{cb}+\left[ L_{bj}^{c}-e_{b}(N_{j}^{c})%
\right] h_{ca}\right\} ,  \nonumber
\end{eqnarray}%
where $\Omega _{jk}^{a}$ are computed as in the second formula in (\ref%
{anhncc}).

For our purposes, it is important to state the conditions when both the
Levi--Civita connection and the canonical d--connection (being of different
geometric nature) may be defined by the same set of coefficients with
respect to a fixed frame of reference. Following formulas (\ref{candcon})
and (\ref{lccon}), we obtain the component equality $_{\shortmid }\Gamma
_{\beta \gamma }^{\alpha }=\mathbf{\Gamma }_{\ \alpha \beta }^{\gamma }$ if%
\begin{equation}
\Omega _{jk}^{c}=0  \label{fols}
\end{equation}%
(there are satisfied the integrability conditions and our manifold admits a
foliation structure),
\begin{equation}
\ _{\shortmid }C_{kb}^{i}=C_{kb}^{i}=0  \label{coef0}
\end{equation}%
and
\begin{equation}
L_{aj}^{c}-e_{a}(N_{j}^{c})=0
\end{equation}%
which, following the second formula in (\ref{candcon}), is equivalent to%
\begin{equation}
\mathbf{e}_{k}h_{bc}-h_{dc}\ e_{b}N_{k}^{d}-h_{db}\ e_{c}N_{k}^{d}=0.
\label{cond3}
\end{equation}

We conclude this section with the remark that if the conditions (\ref{fols}%
), (\ref{coef0}) and (\ref{cond3}) hold true for the metric (\ref{metr}),
equivalently (\ref{dmetr}), the torsion coefficients (\ref{dtors}) vanish.
This results in respective equalities of the coefficients of the Riemann,
Ricci and Einstein tensors (the conditions (\ref{cond1}) being satisfied)
for two different linear connections.

\subsection{Off--diagonal exact solutions}

We consider a five dimensional (5D) manifold $\mathbf{V}$ of necessary
smooth class and conventional splitting of dimensions $\dim \mathbf{V=}$ $%
n+m $ for $n=3$ and $m=2.$ The local coordinates are labelled in the form $%
u^{\alpha }=(x^{i},y^{a})=(x^{1},x^{\widehat{i}},y^{4}=v,y^{5}),$ for $%
i=1,2,3$ and $\widehat{i}=2,3$ and $a,b,...=4,5.$ For our further purposes,
we can consider that any coordinates from a set $u^{\alpha }$ can be of (3D)
space, time, or extra dimension (5th coordinate) type.

\subsubsection{A five dimensional off--diagonal ansatz}

The ansatz of type (\ref{dmetr}) is parametrized in the form
\begin{eqnarray}
\mathbf{g} &=&g_{1}{dx^{1}}\otimes {dx^{1}}+g_{2}(x^{2},x^{3}){dx^{2}}%
\otimes {dx^{2}}+g_{3}\left( x^{2},x^{3}\right) {dx^{3}}\otimes {dx^{3}}
\nonumber \\
&&+h_{4}\left( x^{k},v\right) \ {\delta v}\otimes {\delta v}+h_{5}\left(
x^{k},v\right) \ {\delta y}\otimes {\delta y},  \nonumber \\
\delta v &=&dv+w_{i}\left( x^{k},v\right) dx^{i},\ \delta y=dy+n_{i}\left(
x^{k},v\right) dx^{i}  \label{ans5d}
\end{eqnarray}%
with the coefficients defined by some necessary smooth class functions of
type
\begin{eqnarray}
g_{1} &=&\pm 1,g_{2,3}=g_{2,3}(x^{2},x^{3}),h_{4,5}=h_{4,5}(x^{i},v),  \nonumber
\\
w_{i} &=&w_{i}(x^{i},v),n_{i}=n_{i}(x^{i},v).  \nonumber
\end{eqnarray}%
The off--diagonal terms of this metric, written with respect to the
coordinate dual frame $du^{\alpha }=(dx^{i},dy^{a}),$ can be redefined to
state a N--connection structure $\mathbf{N}=[N_{i}^{4}=w_{i}(x^{k},v),$$%
N_{i}^{5}=n_{i}(x^{k},v)]$ with a N--elongated co--frame (\ref{ddif})
parametrized as
\begin{eqnarray}
c^{1} &=&dx^{1},\ c^{2}=dx^{2},\ c^{3}=dx^{3},  \nonumber \\
\mathbf{c}^{4} &=&\delta v=dv+w_{i}dx^{i},\ \mathbf{c}^{5}=\delta y=dy+n_{i}
dx^{i}.  \label{ddif5}
\end{eqnarray}%
This funfbein is dual to the local basis%
\begin{equation}
\mathbf{e}_{i}=\frac{\partial }{\partial x^{i}}-w_{i}\left( x^{k},v\right)
\frac{\partial }{\partial v}-n_{i}\left( x^{k},v\right) \frac{\partial }{%
\partial y^{5}},e_{4}=\frac{\partial }{\partial v},e_{5}=\frac{\partial }{%
\partial y^{5}}.  \label{dder5}
\end{equation}%
We emphasize that the metric (\ref{ans5d}) does not depend on variable $%
y^{5},$ i.e. it possesses a Killing vector $e_{5}=\partial /\partial y^{5},$
and distinguish the dependence on the so--called ''anisotropic'' variable $%
y^{4}=v.$

Computing the components of the Ricci and Einstein tensors for the metric (%
\ref{ans5d}) (see main formulas in Appendix and details on tensors
components' calculus in Refs. \cite{valg,vsgg}), one proves that the
Einstein equations (\ref{ep1}) for a diagonal with respect to (\ref{ddif5})
and (\ref{dder5}), source%
\begin{eqnarray}
\mathbf{\Upsilon }_{\beta }^{\alpha } &=&[\Upsilon _{1}^{1}=\Upsilon
_{2}+\Upsilon _{4},\Upsilon _{2}^{2}=\Upsilon _{2}(x^{2},x^{3},v),\Upsilon
_{3}^{3}=\Upsilon _{2}(x^{2},x^{3},v),  \nonumber \\
\Upsilon _{4}^{4} &=&\Upsilon _{4}(x^{2},x^{3}),\Upsilon _{5}^{5}=\Upsilon
_{4}(x^{2},x^{3})]  \label{sdiag}
\end{eqnarray}%
transform into this system of partial differential equations:
\begin{eqnarray}
&& R_{2}^{2} =R_{3}^{3}
=\frac{1}{2g_{2}g_{3}}[\frac{g_{2}^{\bullet }g_{3}^{\bullet }}{2g_{2}}+%
\frac{(g_{3}^{\bullet })^{2}}{2g_{3}}-g_{3}^{\bullet \bullet }    \nonumber
\\  && +\frac{%
g_{2}^{^{\prime }}g_{3}^{^{\prime }}}{2g_{3}}+\frac{(g_{2}^{^{\prime }})^{2}%
}{2g_{2}}-g_{2}^{^{\prime \prime }}]=-\Upsilon _{4}(x^{2},x^{3})\label{ep1a} \\
&&S_{4}^{4} =S_{5}^{5}=\frac{1}{2h_{4}h_{5}}\left[ h_{5}^{\ast }\left( \ln
\sqrt{|h_{4}h_{5}|}\right) ^{\ast }-h_{5}^{\ast \ast }\right] =-\Upsilon
_{2}(x^{2},x^{3},v),  \label{ep2a} \\
&&R_{4i} =-w_{i}\frac{\beta }{2h_{5}}-\frac{\alpha _{i}}{2h_{5}}=0,
\label{ep3a} \\
&&R_{5i} =-\frac{h_{5}}{2h_{4}}\left[ n_{i}^{\ast \ast }+\gamma n_{i}^{\ast }%
\right] =0,  \label{ep4a}
\end{eqnarray}%
where, for $h_{4,5}^{\ast }\neq 0,$%
\begin{eqnarray}
\alpha _{i} &=&h_{5}^{\ast }\partial _{i}\phi ,\ \beta =h_{5}^{\ast }\ \phi
^{\ast },\ \gamma =\frac{3h_{5}^{\ast }}{2h_{5}}-\frac{h_{4}^{\ast }}{h_{4}},
\label{coef} \\
\phi &=&\ln |\frac{h_{5}^{\ast }}{\sqrt{|h_{4}h_{5}|}}|,  \label{coefa}
\end{eqnarray}%
when the necessary partial derivatives are written in the form $a^{\bullet
}=\partial a/\partial x^{2},$\ $a^{\prime }=\partial a/\partial x^{3},$\ $%
a^{\ast }=\partial a/\partial v.$ In the vacuum case, we must consider $%
\Upsilon _{2,4}=0.$ We note that we use a source of type (\ref{sdiag}) in
order to show that the anholonomic frame method can be applied also for
non--vacuum configurations, for instance, when $\Upsilon _{2}=\lambda
_{2}=const$ and $\Upsilon _{4}=\lambda _{4}=const,$ defining locally
anisotropic configurations generated by an anisotropic cosmological
constant, which in its turn, can be induced by certain ansatz for the
so--called $H$--field (absolutely antisymmetric third rank tensor field) in
string theory \cite{vesnc,vsgg,valg}, see formulas (\ref{c01}) and (\ref{c02}%
) and related explanations in Appendix. Here we note that the off--diagonal
gravitational interactions can model locally anisotropic configurations even
if $\lambda _{2}=\lambda _{4},$ or both values vanish.

Summarizing the results for an ansatz (\ref{ansatz}) with arbitrary
signatures $\epsilon _{\alpha }=\left( \epsilon _{1},\epsilon _{2},\epsilon
_{3},\epsilon _{4},\epsilon _{5}\right) $ (where $\epsilon _{\alpha }=\pm 1)$
and $h_{4}^{\ast }\neq 0$ and $h_{5}^{\ast }\neq 0,$ one proves \cite%
{vesnc,valg,vsgg} that any off--diagonal metric
\begin{eqnarray}
\ ^{\circ }\mathbf{g} &=&\epsilon _{1}\ dx^{1}\otimes dx^{1}+\epsilon
_{2}g_{2}(x^{\widehat{i}})\ dx^{2}\otimes dx^{2}+\epsilon _{3}g_{3}(x^{%
\widehat{i}})\ dx^{3}\otimes dx^{3} + \epsilon _{4}h_{0}^{2}(x^{i})
\nonumber \\
&&\left[ f^{\ast }\left( x^{i},v\right) \right]
^{2}|\varsigma \left( x^{i},v\right) |\ \delta v\otimes \delta v
+\epsilon _{5}\left[ f\left( x^{i},v\right) -f_{0}(x^{i})\right] ^{2}\
\delta y^{5}\otimes \delta y^{5}  \nonumber \\
\delta v &=&dv+w_{k}\left( x^{i},v\right) dx^{k},\ \delta
y^{5}=dy^{5}+n_{k}\left( x^{i},v\right) dx^{k},  \label{gensol1}
\end{eqnarray}%
with the coefficients being of necessary smooth class and the indices with
''hat'' running the values $\widehat{i},\widehat{j},...=2,3$, where\ \ $g_{%
\widehat{k}}\left( x^{\widehat{i}}\right) $ is a solution of the 2D equation
(\ref{ep1a}) for a given source $\Upsilon _{4}\left( x^{\widehat{i}}\right)
, $%
\begin{equation}
\varsigma \left( x^{i},v\right) =\varsigma _{\lbrack 0]}\left( x^{i}\right) -%
\frac{\epsilon _{4}}{8}h_{0}^{2}(x^{i})\int \Upsilon _{2}(x^{\widehat{k}%
},v)f^{\ast }\left( x^{i},v\right) \left[ f\left( x^{i},v\right)
-f_{0}(x^{i})\right] dv,
\end{equation}%
and the N--connection coefficients $N_{i}^{4}=w_{i}(x^{k},v)$ and $%
N_{i}^{5}=n_{i}(x^{k},v)$ are computed following the formulas
\begin{equation}
w_{i}=-\frac{\partial _{i}\varsigma \left( x^{k},v\right) }{\varsigma ^{\ast
}\left( x^{k},v\right) }  \label{gensol1w}
\end{equation}%
and
\begin{equation}
n_{k}=n_{k[1]}\left( x^{i}\right) +n_{k[2]}\left( x^{i}\right) \int \frac{%
\left[ f^{\ast }\left( x^{i},v\right) \right] ^{2}}{\left[ f\left(
x^{i},v\right) -f_{0}(x^{i})\right] ^{3}}\varsigma \left( x^{i},v\right) dv,
\label{gensol1n}
\end{equation}%
define an exact solution of the system of Einstein equations (\ref{ep1a})--(%
\ref{ep4a}). It should be emphasized that such solutions depend on arbitrary
nontrivial functions $f\left( x^{i},v\right) $ (with $f^{\ast }\neq 0),$ $%
f_{0}(x^{i}),$ $h_{0}^{2}(x^{i})$, $\ \varsigma _{\lbrack 0]}\left(
x^{i}\right) ,$ $n_{k[1]}\left( x^{i}\right) $ and $\ n_{k[2]}\left(
x^{i}\right) ,$ and sources $\Upsilon _{2}(x^{\widehat{k}},v),\Upsilon
_{4}\left( x^{\widehat{i}}\right) .$ Such values for the corresponding
signatures $\epsilon _{\alpha }=\pm 1$ have to be defined by certain
boundary conditions and physical considerations.\footnote{%
Our classes of solutions depending on integration functions are more general
than those for diagonal ansatz depending, for instance, on one radial like
variable like in the case of the Schwarzschild solution (when the Einstein
equations are reduced to an effective nonlinear ordinary differential
equation, ODE). In the case of ODE, the integral varieties depend on
integration constants which can be defined from certain boundary/ asymptotic
and symmetry conditions, for instance, from the constraint that far away
from the horizon the Schwarzschild metric contains corrections from the
Newton potential. Because our ansatz (\ref{ans5d}) results in a system of
nonlinear partial differential equations (\ref{ep1a}--(\ref{ep4a}), the
solutions depend not on integration constants, but on very general classes
of integration functions. A similar situation is considered in the Geroch
method but those solutions are also parametrized by sets of parameters not
treated as local coordinates.}

The ansatz of type (\ref{ans5d}) with $h_{4}^{\ast }=0$ but $h_{5}^{\ast
}\neq 0$ (or, inversely, $h_{4}^{\ast }\neq 0$ but $h_{5}^{\ast }=0)$
consist of more special cases and request a bit different method of
constructing exact solutions. Nevertheless, such type solutions are also
generic off--diagonal and they may be of substantial interest (the length of
paper does not allow to include an analysis of such particular cases).

\subsubsection{Four and five dimensional foliations and the Einstein spaces}

The method of constructing 5D solutions with nontrivial torsion \ induced by
anholonomy coefficients can be restricted to generate 4D nonholonomic
configurations and generic off--diagonal solutions in general relativity. In
order to consider reductions $5D\rightarrow 4D$ for the ansatz (\ref{ans5d})
we can eliminate from the formulas the variable $x^{1}$ and to consider a 4D
space $\mathbf{V}^{4}$ (parametrized by local coordinates $\left(
x^{2},x^{3},v,y^{5}\right) )$ trivially embedded into a 5D spacetime $%
\mathbf{V}$ (parametrized by local coordinates $\left(
x^{1},x^{2},x^{3},v,y^{5}\right) $ with $g_{11}=\pm 1,g_{1\widehat{\alpha }%
}=0,\widehat{\alpha }=2,3,4,5).$ In this case, there are possible \ 4D
conformal and anholonomic transforms depending only on variables $\left(
x^{2},x^{3},v\right) $ of a 4D metric $g_{\widehat{\alpha }\widehat{\beta }%
}\left( x^{2},x^{3},v\right) $ of arbitrary signature. To emphasize that
some coordinates are stated just for a 4D space we might use ''hats'' on the
Greek indices, $\widehat{\alpha },\widehat{\beta },...$ \ and on the Latin
indices from the middle of the alphabet, $\widehat{i},\widehat{j},...=2,3,$
when the local coordinates on $\mathbf{V}^{4}$ are parametrized in the form $%
u^{\widehat{\alpha }}=\left( x^{\widehat{i}},y^{a}\right) =\left(
x^{2},x^{3},y^{4}=v,y^{5}\right) ,$ for $a,b,...=4,5.$ The 4D off--diagonal
ansatz
\begin{equation}
\mathbf{g}=g_{2}\ dx^{2}\otimes dx^{2}+g_{3}\ dx^{3}\otimes dx^{3}+h_{4}\
\delta v\otimes \delta v+h_{5}\ \delta y^{5}\otimes \delta y^{5},
\label{dmetric4}
\end{equation}%
is written with respect to the anholonomic co--frame $\left( dx^{\widehat{i}%
},\delta v,\delta y^{5}\right) ,$ where
\begin{equation}
\delta v=dv+w_{\widehat{i}}dx^{\widehat{i}}\mbox{ and }\delta
y^{5}=dy^{5}+n_{\widehat{i}}dx^{\widehat{i}}  \label{ddif4}
\end{equation}%
is the dual of $\left( \delta _{\widehat{i}},\partial _{4},\partial
_{5}\right) ,$ where
\begin{equation}
\delta _{\widehat{i}}=\partial _{\widehat{i}}+w_{\widehat{i}}\partial
_{4}+n_{\widehat{i}}\partial _{5},  \label{dder4}
\end{equation}%
and the coefficients are necessary smoothly class functions of type:
\begin{eqnarray}
g_{\widehat{j}} &=&g_{\widehat{j}}(x^{\widehat{k}}),h_{4,5}=h_{4,5}(x^{%
\widehat{k}},v),  \nonumber \\
w_{\widehat{i}} &=&w_{\widehat{i}}(x^{\widehat{k}},v),n_{\widehat{i}}=n_{%
\widehat{i}}(x^{\widehat{k}},v);~\widehat{i},\widehat{k}=2,3.  \nonumber
\end{eqnarray}

In the 4D case, a source of type (\ref{sdiag}) should be considered without
the component $\Upsilon _{1}^{1}$ in the form
\begin{equation}
\mathbf{\Upsilon }_{\widehat{\beta }}^{\widehat{\alpha }}=diag[\Upsilon
_{2}^{2}=\Upsilon _{3}^{3}=\Upsilon _{2}(x^{\widehat{k}},v),\ \Upsilon
_{4}^{4}=\Upsilon _{5}^{5}=\Upsilon _{4}(x^{\widehat{k}})].  \label{sdiag4}
\end{equation}%
The Einstein equations with source (\ref{sdiag4}) for the canonical
d--connection (\ref{candcon}) defined by the ansatz (\ref{dmetric4})
transform into a system of nonlinear partial differential equations very
similar to (\ref{ep1a})--(\ref{ep4a}). The difference for the 4D equations
is that the coordinate $x^{1}$ is not contained into the equations and that
the indices of type $i,j,..=1,2,3$ must be changed into the corresponding
indices $\widehat{i},\widehat{j},..=2,3.$ The generated classes of 4D
solutions are defined almost by the same formulas (\ref{gensol1}), (\ref%
{gensol1w}) and (\ref{gensol1n}).

Now we describe how the coefficients of an ansatz (\ref{dmetric4}) defining
an exact vacuum solution for a canonical d--connecton can be constrained to
generate a vacuum solution in Einstein gravity:

We start with the conditions\ (\ref{cond3}) written (for our ansatz) in the
form%
\begin{eqnarray}
\frac{\partial h_{4}}{\partial x^{\widehat{k}}}-w_{\widehat{k}}h_{4}^{\ast
}-2w_{\widehat{k}}^{\ast }h_{4} &=&0,  \label{cond3a} \\
\frac{\partial h_{5}}{\partial x^{\widehat{k}}}-w_{\widehat{k}}h_{5}^{\ast }
&=&0,  \label{cond3b} \\
n_{\widehat{k}}^{\ast }h_{5} &=&0.  \label{cond3c}
\end{eqnarray}%
These equations for nontrivial values of $w_{\widehat{k}}$ and $n_{\widehat{k%
}}$ constructed for some defined values of \ $h_{4}$ and $h_{5}$ must be
compatible with the equations (\ref{ep2a})--(\ref{ep4a}) for $\Upsilon
_{2}=0.$ One can be taken nonzero values for $w_{\widehat{k}}$ in (\ref{ep3a}%
) if and only if $\alpha _{\widehat{i}}=0$ because the the equation (\ref%
{ep2a}) imposes the condition $\beta =0.$ This is possible, for the
sourceless case and $h_{5}^{\ast }\neq 0,$ if and only if
\begin{equation}
\phi =\ln \left| h_{5}^{\ast }/\sqrt{|h_{4}h_{5}|}\right| =const,
\label{eq23}
\end{equation}%
see formula (\ref{coefa}). A very general class of solutions of equations (%
\ref{cond3a}), (\ref{cond3b}) and (\ref{eq23}) can be represented in the
form
\begin{eqnarray}
h_{4} &=&\epsilon _{4}h_{0}^{2}\ \left( b^{\ast }\right) ^{2},h_{5}=\epsilon
_{5}(b+b_{0})^{2},  \label{lcc2} \\
w_{\widehat{k}} &=&\left( b^{\ast }\right) ^{-1}\frac{\partial (b+b_{0})}{%
\partial x^{\widehat{k}}},  \nonumber
\end{eqnarray}%
where $h_{0}=const$ and $b=b(x^{\widehat{k}},v)$ is any function for which $%
b^{\ast }\neq 0$ and $b_{0}=b_{0}(x^{\widehat{k}})$ is an arbitrary
integration function.

The next step is to satisfy the integrability conditions (\ref{fols})
defining a foliated spacetimes provided with metric and N--connection and
d--connection structures \cite{valg,vesnc,vsgg,esv} (we note that (pseudo)
Riemannian foliations are considered in a different manner in Ref. \cite%
{bejf}) for the so--called Schouten -- Van Kampen and Vranceanu connections
not subjected to the condition to generate Einstein spaces). It is very easy
to show that there are nontrivial solutions of the constraints (\ref{fols})
which for the ansatz (\ref{dmetric4}) are written in the form
\begin{eqnarray}
w_{2}^{\prime }-w_{3}^{\bullet }+w_{3}w_{2}^{\ast }-w_{2}w_{3}^{\ast } &=&0,
\label{aux31} \\
n_{2}^{\prime }-n_{3}^{\bullet }+w_{3}n_{2}^{\ast }-w_{2}n_{3}^{\ast } &=&0.
\nonumber
\end{eqnarray}%
We solve these equations for $n_{2}^{\ast }=n_{3}^{\ast }=0$ if we take any
two functions $n_{2,3}(x^{\widehat{k}})$ satisfying
\begin{equation}
n_{2}^{\prime }-n_{3}^{\bullet }=0  \label{aux31b}
\end{equation}%
(it is possible for a particular class of integration functions in (\ref%
{gensol1n}) when $n_{\widehat{k}[2]}\left( x^{\widehat{i}}\right) =0$ and $%
n_{\widehat{k}[1]}\left( x^{\widehat{i}}\right) $ are constraint to satisfy
just the conditions (\ref{aux31b})). Then we can consider any $b(x^{\widehat{%
k}},v)$ for which $w_{\widehat{k}}=\left( b^{\ast }\right) ^{-1}\partial _{%
\widehat{k}}(b+b_{0})$ solve the equation (\ref{aux31}). In a more
particular case, one can be constructed solutions for any $%
b(x^{3},v),b^{\ast }\neq 0,$ and $n_{2}=0$ and $n_{3}=n_{3}(x^{3},v)$ (or,
inversely, for any $n_{2}=n_{2}(x^{2},v)$ and $n_{3}=0).$ Here one should be
also noted that the conditions (\ref{coef0}) are solved in straightforward
form by the ansatz (\ref{dmetric4}).

We conclude that for any sets of $h_{4}(x^{\widehat{k}},v),h_{5}(x^{\widehat{%
k}},v),$ $w_{\widehat{k}}(x^{\widehat{k}},v),$ $n_{2,3}(x^{\widehat{k}})$
respectively generated by functions $b(x^{\widehat{k}},v)$ and $n_{\widehat{k%
}[1]}\left( x^{\widehat{i}}\right) ,$ see (\ref{lcc2}), and satisfying (\ref%
{aux31b}), the generic off--diagonal metric (\ref{dmetric4}) possesses the
same coefficients both for the Levi--Civita and canonical d--connection being
satisfied the conditions (\ref{cond1}) of equality of the Einstein tensors.
Here we note that any 2D metric can be written in a conformally flat form,
i. e. we can chose such local coordinates when
\begin{equation}
g_{2}(dx^{2})^{2}+g_{3}(dx^{3})^{2}=e^{\psi (x^{\widehat{i}})}\left[
\epsilon _{\widehat{2}}(dx^{\widehat{2}})^{2}+\epsilon _{\widehat{3}}(dx^{%
\widehat{3}})^{2}\right] ,
\end{equation}%
for signatures $\epsilon _{\widehat{k}}=\pm 1,$ in (\ref{dmetric4}).

Summarizing the results of this section, we can write down the generic
off--diagonal metric (it is a 4D dimensional reduction of (\ref{gensol1}))
\begin{eqnarray}
\ _{\shortmid }^{\circ }\mathbf{g} &=&e^{\psi (x^{2},x^{3})}\left[ \epsilon
_{2}\ dx^{2}\otimes dx^{2}+\epsilon _{3}\ dx^{3}\otimes dx^{3}\right] +
\epsilon _{4}h_{0}^{2}\
\label{4ds} \\
&&\left[ b^{\ast }\left( x^{i},v\right) \right]
^{2}\ \delta v\otimes \delta v
+\epsilon _{5}\left[ b\left( x^{2},x^{3},v\right) -b_{0}(x^{2},x^{3})%
\right] ^{2}\ \delta y^{5}\otimes \delta y^{5},  \nonumber \\
\delta v &=&dv+w_{2}\left( x^{2},x^{3},v\right) dx^{2}+w_{3}\left(
x^{2},x^{3},v\right) dx^{3},  \nonumber \\
\ \delta y^{5} &=&dy^{5}+n_{2}\left( x^{2},x^{3}\right) dx^{2}+n_{3}\left(
x^{2},x^{3}\right) dx^{3},  \nonumber
\end{eqnarray}%
defining vacuum exact solutions in general relativity if the coefficients
are restricted to solve the equations
\begin{eqnarray}
\epsilon _{2}\psi ^{\bullet \bullet }+\epsilon _{3}\psi ^{^{\prime \prime }}
&=&0,  \label{cond5} \\
w_{2}^{\prime }-w_{3}^{\bullet }+w_{3}w_{2}^{\ast }-w_{2}w_{3}^{\ast } &=&0,
\nonumber \\
n_{2}^{\prime }-n_{3}^{\bullet } &=&0,  \nonumber
\end{eqnarray}%
for $w_{2}=\left( b^{\ast }\right) ^{-1}(b+b_{0})^{\bullet }$ and $%
w_{3}=\left( b^{\ast }\right) ^{-1}(b+b_{0})^{\prime },$ where, for
instance, $n_{3}^{\bullet }=\partial _{2}n_{3}$ and $n_{2}^{\prime
}=\partial _{3}n_{2}.$

We can generalize (\ref{4ds}) similarly to (\ref{gensol1}) in order to
generate solutions for nontrivial sources (\ref{sdiag4}). In general, they
will contain nontrivial anholonomically induced torsions. Such
configurations may be restricted to the case of Levi--Civita connection by
solving the constraints (\ref{cond3a})--(\ref{cond3c}) in order to be
compatible with the equations (\ref{ep2a}) and (\ref{ep3a}) for the
coefficients $\alpha _{\widehat{i}}$ and $\beta $ computed for $h_{5}^{\ast
}\neq 0$ and $\ln \left| h_{5}^{\ast }/\sqrt{|h_{4}h_{5}|}\right| =\phi
(x^{2},x^{3},v)\neq const,$ see formula (\ref{coefa}), resulting in more
general conditions than (\ref{eq23}) and (\ref{lcc2}). Roughly speaking, all
such coefficients are generated by any $h_{4}$ (or $h_{5}$) defined from (%
\ref{ep3a}) for prescribed values $h_{5}$ (or $h_{5}$) and $\Upsilon _{2}(x^{%
\widehat{k}},v).$ The existence of a nontrivial matter source of type (\ref%
{sdiag4}) does not change the condition $n_{\widehat{k}}^{\ast }=0,$ see (%
\ref{cond3c}), necessary for extracting torsionless configurations. This
mean that we have to consider only trivial solutions of (\ref{ep4a}) when
two functions $n_{\widehat{k}}=n_{\widehat{k}}(x^{2},x^{3})$ are subjected
to the condition (\ref{aux31}). We conclude that this class of exact
solutions of the Einstein equations with nontrivial sources (\ref{sdiag4}),
in general relativity, is defined by the ansatz
\begin{eqnarray}
\ _{\shortmid }^{\circ }\mathbf{g} &=&e^{\psi (x^{2},x^{3})}\left[ \epsilon
_{2}\ dx^{2}\otimes dx^{2}+\epsilon _{3}\ dx^{3}\otimes dx^{3}\right] +
\label{es4s} \\
&&h_{4}\left( x^{2},x^{3},v\right) \ \delta v\otimes \delta v+h_{5}\left(
x^{2},x^{3},v\right) \ \delta y^{5}\otimes \delta y^{5},  \nonumber \\
\delta v &=&dv+w_{2}\left( x^{2},x^{3},v\right) dx^{2}+w_{3}\left(
x^{2},x^{3},v\right) dx^{3},  \nonumber \\
\ \delta y^{5} &=&dy^{5}+n_{2}\left( x^{2},x^{3}\right) dx^{2}+n_{3}\left(
x^{2},x^{3}\right) dx^{3},  \nonumber
\end{eqnarray}%
where the coefficients are restricted to satisfy the conditions
\begin{eqnarray}
\epsilon _{2}\psi ^{\bullet \bullet }+\epsilon _{3}\psi ^{^{\prime \prime }}
&=&\Upsilon _{4},  \nonumber \\
h_{5}^{\ast }\phi /h_{4}h_{5} &=&\Upsilon _{2},  \label{ep2b} \\
w_{2}^{\prime }-w_{3}^{\bullet }+w_{3}w_{2}^{\ast }-w_{2}w_{3}^{\ast } &=&0,
\nonumber \\
n_{2}^{\prime }-n_{3}^{\bullet } &=&0,  \nonumber
\end{eqnarray}%
for $w_{\widehat{i}}=\partial _{\widehat{i}}\phi /\phi ^{\ast },$ see (\ref%
{coefa}), being compatible with (\ref{cond3a}) and (\ref{cond3b}), for given
sources $\Upsilon _{4}(x^{\widehat{k}})$ and $\Upsilon _{2}(x^{\widehat{k}%
},v).$ We note that the second equation in (\ref{ep2b}) relates two
functions $h_{4}$ and $h_{5}.$ In references \cite%
{vhep,vt,vp,vs,vesnc,vap2,valg}, we investigated a number of configurations
with nontrivial two and three dimensional solitons, reductions to the Riccati
or Abbel equation, defining off--diagonal deformations of the black hole,
wormhole or Taub NUT spacetimes. Those solutions where constructed to be
with trivial or nontrivial torsions but if the coefficients of the ansatz (%
\ref{es4s}) are restricted to satisfy the conditions (\ref{ep2b}) in a
compatible form with (\ref{cond3a}) and (\ref{cond3b}), for sure, such
metrics will solve the Einstein equations for the Levi--Civita connection.

Finally, we emphasize that the ansatz (\ref{es4s}) defines Einstein spaces
with a cosmological constant $\lambda $ if we put $\Upsilon _{2}=\Upsilon
_{4}=\lambda $ in (\ref{ep2b}).

\setcounter{equation}{0} \renewcommand{\theequation}{3.\arabic{equation}}
\setcounter{section}{0}
\renewcommand{\thesection} {3
}

\section{Anholonomic Transforms and Killing Spa\-ceti\-mes}

Anholonomic deformations can be defined for any primary metric and vielbein
structures on a spacetime $\mathbf{V}$ (as a matter of principle, the
primary metric can be not a solution of the gravitational field equations).
Such deformations always result in a target spacetime possessing one Killing
vector symmetry if the last one is constrained to satisfy the Einstein
equations for the canonical d--connection, or for the Levi--Civita
connection. For such target spacetimes, we can always apply the parametric
transform and generate a set of generic off--diagonal solutions labelled by
a parameter $\theta $ (\ref{ger1t}). There are possible constructions when
the anholonomic frame transforms are applied to a family of metrics
generated by the Geroch method as new exact solutions of the vacuum Einstein
equations, but such primary metrics have to be parametrized by certain type
ansatz admitting anholonomic transforms to other classes of exact solutions.

\subsection{Nonholonomic deformations of metrics}

\label{ssndm}Let us consider a $(n+m)$--dimensional manifold (spacetime) $%
\mathbf{V},$ $n\geq 2,m\geq 1,$ enabled with a metric structure $\mathbf{%
\check{g}}=\check{g}\oplus _{N}\ \check{h}$ distinguished in the form
\begin{eqnarray}
\mathbf{\check{g}} &=&\check{g}_{i}(u)(dx^{i})^{2}+\check{h}_{a}(u)(\mathbf{%
\check{c}}^{a})^{2},  \label{m1} \\
\mathbf{\check{c}}^{a} &=&dy^{a}+\check{N}_{i}^{a}(u)dx^{i}.  \nonumber
\end{eqnarray}%
The local coordinates are parametrized $u=(x,y)=\{u^{\alpha
}=(x^{i},y^{a})\},$ for the indices of type $i,j,k,...=1,2,...,n$ (in brief,
horizontal, or h--indices/ components) and $a,b,c,...=n+1,n+2,...n+m$
(vertical, or v--indices/ components). We suppose that, in general, the
metric (\ref{m1}) is not a solution of the Einstein equations but can be
nonholonomically deformed in order to generate exact solutions. The
coefficients $\check{N}_{i}^{a}(u)$ from (\ref{m1}) state a conventional $%
\left( n+m\right) $--splitting $\oplus _{\check{N}}$ in any point $u\in
\mathbf{V}$ and define a class of 'N--adapted' local bases
\begin{equation}
\mathbf{\check{e}}_{\alpha }=\left( \mathbf{\check{e}}_{i}=\frac{\partial }{%
\partial x^{i}}-\check{N}_{i}^{a}(u)\ \frac{\partial }{\partial y^{a}},e_{a}=%
\frac{\partial }{\partial y^{a}}\right)  \label{b1}
\end{equation}%
and local dual bases (co--frames) $\mathbf{\check{c}}=(c,\check{c}),$ when
\begin{equation}
\mathbf{\check{c}}^{\alpha }=\left( c^{j}=dx^{i},\mathbf{\check{c}}%
^{b}=dy^{b}+\check{N}_{i}^{b}(u)\ dx^{i}\right) ,  \label{cb1}
\end{equation}%
for $\mathbf{\check{c}\rfloor \ \check{e}=I,}$ i.e. $\mathbf{\check{e}}%
_{\alpha }\mathbf{\rfloor }$ $\mathbf{\check{c}}^{\beta }=\delta _{\alpha
}^{\beta },$ where the inner product is denoted by '$\mathbf{\rfloor }$' and
the Kronecker symbol is written $\delta _{\alpha }^{\beta }.$ The vielbeins (%
\ref{b1}) satisfy the nonholonomy (equivalently, anholonomy) relations
\begin{equation}
\mathbf{\check{e}}_{\alpha }\mathbf{\check{e}}_{\beta }-\mathbf{\check{e}}%
_{\beta }\mathbf{\check{e}}_{\alpha }=\mathbf{\check{w}}_{\alpha \beta
}^{\gamma }\mathbf{\check{e}}_{\gamma }
\end{equation}%
with nontrivial anholonomy coefficients
\begin{eqnarray}
\mathbf{\check{w}}_{ji}^{a} &=&-\mathbf{\check{w}}_{ij}^{a}=\mathbf{\check{%
\Omega}}_{ij}^{a}\doteqdot \mathbf{\check{e}}_{j}\left( \check{N}%
_{i}^{a}\right) -\mathbf{\check{e}}_{i}\left( \check{N}_{j}^{a}\right) ,
\label{anhc} \\
\mathbf{\check{w}}_{ia}^{b} &=&-\mathbf{\check{w}}_{ai}^{b}=e_{a}(\check{N}%
_{j}^{b}).  \nonumber
\end{eqnarray}

A metric $\mathbf{g}=g\oplus _{N}h$ parametrized in the form
\begin{eqnarray}
\mathbf{g} &=&g_{i}(u)(c^{i})^{2}+g_{a}(u)(\mathbf{c}^{a}),  \label{m2} \\
\mathbf{c}^{a} &=&dy^{a}+N_{i}^{a}(u)dx^{i}  \nonumber
\end{eqnarray}%
is a nonhlonomic transform (deformation), preserving the $(n+m)$--splitting,
of the metric, $\mathbf{\check{g}}=\check{g}\oplus _{\check{N}}\ \check{h}$
if the coefficients of (\ref{m1}) and (\ref{m2}) are related by formulas
\begin{equation}
g_{i}=\eta _{i}(u)\ \check{g}_{i},\ h_{a}=\eta _{a}(u)\ \check{h}_{a}%
\mbox{
and }N_{i}^{a}=\eta _{i}^{a}(u)\check{N}_{i}^{a},  \label{polf}
\end{equation}%
where the summation rule is not considered for the indices of gravitational
'polarizations' $\eta _{\alpha }=(\eta _{i},\eta _{a})$ \ and $\eta _{i}^{a}$
in (\ref{polf}). For nontrivial values of $\eta _{i}^{a}(u),$ the
nonholonomic frames (\ref{b1}) and (\ref{cb1}) transform correspondingly
into
\begin{equation}
\mathbf{e}_{\alpha }=\left( \mathbf{e}_{i}=\frac{\partial }{\partial x^{i}}%
-N_{i}^{a}(u)\ \frac{\partial }{\partial y^{a}},e_{a}=\frac{\partial }{%
\partial y^{a}}\right)  \label{b1a}
\end{equation}%
and
\begin{equation}
\mathbf{c}^{\alpha }=\left( c^{j}=dx^{i},\mathbf{c}^{a}=dy^{a}+N_{i}^{a}(u)\
dx^{i}\right)  \label{cb1a}
\end{equation}%
with the anholonomy coefficients $\mathbf{W}_{\alpha \beta }^{\gamma }$
defined by $N_{i}^{a}$ (\ref{anhncc}).

We emphasize that in order to generate exact solutions, the gravitational
'polarizations' $\eta _{\alpha }=(\eta _{i},\eta _{a})$ \ and $\eta _{i}^{a}$
in (\ref{polf}) are not arbitrary functions but restricted in  such a form
that the values
\begin{eqnarray}
\pm 1 &=&\eta _{1}(u^{\alpha })\ \check{g}_{1}(u^{\alpha }),\   \label{polf1}
\\
g_{2}(x^{2},x^{3}) &=&\eta _{2}(u^{\alpha })\ \check{g}_{2}(u^{\alpha }),\
g_{3}(x^{2},x^{3})=\eta _{3}(u^{\alpha })\ \check{g}_{3}(u^{\alpha }),
\nonumber \\
h_{4}(x^{i},v) &=&\eta _{4}(u^{\alpha })\ \check{h}_{4}(u^{\alpha }),\
h_{5}(x^{i},v)=\eta _{5}(u^{\alpha })\ \check{h}_{5}(u^{\alpha }),  \nonumber \\
w_{i}(x^{i},v) &=&\eta _{i}^{4}(u^{\alpha })\check{N}_{i}^{4}(u^{\alpha }),\
n_{i}(x^{i},v)=\eta _{i}^{5}(u^{\alpha })\check{N}_{i}^{5}(u^{\alpha }),
\nonumber
\end{eqnarray}%
define an ansatz of type (\ref{gensol1}), or (\ref{4ds}) (for vacuum
configurations) and (\ref{es4s}) for nontrivial matter sources $\Upsilon
_{2}(x^{2},x^{3},v)$ and $\Upsilon _{4}(x^{2},x^{3}).$

Any nonholonomic deformation%
\begin{equation}
\mathbf{\check{g}}=\check{g}\oplus _{\check{N}}\ \check{h}\longrightarrow \
\mathbf{g}=g\oplus _{N}h  \label{nfd}
\end{equation}%
can be described by two vielbein matrices of type (\ref{vt1}),
\begin{equation}
\mathbf{\check{A}}_{\alpha }^{\ \underline{\alpha }}(u)=\left[
\begin{array}{cc}
\delta _{i}^{\ \underline{i}} & -\check{N}_{j}^{b}\delta _{b}^{\ \underline{a%
}} \\
0 & \delta _{a}^{\ \underline{a}}%
\end{array}%
\right] ,  \label{vtn1}
\end{equation}%
generating the d--metric $\mathbf{\check{g}}_{\alpha \beta }=\mathbf{\check{A%
}}_{\alpha }^{\ \underline{\alpha }}\mathbf{\check{A}}_{\beta }^{\
\underline{\beta }}\check{g}_{\underline{\alpha }\underline{\beta }},$ see
formula (\ref{fmt}), and
\begin{equation}
\mathbf{A}_{\alpha }^{\ \underline{\alpha }}(u)=\left[
\begin{array}{cc}
\sqrt{|\eta _{i}|}\delta _{i}^{\ \underline{i}} & -\eta _{i}^{a}\check{N}%
_{j}^{b}\delta _{b}^{\ \underline{a}} \\
0 & \sqrt{|\eta _{a}|}\delta _{a}^{\ \underline{a}}%
\end{array}%
\right] ,  \label{vtn2}
\end{equation}%
generating the d--metric $\mathbf{g}_{\alpha \beta }=\mathbf{A}_{\alpha }^{\
\underline{\alpha }}\mathbf{A}_{\beta }^{\ \underline{\beta }}\check{g}_{%
\underline{\alpha }\underline{\beta }}$ (\ref{polf1}).

If the metric and N--connection coefficients (\ref{polf}) are stated to be
those from an ansatz (\ref{gensol1}) (or (\ref{4ds})), we should write $\
^{\circ }\mathbf{g}=g\oplus _{N}h$ (or $\ _{\shortmid }^{\circ }\mathbf{g}%
=g\oplus _{N}h$) and say that the metric $\mathbf{\check{g}}=\check{g}\oplus
_{N}\ \check{h}$ (\ref{m1}) was nonholonomically deformed in order to
generate an exact solution of the Einstein equations for the canonical
d--connection (or in a restricted case, for the Levi--Civita connection). In
general, such metrics have very different geometrical and (if existing)
physical properties. Nevertheless, at least for some classes of 'small'
nonsingular nonholonomic deformations, it is possible to preserve a similar
physical interpretation by introducing small polarizations of metric
coefficients and deformations of existing horizons, not changing the
singular structure of curvature tensors. We shall construct explicit
examples and discuss the details in Section 4.

\subsection{Superpositions of the parametric transforms and anholonomic
deformations}

As a matter of principle, any first type parametric transform can be
represented as a generalized anholonomic frame transform labelled by an
additional parameter. It should be also noted that there are two
possibilities to define superpositions of the parametic transforms and
anholonomic frame deformations both resulting in new classes of exact
solutions of the vacuum Einstein equations. In the first case, we start with
a parametric transform and, in the second case, the anholonomic deformations
are considered from the beginning. The aim of this section is to examine
such possibilities.

\subsubsection{The parametric transforms as generalized anholonomic
deformations}

We note that any metric $\ _{\shortmid }^{\circ }g_{\alpha \beta }$ defining
an exact solution of the vacuum Einstein equations can be represented in the
form (\ref{m1}). Then, any metric $\ _{\shortmid }^{\circ }\widetilde{g}%
_{\alpha \beta }(\theta )$ (\ref{ger1t}) from a family of new solutions
generated by the first type parametric transform can be written as (\ref{m2}%
) and related via certain polarization functions of type (\ref{polf}), in
the parametric case depending on parameter $\theta ,$ i.e. $\ \eta _{\alpha
}(\theta )=(\eta _{i}(\theta ),\eta _{a}(\theta ))$ \ and $\eta
_{i}^{a}(\theta ).$ Roughly speaking, the parametric transform can be
represented as a generalized class of anholonomic frame transforms
additionally parametrized by $\theta $ and adapted to preserve the $(n+m)$%
--splitting structure.\footnote{%
It should be emphasized that such constructions are not trivial, for usual
coordinate transforms, if at least one of the primary or target metrics is
generic off-diagonal.} The corresponding vielbein transforms (\ref{qe}) and (%
\ref{qef}) are parametrized, respectively, by matrices of type (\ref{vtn1})
and (\ref{vtn2}), also ''labelled'' by $\theta .$ Such nonholonomic
parametric deformations
\begin{equation}
\ _{\shortmid }^{\circ }\mathbf{g}=\ _{\shortmid }^{\circ }g\oplus _{%
\check{N}}\ _{\shortmid }^{\circ }h\longrightarrow \ \ _{\shortmid }^{\circ }%
\widetilde{\mathbf{g}}(\theta )=\ _{\shortmid }^{\circ }\widetilde{g}(\theta
)\oplus _{N(\theta )}\ _{\shortmid }^{\circ }\widetilde{h}(\theta )
\label{gfd}
\end{equation}%
are defined by the vielbein matrices,
\begin{equation}
\ _{\shortmid }^{\circ }\mathbf{A}_{\alpha }^{\ \underline{\alpha }}(u)=%
\left[
\begin{array}{cc}
\delta _{i}^{\ \underline{i}} & -\ _{\shortmid }^{\circ }N_{j}^{b}(u)\delta
_{b}^{\ \underline{a}} \\
0 & \delta _{a}^{\ \underline{a}}%
\end{array}%
\right] ,  \label{qep}
\end{equation}%
generating the d--metric $\ _{\shortmid }^{\circ }\mathbf{g}_{\alpha \beta
}=\ _{\shortmid }^{\circ }\mathbf{A}_{\alpha }^{\ \underline{\alpha }}\
_{\shortmid }^{\circ }\mathbf{A}_{\beta }^{\ \underline{\beta }}\
_{\shortmid }^{\circ }g_{\underline{\alpha }\underline{\beta }}$ and
\begin{equation}
\widetilde{\mathbf{A}}_{\alpha }^{\ \underline{\alpha }}(u,\theta )=\left[
\begin{array}{cc}
\sqrt{|\eta _{i}(u,\theta )|}\delta _{i}^{\ \underline{i}} & -\eta
_{i}^{a}(u,\theta )\ _{\shortmid }^{\circ }N_{j}^{b}(u)\delta _{b}^{\
\underline{a}} \\
0 & \sqrt{|\eta _{a}(u,\theta )|}\delta _{a}^{\ \underline{a}}%
\end{array}%
\right] ,  \label{qefp}
\end{equation}%
generating the d--metric $\ _{\shortmid }^{\circ }\widetilde{\mathbf{g}}%
_{\alpha \beta }(\theta )=\widetilde{\mathbf{A}}_{\alpha }^{\ \underline{%
\alpha }}\widetilde{\mathbf{A}}_{\beta }^{\ \underline{\beta }}\ _{\shortmid
}^{\circ }g_{\underline{\alpha }\underline{\beta }}.$ Using the matrices (%
\ref{qep}) and (\ref{qefp}), we can compute the matrix of parametric
transforms%
\begin{equation}
\widetilde{\mathbf{B}}_{\alpha }^{\ \alpha ^{\prime }}=\widetilde{\mathbf{A}}%
_{\alpha }^{\ \underline{\alpha }}\ \ _{\shortmid }^{\circ }\mathbf{A}_{%
\underline{\alpha }}^{\ \alpha ^{\prime }},  \label{mgt1}
\end{equation}%
like in (\ref{mgt}), but for ''boldfaced' objects, where $\ _{\shortmid
}^{\circ }\mathbf{A}_{\underline{\alpha }}^{\ \alpha ^{\prime }}$ is inverse
to $\ _{\shortmid }^{\circ }\mathbf{A}_{\alpha ^{\prime }}^{\ \underline{%
\alpha }},$ \footnote{%
we use a ''boldface'' symbol because in this case the constructions are
adapted to a $(n+m)$--splitting} and define the target set of metrics in the
form
\begin{equation}
\ _{\shortmid }^{\circ }\widetilde{\mathbf{g}}_{\alpha \beta }=\widetilde{%
\mathbf{B}}_{\alpha }^{\ \alpha ^{\prime }}(u,\theta )\ \widetilde{\mathbf{B}%
}_{\beta }^{\ \beta ^{\prime }}(u,\theta )\ _{\shortmid }^{\circ }\mathbf{g}%
_{\alpha ^{\prime }\beta ^{\prime }}.
\end{equation}

At first site, there are two substantial differences from the case of usual
anholonomic frame transforms (\ref{nfd}) and the case of parametric
deformations (\ref{gfd}). The first one is that the metric $\mathbf{\check{g}%
}$ was not constrained to be an exact solution of the Einstein equations
like it was required for $\ _{\shortmid }^{\circ }\mathbf{g.}$ The second
one is that even $\ \mathbf{g}$ can be restricted to be an exact vacuum
solution, generated by a special type of deformations (\ref{polf1}), in
order to get an ansatz of type (\ref{4ds}), an arbitrary metric from a
family of solutions $\ _{\shortmid }^{\circ }\widetilde{\mathbf{g}}_{\alpha
\beta }(\theta )$ will not be parametrized in a form that the coefficients
will satisfy the conditions (\ref{cond5}). Nevertheless, even in such cases,
we can consider additional nonholonomic frame transforms when $\mathbf{%
\check{g}}$ is transformed into an exact solution and any particular metric
from the set $\left\{ \ _{\shortmid }^{\circ }\widetilde{\mathbf{g}}_{\alpha
\beta }(\theta )\right\} $ will be deformed into an exact solution defined
by an ansatz (\ref{4ds}) with additional dependence on $\theta .$

The first result of this section is that, by superpositions of nonholonomic
deformations, we can always parametrize a solution formally constructed
following the Geroch method (from a primary solution depending on variables $%
x^{2},x^{3})$ in the form
\begin{eqnarray}
&&\ _{\shortmid }^{\circ }\mathbf{\tilde{g}(}\theta \mathbf{)} =e^{\psi
(x^{2},x^{3},\theta )}\left[ \epsilon _{2}\ dx^{2}\otimes dx^{2}+\epsilon
_{3}\ dx^{3}\otimes dx^{3}\right] + \epsilon _{4}h_{0}^{2}\ \label{4dst} \\
&& \left[ b^{\ast }\left( x^{i},v,\theta \right) %
\right] ^{2}\ \delta v\otimes \delta v
 +\epsilon _{5}\left[ b\left( x^{2},x^{3},v,\theta \right)
-b_{0}(x^{2},x^{3},\theta )\right] ^{2}\ \delta y^{5}\otimes \delta y^{5},
\nonumber \\
&&\delta v =dv+w_{2}\left( x^{2},x^{3},v,\theta \right) dx^{2}+w_{3}\left(
x^{2},x^{3},v,\theta \right) dx^{3},  \nonumber \\
&&\ \delta y^{5} =dy^{5}+n_{2}\left( x^{2},x^{3},\theta \right)
dx^{2}+n_{3}\left( x^{2},x^{3},\theta \right) dx^{3},  \nonumber
\end{eqnarray}%
with the coefficients restricted to solve the equations (\ref{cond5}) but
depending additionally on parameter $\theta ,$
\begin{eqnarray}
\epsilon _{2}\psi ^{\bullet \bullet }\mathbf{(}\theta \mathbf{)}+\epsilon
_{3}\psi ^{^{\prime \prime }}\mathbf{(}\theta \mathbf{)} &=&0,
\label{const6} \\
w_{2}^{\prime }\mathbf{(}\theta \mathbf{)}-w_{3}^{\bullet }\mathbf{(}\theta
\mathbf{)}+w_{3}w_{2}^{\ast }\mathbf{(}\theta \mathbf{)}-w_{2}\mathbf{(}%
\theta \mathbf{)}w_{3}^{\ast }\mathbf{(}\theta \mathbf{)} &=&0,  \nonumber \\
n_{2}^{\prime }\mathbf{(}\theta \mathbf{)}-n_{3}^{\bullet }\mathbf{(}\theta
\mathbf{)} &=&0,  \nonumber
\end{eqnarray}%
for $w_{2}\mathbf{(}\theta \mathbf{)}=\left( b^{\ast }\mathbf{(}\theta
\mathbf{)}\right) ^{-1}(b\mathbf{(}\theta \mathbf{)}+b_{0}\mathbf{(}\theta
\mathbf{)})^{\bullet }$ and $w_{3}=\left( b^{\ast }\mathbf{(}\theta \mathbf{)%
}\right) ^{-1}(b\mathbf{(}\theta \mathbf{)}+b_{0}\mathbf{(}\theta \mathbf{)}%
)^{\prime },$ where, for instance, $n_{3}^{\bullet }\mathbf{(}\theta \mathbf{%
)}=\partial _{2}n_{3}\mathbf{(}\theta \mathbf{)}$ and $n_{2}^{\prime
}=\partial _{3}n_{2}\mathbf{(}\theta \mathbf{)}.$

The second result of this section is that if even, in general, any primary
solution $\ _{\shortmid }^{\circ }\mathbf{g}$ can not be parametrized as an
ansatz (\ref{4ds}), it is possible to define nonholonomic deformations to
such a generic off--diagonal ansatz $\ _{\shortmid }^{\circ }\mathbf{%
\check{g}}$ $\ $or any $\mathbf{\check{g}},$ defined by an ansatz (\ref{m1}%
), which in its turn can be transformed into a metric of type (\ref{4dst})
without dependence on $\theta .$\footnote{%
in our formulas we shall not point dependencies on coordinate variables if
that will not result in ambiguities}

Finally, we emphasize that in spite of the fact that both the parametric and
anholonomic frame transforms can be parametrized in very similar forms by
using vielbein transforms there is a criteria distinguishing them one from
another: For a ''pure'' parametric transform, the matrix $\widetilde{\mathbf{%
B}}_{\alpha }^{\ \alpha ^{\prime }}(u,\theta )$ and related $\widetilde{%
\mathbf{A}}_{\alpha }^{\ \underline{\alpha }}\ $and$\ _{\shortmid }^{\circ }%
\mathbf{A}_{\underline{\alpha }}^{\ \alpha ^{\prime }}$are generated by a
solution of the Geroch equations (\ref{eq01}). If the ''pure'' nonholonomic
deformations, or their superposition with a parametric transform, are
introduced into consideration, the matrix $\mathbf{A}_{\alpha }^{\
\underline{\alpha }}(u)$ (\ref{vtn2}), or its generalization to a matrix $%
\widetilde{\mathbf{A}}_{\alpha }^{\ \underline{\alpha }}$ (\ref{qefp}), can
be not derived only from solutions of (\ref{eq01}). Such transforms define
certain, in general, nonintegrable distributions related to new classes of
Einstein equations.

\subsubsection{Parametric transforms of anholonomically generated solutions
and two parameter transforms}

First, let us consider an exact vacuum solution $\ _{\shortmid }^{\circ }%
\mathbf{g}$\ (\ref{4ds}) in Einstein gravity generated following the
anholonomic frame method. Even through it is generic off--diagonal and depends on
various types of integration functions and constants, it is obvious that it
possesses at least a Killing vector symmetry because the metric does not
depend on variable $y^{5}.$ We can apply the first type parametric transform
to a metric generated by anholonomic deforms (\ref{nfd}). If we work in a
coordinate base with the coefficients of $\ _{\shortmid }^{\circ }\mathbf{g}$
defined in the form $\ _{\shortmid }^{\circ }\underline{g}_{\alpha \beta }=\
_{\shortmid }^{\circ }g_{\underline{\alpha }\underline{\beta }},$ we
generate a set of exact solutions
\begin{equation}
\ _{\shortmid }^{\circ }\underline{\widetilde{g}}_{\alpha \beta }\mathbf{(}%
\theta ^{\prime }\mathbf{)}=\widetilde{B}_{\alpha }^{\ \alpha ^{\prime
}}(\theta ^{\prime })\ \widetilde{B}_{\alpha }^{\ \beta ^{\prime }}(\theta
^{\prime })\ _{\shortmid }^{\circ }\underline{g}_{\alpha ^{\prime }\beta
^{\prime }},
\end{equation}%
see (\ref{ger1t}), were the transforms (\ref{mgt}), labelled by a parameter $%
\theta ^{\prime },$ are not adapted to a nonholonomic $(n+m)$--splitting. We
can elaborate N--adapted constructions starting with an exact solution
parametrized in the form (\ref{m2}), for instance, like $\ _{\shortmid
}^{\circ }\mathbf{g_{\alpha ^{\prime }\beta ^{\prime }}=A}_{\alpha ^{\prime
}}^{\ \underline{\alpha }}\mathbf{A}_{\beta ^{\prime }}^{\ \underline{\beta }%
}\check{g}_{\underline{\alpha }\underline{\beta }}$ , with $\mathbf{A}%
_{\alpha }^{\ \underline{\alpha }}$ being of type (\ref{vtn2}) with
coefficients satisfying the conditions (\ref{polf1}). The target 'boldface'
solutions are generated as transforms
\begin{equation}
\ _{\shortmid }^{\circ }\widetilde{\mathbf{g}}_{\alpha \beta }\mathbf{(}%
\theta ^{\prime }\mathbf{)}=\widetilde{\mathbf{B}}_{\alpha }^{\ \alpha
^{\prime }}(\theta ^{\prime })\ \widetilde{\mathbf{B}}_{\alpha }^{\ \beta
^{\prime }}(\theta ^{\prime })\ _{\shortmid }^{\circ }\mathbf{g}_{\alpha
^{\prime }\beta ^{\prime }},  \label{pts}
\end{equation}%
where
\begin{equation}
\widetilde{\mathbf{B}}_{\alpha }^{\ \alpha ^{\prime }}=\widetilde{\mathbf{A}}%
_{\alpha }^{\ \underline{\alpha }}\ \ _{\shortmid }^{\circ }\mathbf{A}_{%
\underline{\alpha }}^{\ \alpha ^{\prime }},
\end{equation}%
like in (\ref{mgt}), but for ''boldfaced' objects, the matrix $\ _{\shortmid
}^{\circ }\mathbf{A}_{\underline{\alpha }}^{\ \alpha ^{\prime }}$ is inverse
to
\begin{equation}
\ _{\shortmid }^{\circ }\mathbf{A}_{\alpha ^{\prime }}^{\ \underline{\alpha }%
}(u)=\left[
\begin{array}{cc}
\sqrt{|\eta _{i^{\prime }}|}\delta _{i^{\prime }}^{\ \underline{i}} & -\eta
_{i^{\prime }}^{b^{\prime }}\check{N}_{j^{\prime }}^{b^{\prime }}\delta
_{b^{\prime }}^{\ \underline{a}} \\
0 & \sqrt{|\eta _{a^{\prime }}|}\delta _{a^{\prime }}^{\ \underline{a}}%
\end{array}%
\right]
\end{equation}%
and the matric  is considered
\begin{equation}
\widetilde{\mathbf{A}}_{\alpha }^{\ \underline{\alpha }}(u,\theta ^{\prime
})=\left[
\begin{array}{cc}
\sqrt{|\eta _{i}\ \widetilde{\eta }_{i}(\theta ^{\prime })|}\delta
_{i^{\prime }}^{\ \underline{i}} & -\eta _{i}^{b}\ \widetilde{\eta }%
_{i}^{b}(\theta ^{\prime })\check{N}_{j}^{b}\delta _{b}^{\ \underline{a}} \\
0 & \sqrt{|\eta _{a}\ \widetilde{\eta }_{a}(\theta ^{\prime })|}\delta
_{a}^{\ \underline{a}}%
\end{array}%
\right] ,
\end{equation}%
where $\widetilde{\eta }_{i}(u,\theta ^{\prime }),\widetilde{\eta }%
_{a}(u,\theta ^{\prime })$ and $\widetilde{\eta }_{i}^{a}(u,\theta ^{\prime
})$ are gravitational polarizations of type (\ref{polf}).\footnote{%
we do not summarize on repeating two indices if they both are of lower/
upper type} Here it should be emphasized that even $\ _{\shortmid }^{\circ }%
\widetilde{\mathbf{g}}_{\alpha \beta }(\theta ^{\prime })$ are exact
solutions of the vacuum Einstein equations they can not be represented by
ansatz of type (\ref{4dst}), with $\theta \rightarrow \theta ^{\prime },$
because the mentioned polarizations were not constrained to be of type (\ref%
{polf1}) and satisfy any conditions of type (\ref{const6}).\footnote{%
As a matter of principle, we can deform nonholonomically any solution from
the family $\ _{\shortmid }^{\circ }\widetilde{\mathbf{g}}_{\alpha \beta
}(\theta ^{\prime })$ to an ansatz of type (\ref{4dst}).}

Now, we prove that by using superpositions of nonholonomic and parametric
transforms we can generate two parameter families of solutions. This is
possible, for instance, if the metric $\ _{\shortmid }^{\circ }\mathbf{g}%
_{\alpha ^{\prime }\beta ^{\prime }}$ form (\ref{pts}), in its turn, was
generated as an ansatz of type (\ref{4dst}) from another exact solution $\
_{\shortmid }^{\circ }\mathbf{g}_{\alpha ^{\prime \prime }\beta ^{\prime
\prime }}.$ We write
\begin{equation}
\ _{\shortmid }^{\circ }\mathbf{g}_{\alpha ^{\prime }\beta ^{\prime
}}(\theta )=\widetilde{\mathbf{B}}_{\alpha ^{\prime }}^{\ \alpha ^{\prime
\prime }}(u,\theta )\ \widetilde{\mathbf{B}}_{\beta ^{\prime }}^{\ \beta
^{\prime \prime }}(u,\theta )\ _{\shortmid }^{\circ }\mathbf{g}_{\alpha
^{\prime \prime }\beta ^{\prime \prime }}
\end{equation}%
and define the superposition of transforms
\begin{equation}
\ _{\shortmid }^{\circ }\widetilde{\mathbf{g}}_{\alpha \beta }\mathbf{(}%
\theta ^{\prime },\theta \mathbf{)}=\widetilde{\mathbf{B}}_{\alpha }^{\
\alpha ^{\prime }}(\theta ^{\prime })\ \widetilde{\mathbf{B}}_{\alpha }^{\
\beta ^{\prime }}(\theta ^{\prime })\ \widetilde{\mathbf{B}}_{\alpha
^{\prime }}^{\ \alpha ^{\prime \prime }}(\theta )\ \widetilde{\mathbf{B}}%
_{\beta ^{\prime }}^{\ \beta ^{\prime \prime }}(\theta )\ _{\shortmid
}^{\circ }\mathbf{g}_{\alpha ^{\prime \prime }\beta ^{\prime \prime }}.
\label{pts1}
\end{equation}%
It can be considered an iteration procedure of nonholonomic parametric
transforms of type (\ref{pts1}) when an exact vacuum solution of the
Einstein equations is related via a multi $\theta $--parameters vielbein map
with another prescribed vacuum solution. Using anholonomic deformations, one
introduces (into chains of such transforms) certain classes of metrics which
are not exact solutions but nonholonomically deformed from, or to, some
exact solutions.

Finally, we briefly discuss the symmetry properties of such anholonomic
multi $\theta $--parameter solutions of the Einstein equations. In the
parameter space, they possess symmetries with infinite dimensional
parametric group structures \cite{geroch1,geroch2} but with respect to
anholonomic deforms one can be considered various types of prescribed Lie
algebroid, solitonic, pp--wave and/or nonholonomic noncommutative symmetries %
\cite{vesnc,valg,vsgg}. In general, many of such way generated solutions do
not have obvious physical interpretation. Nevertheless, if certain small
(non--coordinate) parameters of nonholonomic deformations are introduced
into consideration, it is possible to prescribe various interesting physical
situations for a subset of metrics generated by maps of type (\ref{pts1}),
preserving certain similarities with a primary solution. We construct and
analyze some examples of such solutions in the next section.

\setcounter{equation}{0} \renewcommand{\theequation} {4.\arabic{equation}}
\setcounter{section}{0}
 \renewcommand{\thesection}  {4
 }

\section{Examples of Off--Diagonal Exact Solutions}

\label{sexesol}The purpose of this section is to present explicit examples of
how superpositions of nonholonomic deformations and parametric transforms
can be applied in order to generate new classes of solutions and how
physically valuable configurations can be selected. Some constructions will
be performed for 5D spacetimes with torsion, for instance, related to the
so--called (antisymmetric) "H--fields" in string gravity but the bulk of
them will be restricted to define usual 4D Einstein spacetimes with generic
off--diagonal metrics.

\subsection{Five classes of primary metrics}

We begin with a list of 5D quadratic elements (defined by certain primary
metrics) which will be used for generating new classes of exact solutions
following superpositions of nonholonomic deformations and parametric
transforms:

The fist quadratic element, defined as a particular case of metric (\ref{m1}%
), is
\begin{equation}
\delta s_{[1]}^{2}=\epsilon _{1}d\chi ^{2}-d\xi ^{2}-r^{2}(\xi )\ d\vartheta
^{2}-r^{2}(\xi )\sin ^{2}\vartheta \ d\varphi ^{2}+\varpi ^{2}(\xi )\ dt^{2}
\label{aux1}
\end{equation}%
where the local coordinates and nontrivial metric coefficients are
parametriz\-ed in the form%
\begin{eqnarray}
x^{1} &=&\chi ,x^{2}=\xi ,x^{3}=\vartheta ,y^{4}=\varphi ,y^{5}=t,
\label{aux1p} \\
\check{g}_{1} &=&\epsilon _{1}=\pm 1,\ \check{g}_{2}=-1,\ \check{g}%
_{3}=-r^{2}(\xi ),\ \check{h}_{4}=-r^{2}(\xi )\sin ^{2}\vartheta ,\ \check{h}%
_{5}=\varpi ^{2}(\xi ),  \nonumber
\end{eqnarray}%
for
\begin{equation}
\xi =\int dr\ \left| 1-\frac{2\mu }{r}+\frac{\varepsilon }{r^{2}}\right|
^{1/2}   \mbox{ and \ } \
\varpi ^{2}(r)=1-\frac{2\mu }{r}+\frac{\varepsilon }{r^{2}}.
\end{equation}%
For the constants $\varepsilon \rightarrow 0$ and $\mu $ being a point mass,
the element (\ref{aux1}) defines just a trivial embedding into 5D (with
extra dimension coordinate $\chi )$ of the Schwarzschild solution written in
spacetime spherical coordinates $(r,\vartheta ,\varphi ,t).$\footnote{%
For simplicity, we consider only the case of vacuum solutions, not analyzing
a more general possibility when $\varepsilon =e^{2}$ is related to the
electric charge for the Reissner--Nordstr\"{o}m metric (see, for example, %
\cite{heu}). In our further considerations we shall treat $\varepsilon $ as
a small parameter, for instance, defining a small deformation of a circle
into an ellipse (eccentricity).}

The second quadratic element%
\begin{equation}
\delta s_{[2]}^{2}=-r_{g}^{2}\ d\varphi ^{2}-r_{g}^{2}\ d\check{\vartheta}%
^{2}+\check{g}_{3}(\check{\vartheta})\ d\check{\xi}^{2}+\epsilon _{1}\ d\chi
^{2}+\check{h}_{5}\ (\xi ,\check{\vartheta})\ dt^{2}  \label{aux2}
\end{equation}%
where the local coordinates are
\begin{equation}
x^{1}=\varphi ,x^{2}=\check{\vartheta},x^{3}=\check{\xi},y^{4}=\chi ,y^{5}=t,
\end{equation}%
for
\begin{equation}
d\check{\vartheta}=d\vartheta /\sin \vartheta ,\ d\check{\xi}=dr/r\sqrt{%
|1-2\mu /r+\varepsilon /r^{2}|},
\end{equation}%
and the Schwarzschild radius of a point mass $\mu $ is defined $%
r_{g}=2G_{[4]}\mu /c^{2},$ where $G_{[4]}$ is the 4D Newton constant and $c$
is the light velocity. The nontrivial metric coefficients in (\ref{aux2})
are parametrized%
\begin{eqnarray}
\check{g}_{1} &=&-r_{g}^{2},\ \check{g}_{2}=-r_{g}^{2},\ \check{g}%
_{3}=-1/\sin ^{2}\vartheta ,\   \label{aux2p} \\
\check{h}_{4} &=&\epsilon _{1},\ \check{h}_{5}=\left[ 1-2\mu /r+\varepsilon
/r^{2}\right] /r^{2}\sin ^{2}\vartheta .  \nonumber
\end{eqnarray}%
The quadratic element defined by (\ref{aux2}) and (\ref{aux2p}) is a trivial
embedding into 5D of the Schwarzschild quadratic element multiplied to the
conformal factor $\left( r\sin \vartheta /r_{g}\right) ^{2}.$ We emphasize
that this metric is not a solution of the Einstein equations but it will be
used in order to construct nonholonomic deformations and parametric
transforms to such solutions.

We shall use a quadratic element when the time coordinate is considered to
be ''anisotropic'',
\begin{equation}
\delta s_{[3]}^{2}=-r_{g}^{2}\ d\varphi ^{2}-r_{g}^{2}\ d\check{\vartheta}%
^{2}+\check{g}_{3}(\check{\vartheta})\ d\check{\xi}^{2}+\check{h}_{4}\ (\xi ,%
\check{\vartheta})\ dt^{2}+\epsilon _{1}\ d\chi ^{2}  \label{aux3}
\end{equation}%
where the local coordinates are
\begin{equation}
x^{1}=\varphi ,\ x^{2}=\check{\vartheta},\ x^{3}=\check{\xi},\ y^{4}=t,\
y^{5}=\chi ,
\end{equation}%
and the nontrivial metric coefficients are parametrized%
\begin{eqnarray}
\check{g}_{1} &=&-r_{g}^{2},\ \check{g}_{2}=-r_{g}^{2},\ \check{g}%
_{3}=-1/\sin ^{2}\vartheta ,\   \label{aux3p} \\
\check{h}_{4} &=&\left[ 1-2\mu /r+\varepsilon /r^{2}\right] /r^{2}\sin
^{2}\vartheta ,\ \check{h}_{5}=\epsilon _{1}.  \nonumber
\end{eqnarray}%
The formulas (\ref{aux3}) and (\ref{aux3p}) are respective
reparametrizations of (\ref{aux2}) and (\ref{aux2p}) when the forth and
 fifth coordinates are inverted. Such metrics will be used for constructing new
classes of exact solutions in 5D with explicit dependence on time like
coordinate.

The forth quadratic element is introduced by inverting the forth and fifth
coordinates in (\ref{aux1}) (having the same definitions as in that case)
\begin{equation}
\delta s_{[4]}^{2}=\epsilon _{1}d\chi ^{2}-d\xi ^{2}-r^{2}(\xi )\ d\vartheta
^{2}+\varpi ^{2}(\xi )\ dt^{2}-r^{2}(\xi )\sin ^{2}\vartheta \ d\varphi ^{2}
\label{aux4}
\end{equation}%
where the local coordinates and nontrivial metric coefficients are
parametriz\-ed in the form%
\begin{eqnarray}
x^{1} &=&\chi ,x^{2}=\xi ,x^{3}=\vartheta ,y^{4}=t,y^{5}=\varphi ,
\label{aux4p} \\
\check{g}_{1} &=&\epsilon _{1}=\pm 1,\ \check{g}_{2}=-1,\ \check{g}%
_{3}=-r^{2}(\xi ),\ \check{h}_{4}=\varpi ^{2}(\xi ),\ \check{h}%
_{5}=-r^{2}(\xi )\sin ^{2}\vartheta .  \nonumber
\end{eqnarray}%
Such metrics can be used for constructing exact solutions in 4D gravity with
anisotropic dependence on time coordinate.

Finally, we consider
\begin{equation}
\delta s_{[5]}^{2}=\epsilon _{1}\ d\chi ^{2}-dx^{2}-dy^{2}-2\kappa (x,y,p)\
dp^{2}+\ dv^{2}/8\kappa (x,y,p)  \label{aux5}
\end{equation}%
where the local coordinates are
\begin{equation}
\ x^{1}=\chi ,\ x^{2}=x,\ x^{3}=y,\ y^{4}=p,\ y^{5}=v,
\end{equation}%
and the nontrivial metric coefficients are parametrized%
\begin{eqnarray}
\check{g}_{1} &=&\epsilon _{1}=\pm 1,\ \check{g}_{2}=-1,\ \check{g}_{3}=-1,\
\label{aux5p} \\
\check{h}_{4} &=&-2\kappa (x,y,p),\ \check{h}_{5}=1/\ 8\ \kappa (x,y,p).
\nonumber
\end{eqnarray}%
The metric (\ref{aux5}) is a trivial embedding into 5D of the vacuum
solution of the Einstein equation defining pp--waves \cite{peres} for any $%
\kappa (x,y,p)$ solving
\begin{equation}
\kappa _{xx}+\kappa _{yy}=0,
\end{equation}%
with $p=z+t$ and $v=z-t,$ where $(x,y,z)$ are usual Cartesian coordinates
and $t$ is the time like coordinates. The simplest explicit examples of such
solutions are
\begin{equation}
\kappa =(x^{2}-y^{2})\sin p,
\end{equation}%
defining a plane monochromatic wave, or
\begin{eqnarray*}
\kappa &=&\frac{xy}{\left( x^{2}+y^{2}\right) ^{2}\exp \left[ p_{0}^{2}-p^{2}%
\right] },\mbox{ for }|p|<p_{0}; \\
&=&0,\mbox{ for }|p|\geq p_{0},
\end{eqnarray*}%
defining a wave packet travelling with unit velocity in the negative $z$
direction.

\subsection{Solitonic pp--waves and string torsion}

Pp--wave solutions are intensively exploited for elaborating string models
with nontrivial backgrounds \cite{strpp1,strpp2,strpp3}. A special interest
for pp--waves in general relativity is the fact that any solution in this
theory can be approximated by a pp--wave in vicinity of horizons. Such
solutions can be generalized by introducing nonlinear interactions with
solitonic waves \cite{gravsolit,bv,vs1,vhep,vp} and nonzero sources with
nonhomogeneous cosmological constant induced by an ansatz for the
antisymmetric tensor fields of third rank. A very important property of such
nonlinear wave solutions is that they possess nontrivial limits defining new
classes of generic off--diagonal vacuum Einstein spacetimes.

\subsubsection{Pp--waves and nonholonomic solitonic interactions}

Let us consider the ansatz
\begin{eqnarray}
\delta s_{[5]}^{2} &=&\epsilon _{1}\ d\chi ^{2}-e^{\psi (x,y)}\left(
dx^{2}+dy^{2}\right)          \nonumber   \\
 &&  -2\kappa (x,y,p)\ \eta _{4}(x,y,p)\delta p^{2}+\
 \frac{\eta _{5}(x,y,p)}{8\kappa (x,y,p)}\delta v^{2} \nonumber   \\
\delta p &=&dp+w_{2}(x,y,p)dx+w_{3}(x,y,p)dy,\   \nonumber \\
\delta v &=&dv+n_{2}(x,y,p)dx+n_{3}(x,y,p)dy     \label{sol2}
\end{eqnarray}%
where the local coordinates are
\begin{equation*}
\ x^{1}=\chi ,\ x^{2}=x,\ x^{3}=y,\ y^{4}=p,\ y^{5}=v,
\end{equation*}%
and the nontrivial metric coefficients and polarizations are parametrized%
\begin{eqnarray}
\check{g}_{1} &=&\epsilon _{1}=\pm 1,\ \check{g}_{2}=-1,\ \check{g}_{3}=-1,
 \nonumber \\
\check{h}_{4} &=&-2\kappa (x,y,p),\ \check{h}_{5}=1/\ 8\kappa (x,y,p),
 \nonumber \\
\eta _{1} &=&1,g_{\alpha }=\eta _{\alpha }\check{g}_{\alpha }.
\end{eqnarray}%
For trivial polarizations $\eta _{\alpha }=1$ and $w_{2,3}=0,$ $n_{2,3}=0,$
the metric (\ref{sol2}) is just the pp--wave solution (\ref{aux5}).

\paragraph{Exact solitonic pp--wave solutions in string gravity:\newline
}

Our aim is to define such nontrivial values of polarization functions when $%
\eta _{5}(x,y,p)$ is defined by a 3D soliton $\phi (x,y,p)$, for instance, a
solution of
\begin{equation}
\phi ^{\bullet \bullet }+\epsilon (\phi ^{\prime }+6\phi \ \phi ^{\ast
}+\phi ^{\ast \ast \ast })^{\ast }=0,\ \epsilon =\pm 1,  \label{solit1}
\end{equation}%
see formula (\ref{afa1}) in Appendix, and $\eta _{2}=\eta _{3}=e^{\psi
(x,y)} $ is a solution of (\ref{ep1a}),
\begin{equation}
\psi ^{\bullet \bullet }+\psi ^{\prime \prime }=\frac{\lambda _{H}^{2}}{2}.
\label{lapl}
\end{equation}%
The solitonic deformations of the pp--wave metric will define exact
solutions in string gravity with $H$--fields, see in Appendix the equations (%
\ref{c01}) and (\ref{c02}) for the string torsion ansatz (\ref{ansh}).%
\footnote{%
as a matter of principle we can consider\ that $\phi$ is a solution of any
3D solitonic, or other, nonlinear wave equation.}

Introducing the\ above stated data for the ansatz (\ref{sol2}) into the
equation (\ref{ep2a}), we get an equation relating $h_{4}=\eta _{4}\check{g}%
_{4}$ and $h_{5}=\eta _{5}\check{g}_{5}.$ Such solutions can be constructed
in general form, respectively, following formulas (\ref{solh4}) and (\ref%
{solh5}) (in this section, we take $\Upsilon _{2}=\lambda _{H}^{2}/2).$ We
obtain
\begin{equation}
\eta _{5}=8\ \kappa (x,y,p)\left[ h_{5[0]}(x,y)+\frac{1}{2\lambda _{H}^{2}}%
e^{2\phi (x,y,p)}\right]  \label{sol2h5}
\end{equation}%
and
\begin{equation}
|\eta _{4}|=\frac{e^{-2\phi (x,y,p)}}{2\kappa ^{2}(x,y,p)}\left[ \left(
\sqrt{|\eta _{5}|}\right) ^{\ast }\right] ^{2}  \label{sol2h4}
\end{equation}%
where $h_{5[0]}(x,y)$ is an integration function. Having defined the
coefficients $h_{a},$ we can solve the equations (\ref{ep3a}) and (\ref{ep4a}%
) in a form similar to (\ref{gensol1w}) and (\ref{gensol1n}) but expressing
the solutions through $\eta _{4}$ and $\eta _{5}$ defined by pp-- and
solitonic waves as in (\ref{sol2h4}) and (\ref{sol2h5}). The corresponding
solutions are
\begin{equation}
w_{1}=0,w_{2}=\left( \phi ^{\ast }\right) ^{-1}\partial _{x}\phi
,w_{3}=\left( \phi ^{\ast }\right) ^{-1}\partial _{x}\phi ,  \label{sol2w}
\end{equation}%
for $\phi ^{\ast }=\partial \phi /\partial p,$ and
\begin{equation}
n_{1}=0,n_{2,3}=n_{2,3}^{[0]}(x,y)+n_{2,3}^{[1]}(x,y)\int \left| \eta
_{4}\eta _{5}^{-3/2}\right| dp,  \label{sol2na}
\end{equation}%
where $n_{2,3}^{[0]}(x,y)$ and $n_{2,3}^{[1]}(x,y)$ are integration
functions.

We conclude that the ansatz (\ref{sol2}) with the coefficients computed
following the equations and formulas (\ref{lapl}), (\ref{sol2h4}), (\ref%
{sol2h5}), (\ref{sol2w}) and (\ref{sol2na}) define a class of exact
solutions (depending on integration functions) of gravitational field
equations in string gravity with $H$--field. In a more explicit form,
depending on above stated functions $\psi ,k,\phi $ and $\eta _{5}$ and
respective integration functions, the class of such solutions is
parametrized as
\begin{eqnarray}
\delta s_{[sol2]}^{2} &=&\epsilon _{1}\ d\chi ^{2}-e^{\psi }\left(
dx^{2}+dy^{2}\right) +\ \frac{\eta _{5}}{8\kappa }\delta p^{2}-\kappa ^{-1}\
e^{-2\phi }\left[ \left( \sqrt{|\eta _{5}|}\right) ^{\ast }\right]
^{2}\delta v^{2},  \nonumber \\
\delta p &=&dp+\left( \phi ^{\ast }\right) ^{-1}\partial _{x}\phi \
dx+\left( \phi ^{\ast }\right) ^{-1}\partial _{y}\phi \ dy,\   \label{sol2a}
\\
\delta v &=&dv+\left\{ n_{2}^{[0]}+\widehat{n}_{2}^{[1]}\int k^{-1}e^{2\phi }%
\left[ \left( \left| \eta _{5}\right| ^{-1/4}\right) ^{\ast }\right]
^{2}dp\right\} dx  \nonumber \\
&&+\left\{ n_{3}^{[0]}+\widehat{n}_{3}^{[1]}\int k^{-1}e^{2\phi }\left[
\left( \left| \eta _{5}\right| ^{-1/4}\right) ^{\ast }\right] ^{2}dp\right\}
dy,  \nonumber
\end{eqnarray}%
where some constants and multiples depending on $x$ and $y$ included into $%
\widehat{n}_{2,3}^{[1]}(x,y).$ It should be noted that such spacetimes
possess nontrivial nonholonomically induced torsion (we omit explicit
formulas for the nontrivial components which can be computed by introducing
the coefficients of our ansatz into formulas (\ref{candcon}) and (\ref{dtors}%
)). This is a very general class of solutions describing nonlinear
interactions of pp--waves and 3D solutions in string gravity. The term $%
\epsilon _{1}\ d\chi ^{2}$ can be eliminated in order to describe only 4D
configurations. Nevertheless, in this case, there is not a smooth limit of
such 4D solutions for $\lambda _{H}^{2}\rightarrow 0$ to those in general
relativity.

\paragraph{Pp--waves and solitonic interactions in vacuum Einstein gravity:%
\newline
}

\label{ssaux2}We prove that the anholonomic frame method can be used in a
different form in order to define 4D metrics induced by nonlinear pp--waves
and solitonic interactions for vanishing sources and the Levi--Civita
connection. We can apply the formulas (\ref{eq23}), (\ref{lcc2}), (\ref%
{aux31}) \ and (\ref{aux31b}), for simplicity, considering that $b_{0}=0$
and $b(x,y,p)$ being a function generating solitonic and pp--wave
interactions. For an ansatz of type (\ref{sol2}), we write
\begin{equation}
\eta _{5}=5\kappa b^{2}\mbox{ and }\eta _{4}=h_{0}^{2}(b^{\ast
})^{2}/2\kappa .
\end{equation}%
A 3D solitonic solution can be generated if $b$ is subjected to the
condition to solve a solitonic equation, like $\phi $ in (\ref{solit1}). It
is not possible to satisfy the integrability conditions (\ref{aux31}) for
any $w_{\widehat{k}}=\left( b^{\ast }\right) ^{-1}\partial b/\partial x^{%
\widehat{k}}$ which is necessary for the equality of the coefficients of the
canonical d--connection to those of the Levi--Civita connection.\footnote{%
If such integrability conditions are not satisfied, the solutions may also
exist but they can not be constructed in explicit form.} \ Here we follow a
more simple parametrization when
\begin{equation}
b(x,y,p)=\breve{b}(x,y)q(p)k(p),
\end{equation}%
for any $\breve{b}(x,y)$ and any pp--wave $\kappa (x,y,p)=\breve{\kappa}%
(x,y)k(p)$ (we can take $\breve{b}=\breve{\kappa}),$ where $q(p)=4\tan
^{-1}(e^{\pm p})$ is the solution of ''one dimensional'' solitonic equation
\begin{equation}
q^{\ast \ast }=\sin q.  \label{sol1d}
\end{equation}%
In this case,
\begin{equation}
w_{2}=\left[ (\ln |qk|)^{\ast }\right] ^{-1}\partial _{x}\ln |\breve{b}|%
\mbox{ and }w_{3}=\left[ (\ln |qk|)^{\ast }\right] ^{-1}\partial _{y}\ln |%
\breve{b}|
\end{equation}%
positively solve the conditions (\ref{aux31}). The final step in
constructing such vacuum Einstein solutions is to chose any two functions $%
n_{2,3}(x,y)$ satisfying the conditions $n_{2}^{\ast }=n_{3}^{\ast }=0$ \
and $n_{2}^{\prime }-n_{3}^{\bullet }=0$ (\ref{aux31b}). This mean that in
the integrals of type (\ref{sol2na}) we shall fix the integration functions $%
n_{2,3}^{[1]}=0$ but take such $n_{2,3}^{[0]}(x,y)$ satisfying $%
(n_{2}^{[0]})^{\prime }-(n_{3}^{[0]})^{\bullet }=0.$

We can consider a trivial solution of (\ref{ep1a}), i.e. of (\ref{lapl})
with $\lambda _{H}=0.$

Summarizing the results, we obtain the 4D vacuum metric
\begin{eqnarray}
\delta s_{[sol2a]}^{2} &=&-\left( dx^{2}+dy^{2}\right) -h_{0}^{2}\breve{b}%
^{2}[(qk)^{\ast }]^{2}\delta p^{2}+\breve{b}^{2}(qk)^{2}\delta v^{2},  \nonumber
\\
\delta p &=&dp +\left[ (\ln |qk|)^{\ast }\right] ^{-1}\partial _{x}\ln |%
\breve{b}|\ dx+\left[ (\ln |qk|)^{\ast }\right] ^{-1}\partial _{y}\ln |%
\breve{b}|\ dy,\   \nonumber \\
\delta v &=&dv+n_{2}^{[0]}dx+n_{3}^{[0]}dy,  \label{sol2b}
\end{eqnarray}%
defining nonlinear gravitational interactions of a pp--wave $\kappa =\breve{%
\kappa}k$ and a soliton $q,$ depending on certain type of integration
functions and constants stated above. Such vacuum Einstein metrics can be
generated in a similar form for 3D or 2D solitons but the constructions will
be more cumbersome and for non--explicit functions, see a number of similar
solutions in Refs. \cite{vs,vs1,vsgg}.

\subsubsection{Parametric transforms and solitonic pp--wave solutions}

There are three possibilities: The first is to apply a parametric transform
to a vacuum solution and then to deform it nonholonomically in order to
generate pp--wave solitonic interactions. In the second case, we can subject
the solution (\ref{sol2b}) to one parameter transforms. Finally, in the
third case, we can derive two parameter families of nonholonomic soliton
pp--wave interactions.

\paragraph{First example: nonholonomic solitonic pp--waves from
parametriz\-ed families of solutions \newline
}

Let us consider the metric
\begin{equation}
\delta s_{[5a]}^{2}=-dx^{2}-dy^{2}-2\breve{\kappa}(x,y)\ dp^{2}+\ dv^{2}/8%
\breve{\kappa}(x,y)  \label{auxpw}
\end{equation}%
which is a particular 4D case\ of (\ref{aux5}) when $\kappa
(x,y,p)\rightarrow \breve{\kappa}(x,y).$ It is easy to show that the
nontrivial Ricci components $R_{\alpha \beta }$ for the Levi--Civita
connection are proportional to $\breve{\kappa}^{\bullet \bullet }+\breve{%
\kappa}^{\prime \prime}$ and the non--vanishing components of the curvature
tensor $R_{\alpha \beta \gamma \delta }$ are of type $R_{a1b1}\simeq
R_{a2b2}\simeq \sqrt{\left( \breve{\kappa}^{\bullet \bullet }\right)
^{2}+\left( \breve{\kappa}^{\bullet \prime }\right) ^{2}}.$ So, any function
$\breve{\kappa}$ solving the equation $\breve{\kappa}^{\bullet \bullet }+%
\breve{\kappa}^{\prime \prime }=0$ but with $\left( \breve{\kappa}^{\bullet
\bullet }\right) ^{2}+\left( \breve{\kappa}^{\bullet \prime }\right)
^{2}\neq 0$ defines a vacuum solution of the Einstein equations. In the
simplest case, we can take $\breve{\kappa}=x^{2}-y^{2}$ or $\breve{\kappa}%
=xy/\sqrt{x^{2}+y^{2}}$ like it was suggested in the original work \cite%
{peres}, but for the metric (\ref{auxpw}) we do not consider any multiple $%
q(p) $ depending on $p.$

Subjecting the metric (\ref{auxpw}) to the parametric transform, we get an
off--diagonal metric of type
\begin{eqnarray}
\delta s_{[2p]}^{2} &=&-\eta _{2}(x,y,\theta )dx^{2}+\eta _{3}(x,y,\theta
)dy^{2}    \nonumber \\
&& -2\breve{\kappa}(x,y)\ \eta _{4}(x,y,\theta )\delta p^{2}+\ \frac{\eta
_{5}(x,y,\theta )}{8\breve{\kappa}(x,y)}\delta v^{2}  \nonumber \\
\delta p &=&dp+w_{2}(x,y,\theta )dx+w_{3}(x,y,\theta )dy,\   \nonumber \\
\delta v &=&dv+n_{2}(x,y,\theta )dx+n_{3}(x,y,\theta )dy  \label{auxpwa}
\end{eqnarray}%
which is also a vacuum solution of the Einstein equations if the
coefficients are restricted to satisfy the necessary conditions: This is a
particular case of vierbein transform (\ref{qef}) when the coefficients $g_{%
\underline{\alpha }\underline{\beta }}$ are defined by the coefficients of (%
\ref{auxpw}) and $_{\shortmid }^{\circ }\widetilde{g}_{\alpha \beta }$ are
given by the coefficients (\ref{auxpwa}). The polarizations $\eta _{\widehat{%
\alpha }}(x,y,\theta )$ and N--connection coefficients $w_{\widehat{i}%
}(x,y,\theta )$ and $n_{\widehat{i}}(x,y,\theta )$ determine \ the
coefficients of matrix $\widetilde{A}_{\alpha }^{\ \underline{\alpha }}$
and, in consequence, of the matrix of parametric transforms $\widetilde{B}%
_{\alpha }^{\ \alpha ^{\prime }}$(\ref{mgt}). They can be defined in
explicit form by solving the Geroch equations (\ref{eq01}) which is possible
for any particular parametrization of function $\breve{\kappa}.$ For our
purposes, it is better to preserve a general parametrization but emphasizing
that because the coefficients of metric (\ref{auxpw}) depend only on
coordinates $x$ and $y,$ we can chose such forms of solutions when the
coefficients of the Levi--Civita connection and Ricci and Riemannian tensors
will also depend on such two coordinates. As a result, we can conclude that $%
\eta _{\widehat{\alpha }}$ and $N_{\widehat{i}}^{a}$ depend on variables $%
(x,y,\theta )$ even if we do not restrict our consideration to an explicit
solution of (\ref{eq01}), of type (\ref{germ1c}).

Considering that $\eta _{2}\neq 0,$\footnote{$\eta _{2}\rightarrow 1$ and $%
\eta _{3}\rightarrow 1$ for infinitesimal parameter transforms} we multiply (%
\ref{auxpwa}) on conformal factor $\left( \eta _{2}\right) ^{-1}$ and
redefining the coefficients as $\breve{\eta}_{3}=\eta _{3}/\eta _{2},\breve{%
\eta}_{a}=\eta _{a}/\eta _{2},$ $\breve{w}_{a}=w_{a}$ and $\breve{n}%
_{a}=n_{a},$ for $\hat{\imath}=2,3$ and $a=4,5,$ we obtain
\begin{eqnarray}
\delta s_{[2a]}^{2} &=&-dx^{2}+\breve{\eta}_{3}(x,y,\theta )dy^{2}
-2\breve{\kappa}(x,y)\ \breve{\eta}_{4}(x,y,\theta )\delta p^{2}+\ \frac{%
\breve{\eta}_{5}(x,y,\theta )}{8\breve{\kappa}(x,y)}\delta v^{2}  \nonumber \\
\delta p &=&dp+\breve{w}_{2}(x,y,\theta )dx+\breve{w}_{3}(x,y,\theta )dy,\
\nonumber \\
\delta v &=&dv+\breve{n}_{2}(x,y,\theta )dx+\breve{n}_{3}(x,y,\theta )dy
 \label{auxpwb}
\end{eqnarray}%
which is not an exact solution but can easy nonholonomically deformed into
exact vacuum solutions by multiplying on additional polarization parameters
(it is described in section \ref{ssndm}). We first introduce the
polarizations $\eta _{2}=\exp \psi (x,y,\theta )$ and $\eta _{3}=\breve{\eta}%
_{3}=-\exp \psi (x,y,\theta )$ defined as solutions of $\psi ^{\bullet
\bullet }+\psi ^{\prime \prime }=0.$ Then we redefine $\breve{\eta}%
_{a}\rightarrow \eta _{a}(x,y,p,\theta )$ (for instance, multiplying on
additional multiples) by introducing additional dependencies on
''anisotropic'' coordinate $p$ such a way when the ansatz (\ref{auxpwb})
transform into
\begin{eqnarray}
\delta s_{[2a]}^{2} &=&-e^{\psi (x,y,\theta )}\left( dx^{2}+dy^{2}\right)
 \nonumber \\
 &&-2\breve{\kappa}(x,y)k(p)\ \eta _{4}(x,y,p,\theta )\delta p^{2}+\ \frac{%
\eta _{5}(x,y,p,\theta )}{8\breve{\kappa}(x,y)k(p)}\delta v^{2}  \nonumber \\
\delta p &=&dp+w_{2}(x,y,p,\theta )dx+w_{3}(x,y,p,\theta )dy,\   \nonumber \\
\delta v &=&dv+n_{2}(x,y,\theta )dx+n_{3}(x,y,\theta )dy.  \label{auxpwc}
\end{eqnarray}%
In order to be a vacuum solution for $g_{4}=-2\breve{\kappa}k\eta _{4}$ and $%
g_{5}=\eta _{5}/8\breve{\kappa}k$ and corresponding the Levi--Civita
connection, the metric (\ref{auxpwc}) should have a parametrization of type (%
\ref{4dst}) with the coefficients subjected to constraints (\ref{const6}) if
the coordinates are parametrized as $x^{2}=x,x^{3}=y,y^{4}=p$ and $y^{5}=v.$
It describes a nonholonomic parametric transform from a vacuum metric (\ref%
{auxpw}) to a family of exact solutions depending on parameter $\theta $ and
defining nonlinear superpositions of pp--waves $\kappa =\breve{\kappa}%
(x,y)k(p).$

It is possible to introduce also solitonic waves into the metric (\ref%
{auxpwc}). For instance, we can take $\eta _{5}(x,y,p,\theta )\sim q(p),$
where $q(p)$ is a solution of solitonic equation (\ref{sol1d}). We obtain a
family of vacuum Einstein metrics labelled by parameter $\theta $ and
defining nonlinear interactions of pp--waves and one--dimensional solitons.
Such solutions with prescribed $\psi =0$ can be parametrized in a form very
similar to the ansatz (\ref{sol2b}). We can give them a very simple physical
interpretation: they define families (packages) of nonlinear off--diagonal
interactions of vacuum gravitational pp--waves and solitons parametrized by
the set of solutions of Geroch equations (\ref{eq01}) for a primary vacuum
metric (\ref{auxpw}).

\paragraph{Second example: Parametric transforms of nonholonomic solitonic
pp--waves \newline
}

We begin with the ansatz (\ref{sol2b}) defining a vacuum off--diagonal
solution. That metric does not depend on variable $v$ and possess a Killing
vector $\partial /\partial v.$ It is possible to apply a parametric
transform as it is described by formula (\ref{pts}). In terms of
polarization functions, the new family of metrics is of type
\begin{eqnarray}
\delta s_{[sol2\theta ^{\prime }]}^{2} &=&-\eta _{2}(\theta ^{\prime })\
dx^{2}+\eta _{3}(\theta ^{\prime })\ dy^{2}   \nonumber \\
 && -\eta _{4}(\theta ^{\prime })\
h_{0}^{2}\breve{b}^{2}[(qk)^{\ast }]^{2}\delta p^{2}
+\eta _{5}(\theta ^{\prime })\ \breve{b}^{2}(qk)^{2}\delta v^{2},  \nonumber
\\
\delta p &=&dp +\eta _{2}^{4}(\theta ^{\prime })\ \left[ (\ln |qk|)^{\ast }%
\right] ^{-1}\partial _{x}\ln |\breve{b}|\ dx  \nonumber \\
 && +\eta _{3}^{4}(\theta ^{\prime })\ \left[ (\ln |qk|)^{\ast }\right]
^{-1}\partial _{y}\ln |\breve{b}|\ dy,\   \nonumber \\
\delta v &=&dv+\eta _{2}^{5}(\theta ^{\prime })n_{2}^{[0]}dx+\eta
_{3}^{5}(\theta ^{\prime })n_{3}^{[0]}dy,  \label{sol2bgt}
\end{eqnarray}%
where all polarization functions $\eta _{\widehat{\alpha }}(x,y,p,\theta
^{\prime })$ and $\eta _{\widehat{i}}^{a}(x,y,p,\theta ^{\prime })$ depend
on anisotropic coordinated $p,$ labelled by a parameter $\theta ^{\prime }$
and can be defined in explicit form for any solution of the Geroch equations
(\ref{eq01}) for the vacuum metric (\ref{sol2b}). The new class of solutions
contains the multiples $q(p)$ and $k(p)$ defined respectively by solitonic
and pp--waves and depends on certain integration functions like $n_{\widehat{%
i}}^{[0]}(x,y)$ and integration constant $h_{0}^{2}.$ Such values can
defined by stating an explicit coordinate system and for certain boundary
and initial conditions.

It should be noted that the metric (\ref{sol2bgt}) can not be represented in
the form (\ref{4dst}) because its coefficients do not satisfy the conditions
(\ref{const6}). This is obvious because in our case $\eta _{2}$ and $\eta
_{3}$ may depend on ansiotropic coordinates $p),$ i.e. our ansatz is not
similar to (\ref{dmetric4}) which is necessary for the anholonomic frame
method. Nevertheless, such classes of metrics define exact vacuum solutions
as a consequence of the parametric method. This is the priority to consider
together both methods: we can parametrize different type of transforms by
polarization functions in a unified form and in different cases such
polarizations will be subjected to corresponding type of constraints,
generating anholonomic deformations or parametric transfoms.

\paragraph{Third example: Two parameter nonholonomic solitonic pp--waves
\newline
}

Finally, we give an explicit example of solutions with two parameter $%
(\theta ^{\prime },\theta )$--metrics of type (\ref{pts1}). We begin with
the ansatz metric $\ _{\shortmid }^{\circ }\mathbf{\tilde{g}}_{[2a]}\mathbf{(%
}\theta \mathbf{)}$ (\ref{auxpwc}) having also a parametrization of type (%
\ref{4dst}) with the coefficients subjected to constraints (\ref{const6}) if
the coordinates are parametrized as $x^{2}=x,x^{3}=y,y^{4}=p$ and $y^{5}=v.$
We also consider that the solitonic wave $\phi $ is included as a multiple
in $\eta _{5}$ and that $\kappa =\breve{\kappa}(x,y)k(p)$ is a pp--wave.
This family of vacuum metrics $\ _{\shortmid }^{\circ }\mathbf{\tilde{g}}%
_{[2a]}\mathbf{(}\theta \mathbf{)}$ does not depend on variable $v,$ i.e. it
possess a Killing vector $\partial /\partial v,$ which allows us to apply
the parametric transform as it was described in the previous (second)
example. The resulting two parameter family of solutions, with redefined
polarization functions, is given by
\begin{eqnarray}
\delta s_{[2a]}^{2} &=&-e^{\psi (x,y,\theta )}\left( \overline{\eta }%
_{2}(x,y,p,\theta ^{\prime })dx^{2}+\overline{\eta }_{3}(x,y,p,\theta
^{\prime })dy^{2}\right) - \label{sol2p2} \\
&&2\breve{\kappa}(x,y)k(p)\ \eta _{4}(x,y,p,\theta )\overline{\eta }%
_{4}(x,y,p,\theta ^{\prime })\delta p^{2}  \nonumber  \\
 && +\ \frac{\eta _{5}(x,y,p,\theta )\overline{\eta }_{5}(x,y,p,\theta
^{\prime })}{8\breve{\kappa}(x,y)k(p)}\delta v^{2},  \nonumber \\
\delta p &=&dp +w_{2}(x,y,p,\theta )\overline{\eta }_{2}^{4}(x,y,p,\theta
^{\prime })dx
 +w_{3}(x,y,p,\theta )\overline{\eta }_{3}^{4}(x,y,p,\theta ^{\prime
})dy,\   \nonumber \\
\delta v &=&dv+n_{2}(x,y,\theta )\overline{\eta }_{2}^{5}(x,y,p,\theta
^{\prime })dx+n_{3}(x,y,\theta )\overline{\eta }_{3}^{5}(x,y,p,\theta
^{\prime })dy.  \nonumber
\end{eqnarray}%
The set of multiples in the coefficients are parametrized this way: The
value $\breve{\kappa}(x,y)$ is just that defining an exact vacuum solution
for the primary metric (\ref{auxpw}) stating the first system of Geroch
equations of type (\ref{eq01}). Then we consider the pp--wave component $%
k(p) $ and the solitonic wave included in $\eta _{5}(x,y,p,\theta )$ such
way that the functions $\psi ,\eta _{4,6},w_{2,3}$ and $n_{2,3}$ are
subjected to the condition to define the class of metrics (\ref{auxpwc}).
The metrics are parametrized both by $\theta ,$ following solutions of the
Geroch equations, and by a N--connection splitting with $w_{2,3}$ and $%
n_{2,3},$ all adapted to the corresponding nonholonomic deformation derived
for $g_{2}(\theta )=g_{3}(\theta )=e^{\psi (\theta )}$ and $g_{4}=2\breve{%
\kappa}k\ \eta _{4}$ and $g_{5}=\eta _{5}/8\breve{\kappa}k$ subjected to the
conditions (\ref{const6}). This set of functions also define a new set of
Killing equations (\ref{eq01}), for any metric (\ref{auxpwc}), which, as a
matter of principle, allows to find the ''overlined'' polarizations $%
\overline{\eta }_{\widehat{i}}(\theta ^{\prime })$ and $\overline{\eta }_{%
\widehat{i}}^{a}(\theta ^{\prime }).$ We omit in this work such cumbersome
formulas stating solutions for any particular cases of solutions of the
Geroch equations.

Even the classes of vacuum Einstein metrics (\ref{sol2p2}) depend on certain
classes of general functions (nonholonomic and parametric transform's
polarizatons and integration functions), it is obvious that they define two
parameter nonlinear superpositions of solitonic waves and pp--waves. From
formal point of view, the procedure can be iterated for any finite or
infinite number of $\theta $--parameters not depending on coordinates. We
can construct an infinite number of nonholonomic vacuum states in gravity
constructed from off--diagonal superpositions of nonlinear waves.\footnote{%
this may be very important for investigations in modern quantum gravity}
Like in the ''pure'' Killing case (without nonholonomic deformations) \cite%
{geroch2}, such two transforms do not commute and depend on order of
successive applications. But the nonholonomic deformations not only mix and
relate nonlinearly two different ''Killing'' classes of solutions but
introduce into the formalism new very important and crucial properties. For
instance, the polarization functions can be chosen such ways that the vacuum
solutions will possess noncommutative and/algebroid symmetries even at
classical level, or generalized configurations in order to contain
contributions of torsion, nonmetricity and/or string fields in various
generalized models of string, brane, gauge, metric--affine and
Finsler--Lagrange gravity, such constructions are considered in details in
Refs. \cite{vesnc,valg,vsgg}.

\subsection{Nonholonomic deformations of the Schwarzschild metric}

We shall nonholonomically deform the Schwarzschild metric in order to
construct new classes of generic off--diagonal solutions. There will
analyzed possible physical effects resulting from generic off--diagonal
interactions with solitonic pp--waves and families of such waves generated
by nonholonomic parametric transforms.

\subsubsection{Stationary backgrounds and small deformations}

Following the method outlined in section \ref{ssndm}, we nonholonomically
deform on angular variable $\varphi $ the Schwarzschild type solution (\ref%
{aux1}) into a generic off--diagonal stationary metric. The ansatz is of
type
\begin{eqnarray}
\delta s_{[1]}^{2} &=&\epsilon _{1}d\chi ^{2}-\eta _{2}(\xi )d\xi ^{2}-\eta
_{3}(\xi )r^{2}(\xi )\ d\vartheta ^{2}  \label{sol1} \\
&&-\eta _{4}(\xi ,\vartheta ,\varphi )r^{2}(\xi )\sin ^{2}\vartheta \ \delta
\varphi ^{2}+\eta _{5}(\xi ,\vartheta ,\varphi )\varpi ^{2}(\xi )\ \delta
t^{2},  \nonumber \\
\delta \varphi &=&d\varphi +w_{2}(\xi ,\vartheta ,\varphi )d\xi +w_{3}(\xi
,\vartheta ,\varphi )d\vartheta ,\   \nonumber \\
\delta t &=&dt+n_{2}(\xi ,\vartheta )d\xi +n_{3}(\xi ,\vartheta )d\vartheta ,
\nonumber
\end{eqnarray}%
where we shall use both types of 3D spacial spherical coordinates, $(\xi
,\vartheta ,\varphi )$ or $(r,\vartheta ,\varphi ).$ The nonholonomic
transform generating this off--diagonal metric are defined by $g_{i}=\eta
_{i}\check{g}_{i}$ and $h_{a}=\eta _{a}\check{h}_{a}$ where $(\check{g}_{i},%
\check{h}_{a})$ are given by data (\ref{aux1p}).

\paragraph{Solutions with general nonholonomic polarizations \newline
}

They can be derived as a class of metrics of type (\ref{4ds}) with the
coefficients subjected to the conditions (\ref{cond5}) (in this case for the
ansatz (\ref{sol1}) with coordinates $x^{2}=\xi ,x^{3}=\vartheta
,y^{4}=\varphi ,y^{5}=t).$ The condition (\ref{eq23}) solving (\ref{ep2a}),
in terms of polarization functions, is satisfied if
\begin{equation}
\sqrt{|\eta _{4}|}=h_{0}\sqrt{|\frac{\check{h}_{5}}{\check{h}_{4}}|}\left(
\sqrt{|\eta _{5}|}\right) ^{\ast },  \label{eq23a}
\end{equation}%
where $\check{h}_{a}$ are coefficients stated by the Schwarzschild solution
for the chosen system of coordinates but $\eta _{5}$ can be any function
satisfying the condition $\eta _{5}^{\ast }\neq 0.$ Parametrizations of
solutions like (\ref{lcc2}), with fixed $b_{0}=0,$ when%
\begin{equation}
-h_{0}^{2}(b^{\ast })^{2} = \eta _{4}(\xi ,\vartheta ,\varphi )r^{2}(\xi
)\sin ^{2}\vartheta \ \mbox{\ and \ } \
b^{2} =\eta _{5}(\xi ,\vartheta ,\varphi )\varpi ^{2}(\xi ),
\end{equation}
 will be used in our further considerations.

The polarizations $\eta _{2}$ and $\eta _{3}$ can be taken in a form that $%
\eta _{2}=\eta _{3}r^{2}=e^{\psi (\xi ,\vartheta )},$
\begin{equation*}
\psi ^{\bullet \bullet }+\psi ^{\prime \prime }=0,
\end{equation*}%
defining solutions of (\ref{ep1a}). The solutions of (\ref{ep3a}) and (\ref%
{ep4a}) for vacuum configurations of the Levi--Civita connection are
constructed as those for (\ref{aux31}) as (\ref{aux31b}),
\begin{equation}
w_{2}=\partial _{\xi }(\sqrt{|\eta _{5}|}\varpi )/\left( \sqrt{|\eta _{5}|}%
\right) ^{\ast }\varpi ,\ w_{3}=\partial _{\vartheta }(\sqrt{|\eta _{5}|}%
)/\left( \sqrt{|\eta _{5}|}\right) ^{\ast }
\end{equation}%
and any $n_{2,3}(\xi ,\vartheta )$ for which $n_{2}^{\prime }-n_{3}^{\bullet
}=0.$ For any function $\eta _{5}\sim a_{1}(\xi ,\vartheta )a_{2}(\varphi ),$
the integrability conditions (\ref{aux31}) can be solved in explicit form as
it was discussed in section \ref{ssaux2}.

We conclude that the stationary nonholonomic deformations of the
Sch\-warz\-schild metric are defined by the off--diagonal ansatz
\begin{eqnarray}
\delta s_{[1]}^{2} &=&\epsilon _{1}d\chi ^{2}-e^{\psi }\left( d\xi ^{2}+\
d\vartheta ^{2}\right)
 -h_{0}^{2}\varpi ^{2}\left[ \left( \sqrt{|\eta _{5}|}\right) ^{\ast }%
\right] ^{2}\ \delta \varphi ^{2}+\eta _{5}\varpi ^{2}\ \delta t^{2},  \nonumber
\\
\delta \varphi &=&d\varphi +\frac{\partial _{\xi }(\sqrt{|\eta _{5}|}\varpi )%
}{\left( \sqrt{|\eta _{5}|}\right) ^{\ast }\varpi }d\xi +\frac{\partial
_{\vartheta }(\sqrt{|\eta _{5}|})}{\left( \sqrt{|\eta _{5}|}\right) ^{\ast }}%
d\vartheta ,\   \nonumber \\
\delta t &=&dt+n_{2}d\xi +n_{3}d\vartheta .  \label{sol1a}
\end{eqnarray}%
Such vacuum solutions were constructed mapping a static black hole solution
into Einstein spaces with locally anisotropic backgrounds (on coordinate $%
\varphi )$ defined by an arbitrary function $\eta _{5}(\xi ,\vartheta
,\varphi )$ with $\partial _{\varphi }\eta _{5}\neq 0$, an arbitrary $\psi
(\xi ,\vartheta )$ solving the 2D Laplace equation and certain integration
functions $n_{2,3}(\xi ,\vartheta )$ and integration constant $h_{0}^{2}.$
In general, the solutions from the target set of metrics do not define black
holes and do not describe obvious physical situations. Nevertheless, they
preserve the singular character of the coefficient $\varpi ^{2}$ vanishing
on the horizon of a Schwarzschild black hole. We can also consider a
prescribed physical situation when, for instance, $\eta _{5}$ define 3D, or
2D, solitonic polarizations on coordinates $\xi ,\vartheta ,\varphi ,$ or on
$\xi ,\varphi .$

\paragraph{Solutions with small nonholonomic polarizations \newline
}

\label{sspe}In a more special case, in order to select physically valuable
configurations, it is better to consider decompositions on a small parameter
$0<\varepsilon <1$ in (\ref{sol1a}), when
\begin{eqnarray}
\sqrt{|\eta _{4}|} &=&q_{4}^{\hat{0}}(\xi ,\varphi ,\vartheta )+\varepsilon
q_{4}^{\hat{1}}(\xi ,\varphi ,\vartheta )+\varepsilon ^{2}q_{4}^{\hat{2}%
}(\xi ,\varphi ,\vartheta )...,  \nonumber \\
\sqrt{|\eta _{5}|} &=&1+\varepsilon q_{5}^{\hat{1}}(\xi ,\varphi ,\vartheta
)+\varepsilon ^{2}q_{5}^{\hat{2}}(\xi ,\varphi ,\vartheta )...,  \nonumber
\end{eqnarray}%
where the ''hat'' indices label the coefficients multiplied to $\varepsilon
,\varepsilon ^{2},...$\footnote{%
Of course, this way we construct not an exact solution, but extract from a
class of exact ones (with less clear physical meaning) certain classes
decomposed (deformed) on a small parameter being related to the
Schwarzschild metric.} The conditions (\ref{eq23a}), necessary to generate
an exact solution for the Levi--Civita connection, can are expressed in the
form
\begin{equation}
\varepsilon h_{0}\sqrt{|\frac{\check{h}_{5}}{\check{h}_{4}}|\ }\left( q_{5}^{%
\hat{1}}\right) ^{\ast }=q_{4}^{\hat{0}},\ \varepsilon ^{2}h_{0}\sqrt{|\frac{%
\check{h}_{5}}{\check{h}_{4}}|\ }\left( q_{5}^{\hat{2}}\right) ^{\ast
}=\varepsilon q_{4}^{\hat{1}},...
\end{equation}%
This system can be solved in a form compatible with small decompositions if
we take the integration constant, for instance, to satisfy the condition $%
\varepsilon h_{0}=1$ (choosing a corresponding system of coordinates). For
this class of small deformations, we can prescribe a function $q_{4}^{\hat{0}%
}$ and define $q_{5}^{\hat{1}},$ integrating on $\varphi $ (or inversely,
prescribing $q_{5}^{\hat{1}},$ then taking the partial derivative $\partial
_{\varphi },$ to compute $q_{4}^{\hat{0}}).$ In a similar form, there are
related the coefficients $q_{4}^{\hat{1}}$ and $q_{5}^{\hat{2}}.$ A very
important physical situation is to select the conditions when such small
nonholonomic deformations define rotoid configurations. This is possible,
for instance, if
\begin{equation}
2q_{5}^{\hat{1}}=\frac{q_{0}(r)}{4\mu ^{2}}\sin (\omega _{0}\varphi +\varphi
_{0})-\frac{1}{r^{2}},  \label{aux1sd}
\end{equation}%
where $\omega _{0}$ and $\varphi _{0}$ are constants and the function $%
q_{0}(r)$ has to be defined by fixing certain boundary conditions for
polarizations. In this case, the coefficient before $\delta t^{2}$ is
approximated as
\begin{equation}
\eta _{5}\varpi ^{2}=1-\frac{2\mu }{r}+\varepsilon (\frac{1}{r^{2}}+2q_{5}^{%
\hat{1}}).
\end{equation}%
This coefficient vanishes and defines a small deformation of the
Schwarz\-schild spherical horizon into a an ellipsoidal one (rotoid
configuration) given by
\begin{equation}
r_{+}\simeq \frac{2\mu }{1+\varepsilon \frac{q_{0}(r)}{4\mu ^{2}}\sin
(\omega _{0}\varphi +\varphi _{0})}.
\end{equation}%
Such solutions with ellipsoid symmetry seem to define static black
ellipsoids (they were investigated in details in Refs. \cite{vbe1,vbe2}).
The ellipsoid configurations were proven to be stable under perturbations
and transform into the Schwarzschild solution far away from the ellipsoidal
horizon. This class of vacuum metrics violates the conditions of black hole
uniqueness theorems \cite{heu} because the ''surface'' gravity is not
constant for stationary black ellipsoid deformations. So, we can construct
an infinite number of ellipsoidal locally anisotropic black hole
deformations. Nevertheless, they present physical interest because they
preserve the spherical topology, have the Minkowski asymptotic and the
deformations can be associated to certain classes of geometric spacetime
distorsions related to generic off--diagonal metric terms. Putting $\varphi
_{0}=0,$ in the limit $\omega _{0}\rightarrow 0,$ we get $q_{5}^{\hat{1}%
}\rightarrow 0$ in (\ref{aux1sd}). This allows to state the limits $q_{4}^{%
\hat{0}}\rightarrow 1$ for $\varepsilon \rightarrow 0$ in order to have a
smooth limit to the Schwarzschild solution for $\varepsilon \rightarrow 0.$
Here, one must be emphasized that to extract the spherical static black hole
solution is possible if we parametrize, for instance,
\begin{equation}
\delta \varphi =d\varphi +\varepsilon \frac{\partial _{\xi }(\sqrt{|\eta
_{5}|}\varpi )}{\left( \sqrt{|\eta _{5}|}\right) ^{\ast }\varpi }d\xi
+\varepsilon \frac{\partial _{\vartheta }(\sqrt{|\eta _{5}|})}{\left( \sqrt{%
|\eta _{5}|}\right) ^{\ast }}d\vartheta
\end{equation}%
and
\begin{equation}
\delta t=dt+\varepsilon n_{2}(\xi ,\vartheta )d\xi +\varepsilon n_{3}(\xi
,\vartheta )d\vartheta .
\end{equation}

Certain more special cases can be defined when $q_{5}^{\hat{2}}$ and $%
q_{4}^{\hat{1}}$ (as a consequence) are of solitonic locally anisotropic
nature. In result, such solutions will define small stationary deformations
of the Schwarzschild solution embedded into a background polarized by
anisotropic solitonic waves.

\paragraph{Parametric transforms for nonholonomically deformed
Schwarz\-schild solutions \newline
}

The ansatz (\ref{sol1a}) do not depend on time variable and possess a Killing
vector $\partial /\partial t.$ We can apply the parametric transform and
generate families of new solutions depending on a parameter $\theta .$
Following the same steps as for generating (\ref{sol2bgt}), we construct
\begin{eqnarray}
\delta s_{[1]}^{2} &=&-e^{\psi }\left( \widetilde{\eta }_{2}(\theta )d\xi
^{2}+\ \widetilde{\eta }_{3}(\theta )d\vartheta ^{2}\right) \nonumber \\
&& -h_{0}^{2}\varpi ^{2}\left[ \left( \sqrt{|\eta _{5}|}\right) ^{\ast }%
\right] ^{2}\ \widetilde{\eta }_{4}(\theta )\delta \varphi ^{2}+\eta
_{5}\varpi ^{2}\ \widetilde{\eta }_{5}(\theta )\delta t^{2},  \nonumber \\
\delta \varphi &=&d\varphi +\widetilde{\eta }_{2}^{4}(\theta )\frac{\partial
_{\xi }(\sqrt{|\eta _{5}|}\varpi )}{\left( \sqrt{|\eta _{5}|}\right) ^{\ast
}\varpi }d\xi +\widetilde{\eta }_{3}^{4}(\theta )\frac{\partial _{\vartheta
}(\sqrt{|\eta _{5}|})}{\left( \sqrt{|\eta _{5}|}\right) ^{\ast }}d\vartheta
,\   \nonumber \\
\delta t &=&dt+\widetilde{\eta }_{2}^{5}(\theta )n_{2}(\xi ,\vartheta )d\xi +%
\widetilde{\eta }_{3}^{5}(\theta )n_{3}(\xi ,\vartheta )d\vartheta ,\label{sol1b}
\end{eqnarray}%
where polarizations $\widetilde{\eta }_{\widehat{\alpha }}(\xi ,\vartheta
,\varphi ,\theta )$ and $\widetilde{\eta }_{\widehat{i}}^{a}(\xi ,\vartheta
,\varphi ,\theta )$ are defined by solutions of the Geroch equations (\ref%
{eq01}) for the vacuum metric (\ref{sol1a}). Even this class of metrics does
not satisfy the equations (\ref{ep1a})--(\ref{ep4a}) for an anholonomic
ansatz, they define vacuum exact solutions and we can apply the formalism on
decomposition on a small parameter $\varepsilon $ like we described in
 section \ref{sspe} (one generates not exact solutions, but like in
quantum field theory it can be more easy to formulate a physical
interpretation). For instance, we consider a vacuum background consisting
from solitonic wave polarizations, with components mixed by the parametric
transform, and then to compute nonholonomic deformations of a Schwarzschild
black hole self--consistently imbedded in such a nonperturbative background.

\subsubsection{Exact solutions with anisotropic polarizations on extra
dimension coordinate}

On can be constructed certain classes of exact off--diagonal solutions when
the extra dimension effectively polarizes the metric coefficients and
interaction constants. We take as a primary metric the ansatz (\ref{aux2})
(see the parametrization for coordinates for that quadratic element, with $%
x^{1}=\varphi ,$ $x^{2}=\check{\vartheta},$ $x^{3}=\check{\xi},$ $y^{4}=\chi
,$ $y^{5}=t)$ and consider the off--diagonal target metric
\begin{eqnarray}
\delta s_{[5\chi ]}^{2} &=&-r_{g}^{2}\ d\varphi ^{2}-r_{g}^{2}\ \eta
_{2}(\xi ,\check{\vartheta})d\check{\vartheta}^{2}+\eta _{3}(\xi ,\check{%
\vartheta})\check{g}_{3}(\check{\vartheta})\ d\check{\xi}^{2}   \nonumber  \\
&& +\epsilon _{4}\ \eta _{4}(\xi ,\check{\vartheta},\chi )\delta \chi
^{2}+\eta _{5}(\xi ,\check{\vartheta},\chi )\ \check{h}_{5}\ (\xi ,\check{%
\vartheta})\ \delta t^{2}  \nonumber \\
\delta \chi &=&d\varphi +w_{2}(\xi ,\check{\vartheta},\chi )d\xi +w_{3}(\xi ,%
\check{\vartheta},\chi )d\check{\vartheta},\  \nonumber  \\
\delta t &=&dt+n_{2}(\xi ,\check{\vartheta},\chi )d\xi +n_{3}(\xi ,\check{%
\vartheta},\chi )d\check{\vartheta}.    \label{sol3}
\end{eqnarray}%
The coefficients of this ansatz,
\begin{eqnarray*}
g_{1} &=&-r_{g}^{2},g_{2}=-r_{g}^{2}\ \eta _{2}(\xi ,\check{\vartheta}%
),g_{3}=\eta _{3}(\xi ,\check{\vartheta})\check{g}_{3}(\check{\vartheta}), \\
h_{4} &=&\epsilon _{1}\ \eta _{4}(\xi ,\check{\vartheta},\chi ),h_{5}=\eta
_{5}(\xi ,\check{\vartheta},\chi )\ \check{h}_{5}\ (\xi ,\check{\vartheta})
\end{eqnarray*}%
are subjected to the condition to solve the system of equations (\ref{ep1a}%
)--(\ref{ep4a}) with certain sources (\ref{sdiag}) defined, for instance,
from string gravity by a corresponding ansatz for $H$--fields like (see
formulas (\ref{c01}) and (\ref{c02}) and related explanations in Appendix).

The ansatz (\ref{sol3}) is a particular parametrization, for the mentioned
coordinates (related to spherical coordinates; we prescribe a spherical
topology), see (\ref{gensol1}), with the coefficients computed by formulas (%
\ref{gensol1w}) and (\ref{gensol1n}). The general solution is given by the
data
\begin{equation}
-r_{g}^{2}\ \eta _{2}=\eta _{3}\check{g}_{3}=\exp \psi (\xi ,\check{\vartheta%
}),  \label{auxx1}
\end{equation}%
where $\psi $ is the solution of
\begin{equation}
\psi ^{\bullet \bullet }+\psi ^{\prime \prime }=2\Upsilon _{4}(\xi ,\check{%
\vartheta}),
\end{equation}%
\begin{eqnarray}
\ \eta _{4}\ &=&h_{0}^{2}(\xi ,\check{\vartheta})\left[ f^{\ast }(\xi ,%
\check{\vartheta},\chi )\right] ^{2}|\varsigma (\xi ,\check{\vartheta},\chi
)|\   \label{auxx2} \\
\eta _{5}\ \check{h}_{5}\ &=&\left[ f(\xi ,\check{\vartheta},\chi
)-f_{0}(\xi ,\check{\vartheta})\right] ^{2},  \nonumber
\end{eqnarray}%
where
\begin{eqnarray*}
\varsigma (\xi ,\check{\vartheta},\chi ) &=&\varsigma _{\lbrack 0]}(\xi ,%
\check{\vartheta}) \\
&&-\frac{\epsilon _{4}}{8}h_{0}^{2}(\xi ,\check{\vartheta})\int \Upsilon
_{2}(\xi ,\check{\vartheta},\chi )f^{\ast }(\xi ,\check{\vartheta},\chi )%
\left[ f(\xi ,\check{\vartheta},\chi )-f_{0}(\xi ,\check{\vartheta})\right]
d\chi .
\end{eqnarray*}%
The N--connection coefficients $N_{i}^{4}=w_{i}(\xi ,\check{\vartheta},\chi
) $ and $N_{i}^{5}=n_{i}(\xi ,\check{\vartheta},\chi )$ are computed
following the formulas
\begin{equation}
w_{\widehat{i}}=-\frac{\partial _{\widehat{i}}\varsigma (\xi ,\check{%
\vartheta},\chi )}{\varsigma ^{\ast }(\xi ,\check{\vartheta},\chi )}
\label{auxx3}
\end{equation}%
and
\begin{equation}
n_{\widehat{k}}=n_{\widehat{k}[1]}(\xi ,\check{\vartheta})+n_{\widehat{k}%
[2]}(\xi ,\check{\vartheta})\int \frac{\left[ f^{\ast }(\xi ,\check{\vartheta%
},\chi )\right] ^{2}}{\left[ f(\xi ,\check{\vartheta},\chi )-f_{0}(\xi ,%
\check{\vartheta})\right] ^{3}}\varsigma (\xi ,\check{\vartheta},\chi )d\chi
.  \label{auxx4}
\end{equation}

The solutions depend on arbitrary nontrivial functions $f(\xi ,\check{%
\vartheta},\chi )$ (with $f^{\ast }\neq 0),$ $f_{0}(\xi ,\check{\vartheta}),$
$h_{0}^{2}(\xi ,\check{\vartheta}),$ $\ \varsigma _{\lbrack 0]}(\xi ,\check{%
\vartheta}),$ $n_{k[1]}(\xi ,\check{\vartheta})$ and $\ n_{k[2]}(\xi ,\check{%
\vartheta}),$ and sources $\Upsilon _{2}(\xi ,\check{\vartheta},\chi
),\Upsilon _{4}(\xi ,\check{\vartheta}).$ Such values have to be defined by
certain boundary conditions and physical considerations. For instance, if
the sources are taken for a cosmological constant $\lambda _{H}$ induced
from string gravity, we have to put $\Upsilon _{2}=\Upsilon _{4}=-\lambda
_{H}^{2}/2$ into above formulas for $\ \eta _{4},\varsigma ,w_{\widehat{i}}$
and $n_{\widehat{i}}.$ In general, we can consider arbitrary matter field
sources (with locally anisotropic presure, mass density, ...) with
nontrivial components of type $\Upsilon _{2}(\xi ,\check{\vartheta},\chi )$
and $\Upsilon _{4}(\xi ,\check{\vartheta})$ stated with respect to the
locally N--adapted basis. In the sourceless case, $\varsigma _{\lbrack
0]}\rightarrow 1;$\ for the Levi--Civita connection, we have to consider $%
h_{0}^{2}(\xi ,\check{\vartheta})\rightarrow const$, in order to satisfy the
conditions (\ref{eq23}), and have to prescribe the integration functions of
type $n_{\widehat{k}[2]}=0$ and $n_{\widehat{k}[1]}$ solving the equation $%
\partial _{\check{\vartheta}}n_{2[1]}=\partial _{\xi }n_{3[1]},$ in order to
satisfy the conditions (\ref{aux31}) and (\ref{aux31b}).

The class of solutions (\ref{sol3}) define self--consistent nonholonomic
maps of the Schwarzschild solution into a 5D backgrounds with nontrivial
sources, depending on a general function $f(\xi ,\check{\vartheta},\chi )$
and mentioned integration functions and constants. Fixing $f(\xi ,\check{%
\vartheta},\chi )$ to be a 3D soliton (we can consider also solitonic
pp--waves as in previous sections) running on extra dimension $\chi ,$ we
describe self-consisted embedding of the Schwarzschild solutions into
nonlinear wave 5D curved spaces. In general, it is not clear if any target
solutions preserve the black hole character of primary solution. It is
necessary a very rigorous analysis of geodesic configurations on such
spacetimes, definition of horizons, singularities and so on. Nevertheless,
for small nonholonomic deformations (by introducing a small parameter $%
\varepsilon ,$ like in the section \ref{sspe}), we can select classes of
''slightly'' deformed solutions preserving the primary black hole character.
In 5D, such solutions are not subjected to the conditions of black hole
uniqueness theorems.

The ansatz (\ref{sol3}) possesses two Killing vector symmetry, $\partial
/\partial t$ and $\partial /\partial \varphi .$ In the sourceless case, we
can apply the parametric transform and generate new families depending on a
parameter $\theta ^{\prime }.$ The constructions are similar to those
generating (\ref{sol1b}) (we omit here such details). Finally, we emphasize
that we can not apply the parametric transform to the primary metric (\ref%
{aux2}) (it is not a vacuum solution) in order to generate families of
parametrized solutions with the aim to subject them to further anholonomic
transforms.

\subsubsection{5D solutions with nonholonomic time like coordinate}

We use the primary metric (\ref{aux3}) (which is not a vacuum solution and
does not admit parametric transforms but can be nonholonomically deformed)
resulting in a target off--diagonal ansatz,
\begin{eqnarray}
\delta s_{[3t]}^{2} &=&-r_{g}^{2}\ d\varphi ^{2}-r_{g}^{2}\eta _{2}(\xi ,%
\check{\vartheta})\ d\check{\vartheta}^{2}+\eta _{3}(\xi ,\check{\vartheta})%
\check{g}_{3}(\check{\vartheta})\ d\check{\xi}^{2}   \nonumber \\
&& +\eta _{4}(\xi ,\check{\vartheta},t)\ \check{h}_{4}\ (\xi ,\check{\vartheta%
})\ \delta t^{2}+\epsilon _{5}\ \eta _{5}(\xi ,\check{\vartheta},t)\ \delta
\chi ^{2},  \nonumber \\
\delta t &=&dt+w_{2}(\xi ,\check{\vartheta},t)d\xi +w_{3}(\xi ,\check{%
\vartheta},t)d\check{\vartheta}, \nonumber  \\
\delta \chi &=&d\varphi +n_{2}(\xi ,\check{\vartheta},t)d\xi +n_{3}(\xi ,%
\check{\vartheta},t)d\check{\vartheta},\   \label{sol4}
\end{eqnarray}%
where the local coordinates are established $x^{1}=\varphi ,\ x^{2}=\check{%
\vartheta},\ x^{3}=\check{\xi},\ y^{4}=t,\ y^{5}=\chi $ and the polarization
functions and coefficients of the N--connection are chosen to solve the
system of equations (\ref{ep1a})--(\ref{ep4a}). Such solutions are generic
5D and emphasize the anisotropic dependence on time like coordinate $t.$ The
coefficients can be computed by the same formulas (\ref{auxx1})--(\ref{auxx4}%
) as in the previous section, for the ansatz (\ref{sol3}), by changing the
coordinate $t$ into $\chi $ and, inversely, $\chi $ into $t.$ This class of
solutions depends on a function $f(\xi ,\check{\vartheta},t), $ with $%
\partial _{t}f\neq 0,$ and on integration functions (depending on $\xi $ and
$\check{\vartheta}$) and constants. We can consider more particular physical
situations when $f(\xi ,\check{\vartheta},t)$ defines a 3D solitonic wave,
or a pp--wave, or their superpositions, and analyze configurations when a
Schwarzschild black hole is self--consistently embedded into a dynamical 5D
background. We analyzed certain similar physical situations in Ref. \cite%
{vs1} when an extra dimension soliton ''running'' away a 4D black hole.

The set of 5D solutions (\ref{sol4}) also posseses two Killing vector
symmetry, $\partial /\partial t$ and $\partial /\partial \chi ,$ like in the
previous section, but with another types of vectors. For the vacuum
configurations, it is possible to perform a 5D parametric transform and to
generate parametric (on $\theta ^{\prime }$) 5D solutions (labelling, for
instance, packages of nonlinear waves).

\subsubsection{Dynamical anholonomic deformations of Schwarzschild me\-trics}

As a primary metric we use the ansatz (\ref{aux4}) by eliminating the extra
dimension term $\epsilon _{1}d\chi ^{2}$ and, firstly, subject it to a
parametric transform with parameter $\theta $ (which is allowed because the
primary metric is a vacuum black hole solution described in terms of
coordinates $x^{2}=\xi ,$ $x^{3}=\vartheta ,$ $y^{4}=t,$ $y^{5}=\varphi )$
and, secondly, nonholonomically deform the family of solutions. This results
in an ansatz of type
\begin{eqnarray}
\delta s_{[4d]}^{2} &=&-\check{\eta}_{2}(\xi ,\vartheta ,\theta )\eta
_{2}(\xi ,\vartheta )d\xi ^{2}-r^{2}(\xi )\ \check{\eta}_{3}(\xi ,\vartheta
,\theta )\eta _{3}(\xi ,\vartheta )d\vartheta ^{2} +\varpi ^{2}(\xi )\
 \nonumber \\
&&\check{\eta}_{4}(\xi ,\vartheta ,\theta )\eta _{4}(\xi
,\vartheta ,t)\delta t^{2}
-\check{\eta}_{5}(\xi ,\vartheta ,\theta )\eta _{5}(\xi ,\vartheta
,t)r^{2}(\xi )\sin ^{2}\vartheta \ \delta \varphi ^{2},  \nonumber \\
\delta t &=&dt+\check{\eta}_{2}^{4}(\xi ,\vartheta ,\theta )w_{2}(\xi
,\vartheta ,t)d\xi +\check{\eta}_{3}^{4}(\xi ,\vartheta ,\theta )w_{3}(\xi
,\vartheta ,t)d\vartheta ,  \nonumber \\
\ \delta \varphi &=&d\varphi +\check{\eta}_{2}^{5}(\xi ,\vartheta ,\theta
)n_{2}(\xi ,\vartheta )d\xi +\check{\eta}_{3}^{5}(\xi ,\vartheta ,\theta
)n_{3}(\xi ,\vartheta )d\vartheta .  \label{sol5}
\end{eqnarray}%
The polarization functions $\check{\eta}_{\widehat{\alpha }}$ and $\check{%
\eta}_{\widehat{i}}^{a}$ are defined by a solution of the Geroch equations (%
\ref{eq01}) for the Schwarzschild solution. We do not consider such
solutions in explicit form, but emphasize that because the coefficients of
the metric (\ref{aux4}) depend only on two coordinates $\xi $ and $\vartheta
,$ the polarizations for the parametric transforms also depend on such two
coordinates and on parameter $\theta .$ The polarizations $\eta _{\widehat{%
\alpha }}$ and coefficients $w_{\widehat{i}}$ and $n_{\widehat{i}}$ have to
be defined as the coefficients of the metric (\ref{sol5}) will generate new
classes of vacuum solutions depending both on $\theta $ and integration
functions, like we considered for the general metric (\ref{4dst}) with the
coefficients subjected to the conditions (\ref{const6}). We relate both
cases, if we take $\eta _{\widehat{i}}$ such that
\begin{equation}
\check{\eta}_{2}(\xi ,\vartheta ,\theta )\eta _{2}(\xi ,\vartheta
)=r^{2}(\xi )\ \check{\eta}_{3}(\xi ,\vartheta ,\theta )\eta _{3}(\xi
,\vartheta )=e^{\psi (\xi ,\vartheta ,\theta )}  \label{aux5s3}
\end{equation}%
for $\psi $ being a solution of $\psi ^{\bullet \bullet }+\psi ^{\prime
\prime }=0,$ and than define $\eta _{a}$ to have
\begin{eqnarray*}
\varpi ^{2}(\xi )\ \check{\eta}_{4}(\xi ,\vartheta ,\theta )\eta _{4}(\xi
,\vartheta ,t) &=&h_{0}^{2}\ \left[ b^{\ast }(\xi ,\vartheta ,t,\theta )%
\right] ^{2}, \\
\check{\eta}_{5}(\xi ,\vartheta ,\theta )\eta _{5}(\xi ,\vartheta
,t)r^{2}(\xi )\sin ^{2}\vartheta &=&\left[ b(\xi ,\vartheta ,t,\theta
)-b_{0}(\xi ,\vartheta ,\theta )\right] ^{2}.
\end{eqnarray*}%
The N--connection coefficients have to satisfy the constraints
\begin{eqnarray}
(\check{\eta}_{2}^{4}\ w_{2})^{\prime }-(\check{\eta}_{3}^{4}\
w_{3})^{\bullet }+(\check{\eta}_{3}^{4}\ w_{3})(\check{\eta}_{2}^{4}\
w_{2})^{\ast }-(\check{\eta}_{2}^{4}\ w_{2})(\check{\eta}_{3}^{4}\
w_{3})^{\ast } &=&0,  \label{aux5s2} \\
\left( \check{\eta}_{2}^{5}n_{2}\right) ^{\prime }-(\check{\eta}%
_{3}^{5}n_{3})^{\bullet } &=&0,  \nonumber
\end{eqnarray}%
for
\begin{equation}
\check{\eta}_{2}^{4}w_{2}=\left( b^{\ast }\right) ^{-1}(b+b_{0})^{\bullet }%
\mbox{ and }\check{\eta}_{3}^{4}w_{3}=\left( b^{\ast }\right)
^{-1}(b+b_{0})^{\prime },  \label{aux5s1}
\end{equation}%
where, for instance, $(\check{\eta}_{2}^{4}\ w_{2})^{\prime }=\partial
_{\vartheta }(\check{\eta}_{2}^{4}\ w_{2}),$ $(\check{\eta}_{3}^{4}\
w_{3})^{\bullet }=\partial _{\xi }(\check{\eta}_{3}^{4}\ w_{3})$ and $(%
\check{\eta}_{3}^{4}\ w_{3})^{\ast }=\partial _{t}(\check{\eta}_{3}^{4}\
w_{3}).$ We note that the functions (\ref{aux5s1}) satisfy the integrability
condition (\ref{aux5s2}) in explicitly form if $b_{0}=0$ and $b(\xi
,\vartheta ,t,\theta )$ can be parametrized in the form $b\sim b(\xi
,\vartheta ,\theta )\widetilde{b}(t)$ ( it was discussed in section \ref%
{ssaux2}).

The set of vacuum solutions (\ref{aux4}) with the coefficients satisfying
the conditions (\ref{aux5s3})--(\ref{aux5s1}) contains two families of
generating functions \newline $b(\xi ,\vartheta ,t,\theta )$ and $\psi (\xi
,\vartheta ,\theta )$ and certain integration functions (roughly speaking,
such solutions define self--consistent embedding of Schwarzschild black
holes into nontrivial backgrounds labelled by parameter $\theta $). The
solutions possess spherical symmetry and the coefficient $h_{4}(\theta )$
vanishes on the horizon of the primary black hole solution but, in general,
they do not define off--diagonal black hole solutions. It is possible to
prescribe a physical situation describing nonlinear interactions with
solitonic pp--waves if $b(\xi ,\vartheta ,t,\theta )$ is considered to be a
package of solutions of such wave equations. In more special cases, by
considering decompositions on a small parameter $\varepsilon $ (as we
discussed in section \ref{sspe}), we can treat such metrics as Schwarzschild
black holes in a background of small perturbations by solitonic pp--waves of
the Minkowski spacetime. For certain configurations, we can say that we
consider propagation of packages of small locally anisotropic solitonic
pp--waves in a Schwarzschild background.

Finally, we note that the ansatz (\ref{sol5}) posseses a Killing vector
symmetry because it does not depend on coordinate $\varphi .$ We can perform
another transform parametrized by a second parameter $\theta ^{\prime },$
resulting in two--parameter families of exact solutions, like we discussed
in general form deriving the transform (\ref{pts1}) and constructing the set
of solutions (\ref{sol2p2}). Such vacuum solutions are parametrized by this
type ansatz:
\begin{eqnarray*}
\delta s_{[4b]}^{2} &=&-\overline{\eta }_{2}(\xi ,\vartheta ,t,\theta
^{\prime })\check{\eta}_{2}(\xi ,\vartheta ,\theta )\eta _{2}(\xi ,\vartheta
)d\xi ^{2}   \\
&&-r^{2}(\xi )\ \overline{\eta }_{3}(\xi ,\vartheta ,t,\theta ^{\prime })%
\check{\eta}_{3}(\xi ,\vartheta ,\theta )\eta _{3}(\xi ,\vartheta
)d\vartheta ^{2} \\
&&+\varpi ^{2}(\xi )\ \overline{\eta }_{4}(\xi ,\vartheta ,t,\theta ^{\prime
})\check{\eta}_{4}(\xi ,\vartheta ,\theta )\eta _{4}(\xi ,\vartheta
,t)\delta t^{2}   \\
&& -\overline{\eta }_{5}(\xi ,\vartheta ,t,\theta ^{\prime })\check{\eta}%
_{5}(\xi ,\vartheta ,\theta )\eta _{5}(\xi ,\vartheta ,t)r^{2}(\xi )\sin
^{2}\vartheta \ \delta \varphi ^{2}, \\
\delta t &=&dt+\overline{\eta }_{2}^{4}(\xi ,\vartheta ,t,\theta ^{\prime })%
\check{\eta}_{2}^{4}(\xi ,\vartheta ,\theta )w_{2}(\xi ,\vartheta ,t)d\xi
\\
&& +\overline{\eta }_{3}^{4}(\xi ,\vartheta ,t,\theta ^{\prime })\check{\eta}%
_{3}^{4}(\xi ,\vartheta ,\theta )w_{3}(\xi ,\vartheta ,t)d\vartheta , \\
\ \delta \varphi &=&d\varphi +\overline{\eta }_{2}^{5}(\xi ,\vartheta
,t,\theta ^{\prime })\check{\eta}_{2}^{5}(\xi ,\vartheta ,\theta )n_{2}(\xi
,\vartheta )d\xi    \\
 &&
 +\overline{\eta }_{3}^{5}(\xi ,\vartheta ,t,\theta ^{\prime })\check{\eta}%
_{3}^{5}(\xi ,\vartheta ,\theta )n_{3}(\xi ,\vartheta )d\vartheta .
\end{eqnarray*}%
The ''overlined'' polarization functions $\overline{\eta }_{\widehat{\alpha }%
}(\theta ^{\prime })$ and $\overline{\eta }_{\widehat{i}}^{a}$ $(\theta
^{\prime })$ can be computed in explicit form (solving a system of algebraic
equations) for any solution of the Geroch equations (\ref{eq01}) for the
ansatz (\ref{sol5}). The generated vacuum Einstein solutions may define two
parameter nonlinear solitonic pp--wave interactions with nonholonomic
deformations of a primary Schwarzschild background. For small amplitudes of
waves, using decompositions on small parameters, we can say that the black
hole character of solutions is preserved but packages of nonlinear waves
define certain off--diagonal interactions self--consistently propagating in
a spacetime of spherical topology.

\section{Discussion}

In this work, we have developed an unified geometric approach to
constructing exact solutions in gravity following superpositions of the
parametric and anholonomic frame transforms. This provides a method for
generating and classifying exact off--diagonal solutions in vacuum Einstein
gravity and in higher dimensional theories of gravity. A classification of
solutions is possible in terms of oriented chains of nonholonomic parametric
maps.

Following the Geroch ideas, the scheme can be elaborated to be iterative on
certain $\theta $--parameters. The techniques being generalized with
nonholonomic transforms states a number of possibilities to construct
''target'' families of exact solutions starting with primary metrics not
subjected to the conditions to solve the Einstein equations. The new classes
of solutions depend on sets of integration functions and constants resulting
from the procedure of integrating systems of partial differential equations
to which the field equations are reduced for certain off--diagonal metric
ansatz and generalized connections. Constraining the integral varieties, for
a corresponding subset of integration functions, the target solutions are
determined to define vacuum Einstein spacetimes.

The freedom in the choice of integration functions considered in this paper
is a universal intrinsic feature of generic off--diagonal solutions
depending on three/ four coordinates in vacuum and nonvacuum gravity. For
diagonalizable ansatz depending on one coordinate (for instance, an ansatz
depending on radial coordinate and generating the Schwarzschild solution),
the Einstein equations are transformed effectively into an ordinary
nonlinear differential equation which can be solved in general form and
contains integration constants. The physical meaning of such constants is
defined from certain prescribed spherical topology and asymptotic conditions
(to get the Newton potential for large distances and embedding into the
Minkowski spacetime). For more general off--diagonal ansatz, it is a very
difficult task to elaborate general principles for generating solutions of
the gravitational and matter field equations with clear physical
significance. Such generalized solutions depend on different classes of
integration functions and constants and can  only be tested if certain
physical situations (with prescribed topology and symmetries) can be
extracted.

Having represented the parametrized transforms as matrices of vielbein maps,
it is possible to answer a lot of questions about geometric and physical
properties of new generated classes of solutions (Is a solution
asymptotically flat? Static or stationary? Deformed to an ellipsoidal
configuration? Defines interactions with nonlinear waves? There are possible
singularities and/or horizons? Can be nontrivial generalizations to extra
dimensions?...) even the Geroch equations are not solved in explicit form.
For instance, in ref. \cite{geroch2}, for certain cases of spacetimes with
two commuting Killing vectors, the parametric transforms are labelled by
some sets of curves and boundary conditions on a hypersurface. It is
possible to define an iteration procedure on $\theta$--parameters, and to
generate an infinite set of new solutions. All such parametric type
solutions can parametrized as certain multiples in "gravitational"
polarizations like in the anholonomic frame method but subjected to other
type of constraints. As a result, we can analyze if a solitonic pp--wave
configuration can be generated (or not) following certain superpositions of
the parametric transforms and nonhlonomic deformations (for instance, of a
black hole background).

Of course, it is not only general formulas and description of possible
physical implications of exact solutions which are of interest in gravity
theories. The unified version of the parametric and anholonomic frame
methods helps to understand more deeply the structure of the gravitational
and matter field equations, to define new generalized symmetries of
nonlinear gravitational field interactions and to consider their nonlinear
superposition as solitonic pp--wave packages, or (in particular cases) as
small self--consistent deformations. Alternatively, one can take the
viewpoint that some prescribed topological and geometrical configurations
are fundamental, so that nonlinear wave deformations being somehow
self--consistently created out of fundamental gravitational fields and
superpositions of nonlinear waves parametrized by certain parameters and
classes of integration functions.

\section*{Acknowledgments}
 The work is performed during a visit at
Fields Institute and Brock University, Canada.

\appendix

\setcounter{equation}{0} \renewcommand{\theequation}
{A.\arabic{equation}} \setcounter{subsection}{0}
\renewcommand{\thesubsection}
{A.\arabic{subsection}}

 \section{The Einstein Equations for d--Con\-nec\-ti\-ons}

The coefficients of curvature (\ref{curva}), $\mathbf{R}_{\ \beta \gamma
\tau }^{\alpha }=(R_{\ hjk}^{i},R_{\ bjk}^{a},P_{\ jka}^{i},P_{\ bka}^{c},$ $%
S_{\ jbc}^{i},$ $S_{\ bcd}^{a}),$ i.e. d--curvatures, of a
d--con\-nec\-ti\-on $\mathbf{\Gamma }_{\alpha \beta }^{\gamma }$ with the
coefficients (\ref{candcon}), defined by a d--metric (\ref{ans5d}), can be
computed following a N--adapted differential form calculus, see 2--form (\ref%
{curva}), with respect to (\ref{dder}) (with $e_{k}=\delta /\partial x^{k}$)
and (\ref{ddif}),
\begin{eqnarray}
R_{\ hjk}^{i} &=&\frac{\delta L_{.hj}^{i}}{\partial x^{k}}-\frac{\delta
L_{.hk}^{i}}{\partial x^{j}}%
+L_{.hj}^{m}L_{mk}^{i}-L_{.hk}^{m}L_{mj}^{i}-C_{.ha}^{i}\Omega _{.jk}^{a},
\label{dcurv} \\
R_{\ bjk}^{a} &=&\frac{\delta L_{.bj}^{a}}{\partial x^{k}}-\frac{\delta
L_{.bk}^{a}}{\partial x^{j}}%
+L_{.bj}^{c}L_{.ck}^{a}-L_{.bk}^{c}L_{.cj}^{a}-C_{.bc}^{a}\ \Omega
_{.jk}^{c},  \nonumber \\
P_{\ jka}^{i} &=&\frac{\partial L_{.jk}^{i}}{\partial y^{a}}-\left( \frac{%
\partial C_{.ja}^{i}}{\partial x^{k}}%
+L_{.lk}^{i}C_{.ja}^{l}-L_{.jk}^{l}C_{.la}^{i}-L_{.ak}^{c}C_{.jc}^{i}\right)
+C_{.jb}^{i}P_{.ka}^{b},  \nonumber \\
P_{\ bka}^{c} &=&\frac{\partial L_{.bk}^{c}}{\partial y^{a}}-\left( \frac{%
\partial C_{.ba}^{c}}{\partial x^{k}}+L_{.dk}^{c%
\,}C_{.ba}^{d}-L_{.bk}^{d}C_{.da}^{c}-L_{.ak}^{d}C_{.bd}^{c}\right)
+C_{.bd}^{c}P_{.ka}^{d},  \nonumber \\
S_{\ jbc}^{i} &=&\frac{\partial C_{.jb}^{i}}{\partial y^{c}}-\frac{\partial
C_{.jc}^{i}}{\partial y^{b}}+C_{.jb}^{h}C_{.hc}^{i}-C_{.jc}^{h}C_{hb}^{i},
\nonumber \\
S_{\ bcd}^{a} &=&\frac{\partial C_{.bc}^{a}}{\partial y^{d}}-\frac{\partial
C_{.bd}^{a}}{\partial y^{c}}+C_{.bc}^{e}C_{.ed}^{a}-C_{.bd}^{e}C_{.ec}^{a}.
\nonumber
\end{eqnarray}%
Details of such computations are given in Refs. \cite{vsgg,vesnc,valg,ma}.

The Ricci tensor
\begin{equation}
\mathbf{R}_{\alpha \beta }\doteqdot \mathbf{R}_{\ \alpha \beta \tau }^{\tau }
\end{equation}%
is characterized by four d--tensor components $\mathbf{R}_{\alpha \beta
}=(R_{ij},R_{ia},R_{ai},S_{ab}),$ where%
\begin{eqnarray}
R_{ij} &\doteqdot &R_{\ ijk}^{k},\quad R_{ia}\doteqdot -\ ^{2}P_{ia}=-P_{\
ika}^{k},  \label{dricci} \\
R_{ai} &\doteqdot &\ ^{1}P_{ai}=P_{\ aib}^{b},\quad S_{ab}\doteqdot S_{\
abc}^{c}.  \nonumber
\end{eqnarray}%
It should be emphasized that because, in general, $^{1}P_{ai}\neq
~^{2}P_{ia},$ the Ricci d--tensors are non symmetric (this a nonholonmic
frame effect).\ Such a tensor became symmetric with respect to holonomic
vielbeins and for the Levi--Civita connection.

Contracting with the inverse to the d--metric (\ref{ans5d}) in $\mathbf{V}, $
we can introduce the scalar curvature of a d--connection $\mathbf{D,}$
\begin{equation}
{\overleftarrow{\mathbf{R}}}\doteqdot \mathbf{g}^{\alpha \beta }\mathbf{R}%
_{\alpha \beta }\doteqdot R+S,  \label{dscal}
\end{equation}%
where $R\doteqdot g^{ij}R_{ij}$ and $S\doteqdot h^{ab}S_{ab}$ and compute
the Einstein tensor
\begin{equation}
\mathbf{G}_{\alpha \beta }\doteqdot \mathbf{R}_{\alpha \beta }-\frac{1}{2}%
\mathbf{g}_{\alpha \beta }{\overleftarrow{\mathbf{R}}.}  \label{deinst}
\end{equation}%
In the vacuum case, $\mathbf{G}_{\alpha \beta }=0,$ which mean that all
Ricci d--tensors (\ref{dricci}) vanish.

The Einstein equations for the canonical d--connection $\mathbf{\Gamma }_{\
\alpha \beta }^{\gamma }$ (\ref{candcon}),
\begin{equation}
\mathbf{R}_{\alpha \beta }-\frac{1}{2}\mathbf{g}_{\alpha \beta }%
\overleftarrow{\mathbf{R}}=\kappa \mathbf{\Upsilon }_{\alpha \beta },
\label{einsteq}
\end{equation}%
are defined for a general source of matter fields and, for instance,
possible string corrections, $\mathbf{\Upsilon }_{\alpha \beta }.$ It should
be emphasized that there is a nonholonomically induced torsion $\mathbf{T}%
_{\ \alpha \beta }^{\gamma }$ with d--torsions computed by introducing
consequently the coefficients of d--metric (\ref{ans5d}) into (\ref{candcon}%
) and than into formulas (\ref{dtors}). The gravitational field equations (%
\ref{einsteq}) can be decomposed into h-- and v--components following
formulas (\ref{dricci}) and (\ref{dscal}),%
\begin{eqnarray}
R_{ij}-\frac{1}{2}g_{ij}\left( R+S\right) &=&\mathbf{\Upsilon }_{ij},
\label{ep1} \\
S_{ab}-\frac{1}{2}h_{ab}\left( R+S\right) &=&\mathbf{\Upsilon }_{ab},  \nonumber
\\
\ ^{1}P_{ai} &=&\mathbf{\Upsilon }_{ai},  \nonumber \\
\ -\ ^{2}P_{ia} &=&\mathbf{\Upsilon }_{ia}.  \nonumber
\end{eqnarray}%
The vacuum equations, in terms of the Ricci tensor $R_{\ \beta }^{\alpha
}=g^{\alpha \gamma }R_{\gamma \beta },$ are
\begin{equation}
R_{j}^{i}=0,S_{b}^{a}=0,\ ^{1}P_{i}^{a}=0,\ ^{2}P_{a}^{i}=0.  \label{ep2}
\end{equation}%
If the conditions (\ref{fols}), (\ref{coef0}) and (\ref{cond3}) are
satisfied, the equations (\ref{ep1}) and (\ref{ep2}) are equivalent to those
derived for the Levi--Civita connection.

In string gravity the nontrivial torsion components (\ref{dtors}) and source
$\kappa \mathbf{\Upsilon }_{\alpha \beta }$ can be related to certain
effective interactions with the strength (torsion)
\begin{equation}
H_{\mu \nu \rho }=\mathbf{e}_{\mu }B_{\nu \rho }+\mathbf{e}_{\rho }B_{\mu
\nu }+\mathbf{e}_{\nu }B_{\rho \mu }
\end{equation}%
of an antisymmetric field $B_{\nu \rho },$ when%
\begin{equation}
R_{\mu \nu }=-\frac{1}{4}H_{\mu }^{\ \nu \rho }H_{\nu \lambda \rho }
\label{c01}
\end{equation}%
and
\begin{equation}
D_{\lambda }H^{\lambda \mu \nu }=0,  \label{c02}
\end{equation}%
see details on string gravity, for instance, in Refs. \cite{string1,string2}%
. The conditions (\ref{c01}) and (\ref{c02}) are satisfied by the ansatz
\begin{equation}
H_{\mu \nu \rho }=\widehat{Z}_{\mu \nu \rho }+\widehat{H}_{\mu \nu \rho
}=\lambda _{\lbrack H]}\sqrt{\mid g_{\alpha \beta }\mid }\varepsilon _{\nu
\lambda \rho }  \label{ansh}
\end{equation}%
where $\varepsilon _{\nu \lambda \rho }$ is completely antisymmetric and the
distorsion (from the Levi--Civita connection) and
\begin{equation*}
\widehat{Z}_{\mu \alpha \beta }\mathbf{c}^{\mu }=\mathbf{e}_{\beta }\rfloor
\mathcal{T}_{\alpha }-\mathbf{e}_{\alpha }\rfloor \mathcal{T}_{\beta }+\frac{%
1}{2}\left( \mathbf{e}_{\alpha }\rfloor \mathbf{e}_{\beta }\rfloor \mathcal{T%
}_{\gamma }\right) \mathbf{c}^{\gamma }
\end{equation*}%
is defined by the torsion tensor (\ref{torsa}) with coefficients (\ref{tors}%
). We emphasize that our $H$--field ansatz is different from those already
used in string gravity when $\widehat{H}_{\mu \nu \rho }=\lambda _{\lbrack
H]}\sqrt{\mid g_{\alpha \beta }\mid }\varepsilon _{\nu \lambda \rho }.$ \ In
our approach, we define $H_{\mu \nu \rho }$ and $\widehat{Z}_{\mu \nu \rho }$
from the respective ansatz for the $H$--field and nonholonomically deformed
metric, compute the torsion tensor for the canonical distinguished
connection and, finally, define the 'deformed' H--field as $\widehat{H}_{\mu
\nu \rho }=\lambda _{\lbrack H]}\sqrt{\mid g_{\alpha \beta }\mid }%
\varepsilon _{\nu \lambda \rho }-\widehat{Z}_{\mu \nu \rho }.$

\setcounter{equation}{0} \renewcommand{\theequation}
{B.\arabic{equation}} \setcounter{subsection}{0}
\renewcommand{\thesubsection}
{B.\arabic{subsection}}

\section{A Solution for v--Components in Einstein Equations}

We give a new method of constructing the general solution of the equation (%
\ref{ep2a}) for a general non--vanishing source $\Upsilon
_{2}(x^{2},x^{3},v) $ and $h_{5}^{\ast }\neq 0.$\footnote{%
It is more simple than that elaborated in Ref. \cite{valg}.} Introducing the
function
\begin{equation}
\phi (x^{2},x^{3},v)=\ln \left| \frac{h_{5}^{\ast }}{\sqrt{|h_{4}h_{5}|}}%
\right| ,  \label{afa1}
\end{equation}%
we write that equation in the form%
\begin{equation}
\left( \sqrt{|h_{4}h_{5}|}\right) ^{-1}\left( e^{\phi }\right) ^{\ast
}=2\Upsilon _{2}.  \label{afa2}
\end{equation}%
Using (\ref{afa1}), we express $\sqrt{|h_{4}h_{5}|}$ as a function of $\phi $
and $h_{5}^{\ast }$ and obtain%
\begin{equation}
h_{5}^{\ast }=(e^{\phi })^{\ast }/4\Upsilon _{2}
\end{equation}%
which can integrated in general form
\begin{equation}
h_{5}=h_{5[0]}(x^{2},x^{3})+\frac{1}{4}\int dv\ \frac{\left[ e^{2\phi
(x^{2},x^{3},v)}\right] ^{\ast }}{\Upsilon _{2}(x^{2},x^{3},v)},
\label{solh4}
\end{equation}%
where $h_{5[0]}(x^{2},x^{3})$ is the integration function. Having defined $%
h_{5}$ and using again (\ref{afa1}), we can express $h_{4}$ via $h_{5}$ and $%
\phi ,$
\begin{equation}
|h_{4}|=4e^{-2\phi (x^{2},x^{3},v)}\left[ \left( \sqrt{|h_{5}|}\right)
^{\ast }\right] ^{2}.  \label{solh5}
\end{equation}%
The conclusion is that prescribing any two functions $\phi (x^{2},x^{3},v)$
and $\Upsilon _{2}(x^{2},$ $x^{3},$ $\ v)$ we can always find the
corresponding metric coefficients $h_{4}$ and $h_{5}$ solving (\ref{ep2a}).

Finally, we note that if $\Upsilon _{2}=0,$ we can relate $h_{4}$ and $h_{5}$
by solving (\ref{afa2}) as $\left( e^{\phi }\right) ^{\ast }=0.$ Such
solutions can be written, for instance, in the form (\ref{eq23}) and (\ref%
{lcc2}) being defined by an arbitrary function $b(x^{2},x^{3},v),$
integration function $b_{0}(x^{2},x^{3})$ and constant $h_{0}.$

\end{document}